\pdfoutput=1
\documentclass[aps,prd,onecolumn,superscriptaddress,nofootinbib,longbibliography]{revtex4-2}

\usepackage{amsmath}
\usepackage{amssymb}
\usepackage{graphicx}
\usepackage{xcolor}
\usepackage{tikz}
\usetikzlibrary{positioning,arrows.meta,decorations.pathmorphing,decorations.markings,calc,fit,patterns,shapes.geometric}
\usepackage{hyperref}
\hypersetup{colorlinks=true,linkcolor=blue!55!black,citecolor=blue!55!black,urlcolor=blue!55!black}
\usepackage[capitalise]{cleveref}

\IfFileExists{orcidlink.sty}{\usepackage{orcidlink}}{\newcommand{\orcidlink}[1]{}}

\providecommand{\citep}{\cite}
\providecommand{\citet}{\cite}

\newcommand\SU{\mathrm{SU}}
\newcommand\LQCD{\Lambda_\text{QCD}}
\newcommand\minisec[1]{\paragraph*{{#1}---}}

\newcommand{\contribution}[3]{%
  \par\addvspace{0.8ex}%
  \begingroup
    \centering
    {\normalsize #1\par}%
    {\footnotesize\itshape #2\par}%
    \ifx\relax#3\relax\else{\footnotesize\texttt{#3}\par}\fi
  \endgroup
  \addvspace{1.0ex}\par\noindent\ignorespaces}

\NewDocumentCommand{\vb}{s m}{\IfBooleanTF{#1}{\boldsymbol{#2}}{\mathbf{#2}}}
\providecommand{\ket}[1]{\left|#1\right\rangle}
\providecommand{\braket}[1]{\left\langle #1\right\rangle}
\providecommand{\abs}[1]{\left|#1\right|}
\newcommand{\ham}{\hat{H}}
\newcommand{\avg}[1]{\langle#1\rangle}
\newcommand{\hc}{{\rm{h.c.}}}
\newcommand{\spin}{\mathrm{j}}
\newcommand{\jmax}{\spin_{\max}}
\newcommand{\jonehalf}{\hspace{-0.7pt}{\frac12}\hspace{-0.7pt}}
\newcommand{\gaugegroup}{\mathcal{G}}
\newcommand{\casimir}{\hat{E}^{2}}
\newcommand{\NApara}{\hat{U}^{\alpha\beta}}
\newcommand{\hpsi}{\hat{\psi}}
\newcommand{\vecsite}{\vb{n}}
\newcommand{\latvec}[1][]{\vb*{\mu}_{#1}}
\newcommand{\phdag}{\vphantom{dagger}}
\NewDocumentCommand{\lattf}{smmm}{\hat{#2}_{\vb{#3}}^{\IfBooleanTF{#1}{\dagger\,}{\phdag}#4}}
\NewDocumentCommand{\plaq}{s}{\IfBooleanTF{#1}{\lattf*}{\lattf}{U}{\square}{}}

\AtBeginDocument{}

\newcommand{\lqcd}{\Lambda_{\rm QCD}}
\newcommand{\cO}{{\cal O}}
\newcommand{\cE}{{\cal E}}
\providecommand{\bra}[1]{\left\langle #1\right|}

\usepackage{xspace}

\def\beq{\begin{equation}}
\def\enq{\end{equation}}
\def\bpl{\beta_{pl.}}
\def\bl{\beta_{l.}}

\def\dmu{\partial_\mu}

\makeatletter
\newcommand{\appendixcontents}{%
  \begingroup
  \par\addvspace{1ex}
  {\centering\normalsize\bfseries Contents of the appendices\par}
  \addvspace{1ex}
  \footnotesize
  \setlength{\parindent}{0pt}\setlength{\parskip}{0.35ex}
  \appendixcontentslist
  \endgroup
  \par\addvspace{1.5ex}}
\newcommand{\appendixcontentslist}{}
\newcommand{\appendixitem}[3]{%
  \g@addto@macro\appendixcontentslist{%
    \makebox[1.6em][l]{\ref{#1}}\parbox[t]{\dimexpr\linewidth-1.6em-2.5em\relax}{#2\\{\itshape #3}}\hfill\makebox[2.5em][r]{\pageref{#1}}\par}}
\makeatother

\newcommand{\editormark}{\textsuperscript{\(\diamondsuit\)}}

\begin{document}

\begin{figure}
\vskip -1.cm
\leftline{\includegraphics[width=0.15\textwidth]{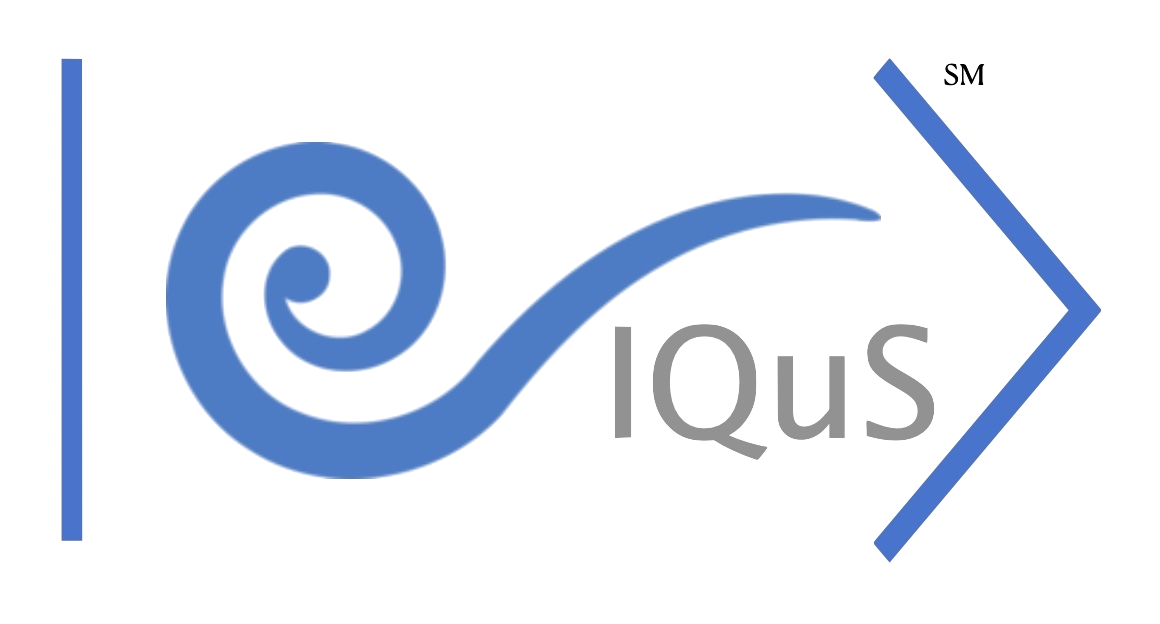}}
\vskip -1.cm
\end{figure}

\title{\includegraphics[width=0.32\linewidth]{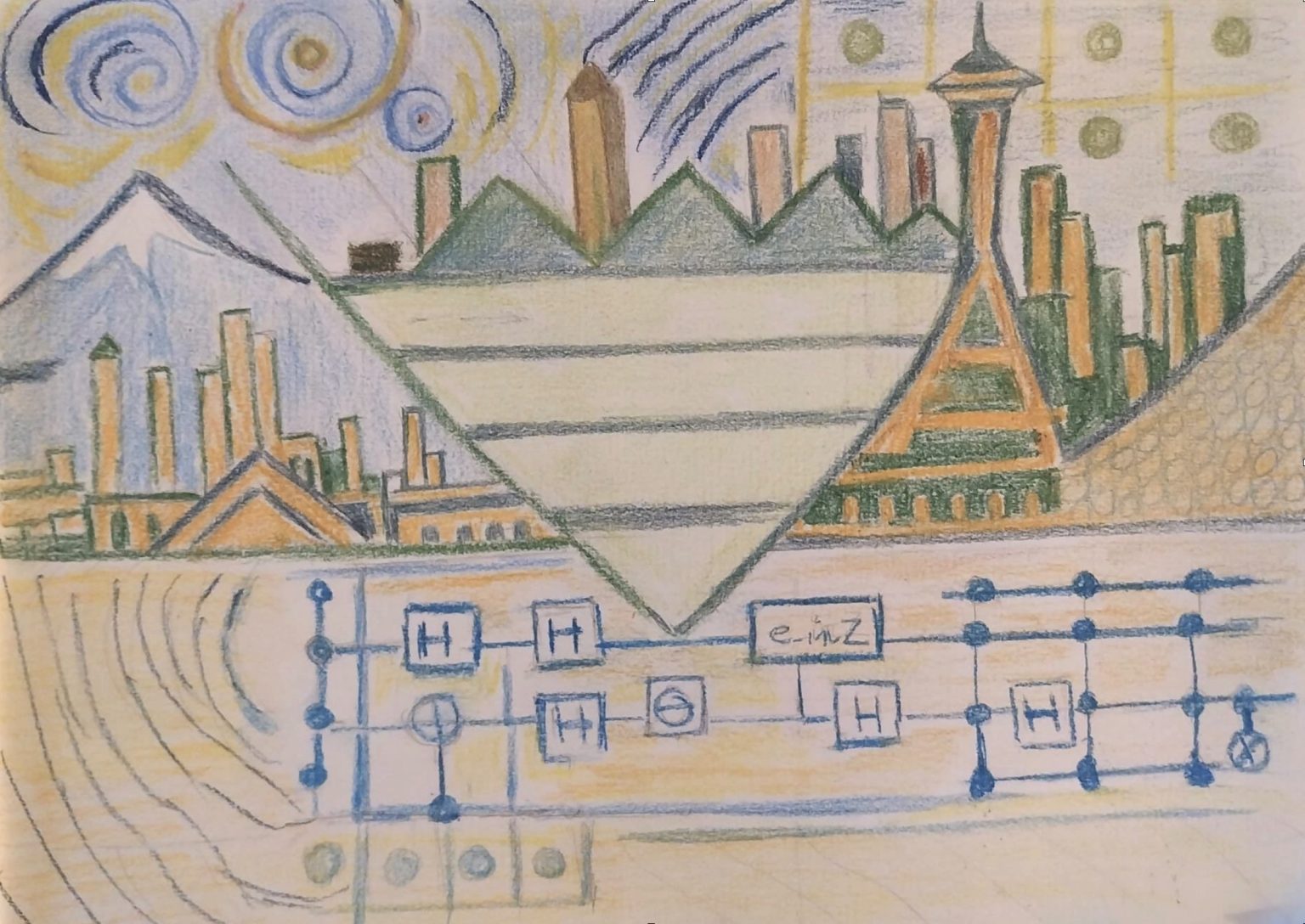}\\[1.5ex]
Quantum Simulation of Hadronization:\\
From String Breaking to Event Generation}

\author{Christian W. Bauer\editormark\,\orcidlink{0000-0001-9820-5810}}
\email{cwbauer@lbl.gov}
\affiliation{Physics Division, Lawrence Berkeley National Laboratory, Berkeley, CA 94720, USA}
\affiliation{Department of Physics, University of California, Berkeley, CA 94720, USA}

\author{Anthony N. Ciavarella\editormark\,\orcidlink{0000-0003-3918-4110}}
\email{anciavarella@lbl.gov}
\affiliation{Applied Math and Computational Research Division, Lawrence Berkeley National Laboratory, Berkeley, California 94720, USA}

\author{Jad C. Halimeh\editormark\,\orcidlink{0000-0002-0659-7990}}
\email{jad.halimeh@lmu.de}
\affiliation{Max Planck Institute of Quantum Optics, 85748 Garching, Germany}
\affiliation{Department of Physics and Arnold Sommerfeld Center for Theoretical Physics (ASC), Ludwig-Maximilians-Universit\"at M\"unchen, 80333 M\"unchen, Germany}
\affiliation{Munich Center for Quantum Science and Technology (MCQST), 80799 M\"unchen, Germany}
\affiliation{Department of Physics, College of Science, Kyung Hee University, Seoul 02447, Republic of Korea}

\author{Yannick Meurice\editormark\,\orcidlink{0000-0002-0995-9694}}
\email{yannick-meurice@uiowa.edu}
\affiliation{Department of Physics and Astronomy, University of Iowa, Iowa City, IA 52242, USA}

\author{Stephen Mrenna\editormark\,\orcidlink{0000-0001-8731-160X}}
\email{mrenna@fnal.gov}
\affiliation{Fermi National Accelerator Laboratory, Batavia, Illinois 60510, USA}

\author{Martin J. Savage\,\orcidlink{0000-0001-6502-7106}}
\affiliation{InQubator for Quantum Simulation (IQuS), Department of Physics, University of Washington, Seattle, WA 98195, USA}

\author{Giovanni Cataldi\,\orcidlink{0000-0002-9073-8978}}
\affiliation{Max Planck Institute of Quantum Optics, 85748 Garching, Germany}
\affiliation{Munich Center for Quantum Science and Technology (MCQST), 80799 M\"unchen, Germany}
\affiliation{Department of Physics and Arnold Sommerfeld Center for Theoretical Physics (ASC), Ludwig-Maximilians-Universit\"at M\"unchen, 80333 M\"unchen, Germany}

\author{Henry Froland\,\orcidlink{0009-0008-4356-0602}}
\affiliation{InQubator for Quantum Simulation (IQuS), Department of Physics, University of Washington, Seattle, WA 98195, USA}

\author{Dorota M. Grabowska\,\orcidlink{0000-0002-0760-4734}}
\affiliation{InQubator for Quantum Simulation (IQuS), Department of Physics, University of Washington, Seattle, WA 98195, USA}

\author{Sebastian Grieninger\,\orcidlink{0000-0002-9523-5819}}
\affiliation{InQubator for Quantum Simulation (IQuS), Department of Physics, University of Washington, Seattle, WA 98195, USA}

\author{Kenneth Heitritter\,\orcidlink{0000-0002-8901-6951}}
\affiliation{Department of Physics and Astronomy, University of Iowa, Iowa City, IA 52242, USA}
\affiliation{Fermi National Accelerator Laboratory, Batavia, Illinois 60510, USA}
\affiliation{qBraid Co.}

\author{Michael Hite}
\affiliation{Department of Physics, University of Arizona, Tucson, AZ 85721, USA}

\author{Marc Illa\,\orcidlink{0000-0003-3570-2849}}
\affiliation{Physical Sciences Division, Pacific Northwest National Laboratory, Richland, WA 99354, USA}

\author{Saurabh V. Kadam\,\orcidlink{0000-0001-9218-1600}}
\affiliation{Lawrence Livermore National Laboratory, Livermore, CA 94550, USA}

\author{Henry Lamm\,\orcidlink{0000-0003-3033-0791}}
\affiliation{Fermi National Accelerator Laboratory, Batavia, Illinois 60510, USA}

\author{Zhiyao Li\,\orcidlink{0000-0002-7614-8496}}
\affiliation{InQubator for Quantum Simulation (IQuS), Department of Physics, University of Washington, Seattle, WA 98195, USA}

\author{Zane Ozzello\,\orcidlink{0009-0003-9773-3808}}
\affiliation{Department of Physics and Astronomy, University of Iowa, Iowa City, IA 52242, USA}

\author{Vincent R. Pascuzzi\,\orcidlink{0000-0003-3167-8773}}
\affiliation{IBM Quantum, IBM T.\,J. Watson Research Center, Yorktown Heights, NY 10598, USA}

\author{Blake Senseman\,\orcidlink{0009-0003-4196-0147}}
\affiliation{Department of Physics and Astronomy, University of Iowa, Iowa City, IA 52242, USA}

\author{Kaidi Xu\,\orcidlink{0000-0003-2184-0829}}
\affiliation{Max Planck Institute of Quantum Optics, 85748 Garching, Germany}
\affiliation{Department of Physics and Arnold Sommerfeld Center for Theoretical Physics (ASC), Ludwig-Maximilians-Universit\"at M\"unchen, 80333 M\"unchen, Germany}
\affiliation{Munich Center for Quantum Science and Technology (MCQST), 80799 M\"unchen, Germany}

\author{Xiaojun Yao\,\orcidlink{0000-0002-8377-2203}}
\affiliation{InQubator for Quantum Simulation (IQuS), Department of Physics, University of Washington, Seattle, WA 98195, USA}

\author{Nikita A. Zemlevskiy\,\orcidlink{0000-0002-0794-2389}}
\affiliation{InQubator for Quantum Simulation (IQuS), Department of Physics, University of Washington, Seattle, WA 98195, USA}
\affiliation{Institute for Quantum Information and Matter, California Institute of Technology}
\affiliation{Department of Physics, California Institute of Technology}

\date{\today}

\begin{abstract}
High-energy collisions produce large numbers of quarks and gluons that are bound into hadrons by the confining interactions of quantum chromodynamics (QCD). This hadronization process is intrinsically real-time and non-perturbative. Event generators describe hadronization using phenomenological models with parameters fitted to data rather than calculated from QCD. Uncertainties in these models limit the precision of certain measurements at the Large Hadron Collider. Quantum computers could enable first-principles simulations of these dynamics. Existing quantum simulations of gauge theories have demonstrated string-breaking dynamics. This white paper summarizes the conclusions of the InQubator for Quantum Simulation workshop \textit{From String Dynamics to Event Generation with Quantum Simulation}. We develop a roadmap for extending these simulations to inform phenomenological models of hadronization. A sequence of toy models is proposed for connecting current quantum simulations to QCD. The observables that should be extracted are listed, and the challenges that separate present simulations from quantitative predictions are identified. We propose a staged program in which quantum simulations first provide qualitative modifications to event generators and later provide quantitative input.

\end{abstract}

\maketitle
{\renewcommand\thefootnote{\(\diamondsuit\)}\footnotetext{Editors of the white paper and organizers of the workshop.}}

\section{Introduction}
\label{sec:intro}

High-energy physics is directly probed at particle colliders where large numbers of quarks and gluons are produced. As these particles propagate, the confining interactions of QCD bind them together into hadrons. This hadronization process is intrinsically real-time and non-perturbative. In event generators, it is described by phenomenological models such as the Lund string model and the cluster model, whose parameters are fit to data~\cite{Andersson:1983ia,Andersson:1997xwk,Webber:1983if,Bierlich:2022pfr,Bellm:2025pcw,Sherpa:2024mfk}. Direct simulations of these dynamics on classical computers suffer from a sign problem and require exponentially scaling resources~\cite{Banuls:2019rao,Banuls:2019bmf}. Quantum computers offer the potential to directly probe these dynamics, and simulations of string breaking in low-dimensional gauge theories have recently been performed on quantum hardware~\cite{De:2024smi,Farrell:2024fit,Davoudi:2024wyv,Ciavarella:2024lsp,Gonzalez-Cuadra:2024xul,Cobos:2025krn,Xu:2026ibi,Joshi:2026hfe,Ciavarella:2026tau}.

The uncertainties of the phenomenological models are now a limiting systematic at the Large Hadron Collider (LHC).
The default \textsc{Pythia} tune depends on $\mathcal{O}(20)$ fitted non-perturbative parameters and under-predicts LHC baryon yields~\cite{Skands:2014pea}.
Charm fragmentation is not universal across collision systems.
The $\Lambda_c^+/D^0$ ratio measured in $pp$ collisions is several times the value expected from $e^+e^-$ tuning, and it is under-predicted by the generator tunes available at the time of the measurement~\cite{ALICE:2017thy}.
The ratio is described once string-junction fragmentation is added, but charm-strange baryon ratios remain under-predicted~\cite{Altmann:2024odn}.
The spread between models is commonly used as an estimate of the hadronization uncertainty, although it does not provide a systematically improvable error bound.
In a recent comparison of three generator tunes, none described the identified-particle transverse-momentum spectra measured in $7~\mathrm{TeV}$ proton-proton collisions at the LHC, and the predictions of the different hadronization models differed by up to $50\%$~\cite{Divisova:2025xqp}.
Retuning the flavor parameters improved one spectrum at the expense of another, which suggests that improvements to the underlying models are required~\cite{Divisova:2025xqp}.
Non-perturbative modeling is one of the dominant systematic uncertainties on the top quark mass, and hadronization limits the extraction of $\alpha_s$ from event shapes~\cite{Divisova:2025xqp}.

Current hadronization models rely on relatively simplistic classical models. In the Lund string model, the color field between separating charges is treated as a one-dimensional flux tube of constant tension $\kappa \approx 1$~GeV/fm. The tube breaks through the production of $q\bar q$ pairs with a Schwinger-like probability proportional to $\exp(-\pi m_\perp^2/\kappa)$~\cite{Schwinger:1951nm,Casher:1978wy}. Here $m_\perp^2=m_q^2+p_\perp^2$ is the squared transverse mass of a quark with mass $m_q$ and transverse momentum $p_\perp$. Successive breaks are assumed to be causally disconnected, so that they can be ordered arbitrarily. Longitudinal momenta are drawn from a simple fragmentation function $f(z)$, with $z$ being the light-cone momentum fraction carried by the hadron. Transverse momenta are drawn from a Gaussian and compensated locally between neighboring hadrons~\cite{Andersson:1983ia,Andersson:1983jt,Andersson:1997xwk,Bierlich:2022pfr}. Each break is assigned a definite energy, momentum, and production vertex, so the description is classical even though it is motivated by quantum mechanics. The cluster model replaces the string with preconfined color-singlet clusters that decay according to phase space~\cite{Webber:1983if,Bellm:2025pcw,Sherpa:2024mfk}. Over time, cluster models have acquired string-like properties and string models have acquired cluster-like properties. The \textsc{Pythia} manual records further limitations. Strangeness suppression is a fitted number, the transverse momentum width exceeds the Schwinger value, hadron spin enters only through fitted suppression factors, and glueballs are absent~\cite{Bierlich:2022pfr}. The Hagedorn temperature is the temperature at which the exponentially growing density of hadronic states makes the partition function of a hadron gas diverge. Thermodynamical fragmentation near this temperature describes identified spectra better in several respects~\cite{Fischer:2016zzs}. A time-dependent tension $\kappa(\tau)$ has been proposed as an extension of the Lund model, but is constrained by the lattice value only in the static, late-time limit~\cite{Hunt-Smith:2020lul}. The rope and shoving models introduced for overlapping flux tubes at LHC densities contain new ingredients that cannot currently be calculated from first principles~\cite{Bierlich:2014xba}. None of these dynamical assumptions has been tested directly against QCD itself.

Testing the models against data rather than against QCD has three structural limitations. A color string cannot be prepared experimentally, so hadronization is never separated from the short-distance process and the parton shower that precede it. Detectors observe only the asymptotic final state, so the dynamics of hadron formation is not directly visible. No rigorous systematic uncertainty can be assigned to a phenomenological model, and the spread between models is the usual substitute. Therefore, additional data alone cannot establish whether the assumptions listed above hold, and it will be necessary to compare directly against QCD.

Performing this comparison requires a first-principles calculation of real-time confining dynamics. Euclidean lattice QCD determines static quantities to high precision, and static string breaking has been observed as a flattening of the static potential in QCD with two and with three dynamical flavors~\cite{Bali:2005fu,Bulava:2019iut}. Some dynamical quantities are also accessible from Euclidean correlators. The finite-volume formalism of L\"uscher gives two-particle scattering amplitudes below inelastic thresholds~\cite{Luscher:1990ux}, and large-momentum effective theory (LaMET) gives parton distributions from equal-time matrix elements in boosted hadrons~\cite{Ji:2013dva,Ji:2020ect}. However, the inclusive real-time evolution of a breaking string is not something that can be achieved with Euclidean lattice QCD. Its extraction requires an analytic continuation that is exponentially ill-conditioned~\cite{Alexandru:2016gsd,Nagata:2021ugx,Banuls:2019rao,Bauer:2025nzf}. Additionally, Euclidean methods do not give the matrix elements of the soft functions defined in soft-collinear effective theory (SCET), which contain several light-like directions and need an inclusive sum over final states~\cite{Bauer:2000ew,Bauer:2000yr,Bauer:2001ct,Bauer:2001yt,Bauer:2002nz,Bauer:2025nzf}. Classical simulation of the real-time quantum state requires resources that scale exponentially with the volume, and tensor-network methods are limited by the growth of entanglement. Jordan, Lee, and Preskill showed that quantum computers can simulate the dynamics of a quantum field theory with resources polynomial in the number of lattice sites~\cite{Jordan:2011ci,Jordan:2012xnu}, and finite density carries no sign problem in a Hamiltonian simulation. Community planning documents identify hadronization and final-state dynamics as targets for quantum simulation~\cite{Bauer:2022hpo,Bauer:2023qgm,Humble:2022vtm}. However, a direct simulation of a collider event is not practical. The lattice spacing must resolve the largest energy scale in the event and the lattice extent must contain the smallest, so the number of sites scales as $n_L \sim (E_{\max}/E_{\min})^3$. This gives $n_L \gtrsim 10^9$ for the Large Electron-Positron Collider (LEP) and $n_L \gtrsim 10^{15}$ for the LHC at $\mathcal{O}(10$--$100)$ qubits per site, and most of those sites would describe physics far above $\Lambda_{\rm QCD}$ that perturbation theory already handles~\cite{Bauer:2021gup,Bauer:2025nzf}. Therefore, the relevant question is which non-perturbative object a quantum computer should compute. The organizing principle adopted in this white paper is to identify the smallest quantum object that is inaccessible to classical methods, and compute it directly rather than produce it in a simulated collision.

This white paper summarizes the conclusions of the InQubator for Quantum Simulation workshop \textit{From String Dynamics to Event Generation with Quantum Simulation}. Section~\ref{sec:generators} reviews the hadronization models used in event generators and the uncertainties of their parameters. Section~\ref{sec:toy} describes the toy models that connect present simulations to QCD and the physics that each of them can provide. Section~\ref{sec:observables} lists the observables that should be extracted. Section~\ref{sec:challenges} identifies the challenges that separate present simulations from quantitative predictions. Section~\ref{sec:program} summarizes the state of hardware and describes a staged program. The individual contributions on which these conclusions are based are reproduced in the appendices.

\section{Hadronization models in event generators}
\label{sec:generators}

\textit{The role of hadronization.} Monte Carlo event generators such as
\textsc{Pythia}~\cite{Sjostrand:2014zea}, \textsc{Herwig}~\cite{Bahr:2008pv}, and
\textsc{Sherpa}~\cite{Sherpa:2019gpd} simulate the full chain of physics in a
high-energy collision: the hard scattering cross section, initial- and
final-state parton showers, multiple parton interactions, and finally the
non-perturbative hadronization of the resulting colored parton ensemble
into observable hadrons. Hadronization is the last and least theoretically
controlled step. However, it is unavoidable, and every observable quantity at a hadron
collider is sensitive to it. Improving hadronization models is therefore
not only scientifically important but computationally urgent. Hadronization
should be treated as interesting science in its own right, not merely a
nuisance absorbed into tuned parameters.

\begin{figure}[!t]
  \centering
  \includegraphics[width=\linewidth]{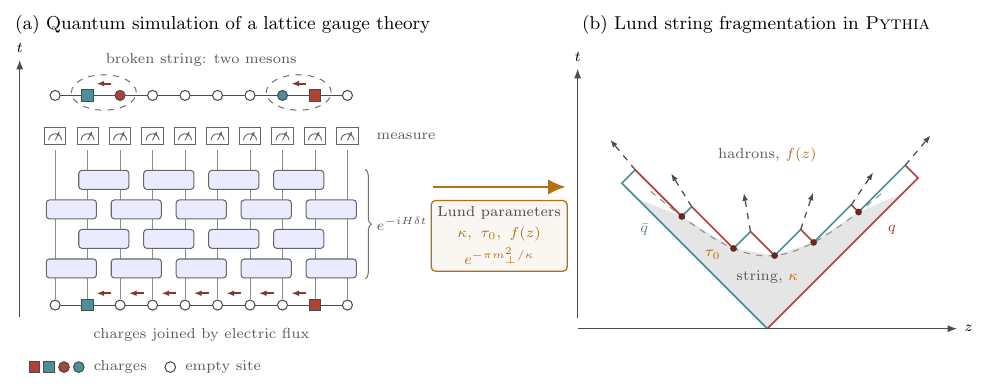}
  \caption{Schematic comparison of string breaking in a quantum simulation and in the
  Lund string model. (a) A lattice gauge theory is mapped onto qubits. The initial state
  contains two charges $Q$ and $\bar Q$ connected by electric flux. The state is
  evolved with a Trotterized circuit and then measured. In the final state, a dynamical
  quark-antiquark pair screens the static charges, the flux between them is absent, and
  two mesons are formed. (b) The Lund model as implemented in \textsc{Pythia}. The
  endpoint quark and antiquark separate along the light cone, and the string between
  them breaks at vertices (dots) near the hyperbola $t^2-z^2=\tau_0^2$ (dashed). Quarks
  from adjacent breaks combine into hadrons. The arrow indicates the proposed use of
  quantum simulations to inform the parameters of the Lund model, which are currently fit
  to data. These are the string tension $\kappa$, the breakup proper time $\tau_0$, the
  fragmentation function $f(z)$, and the tunneling factor of Eq.~(\ref{eq:schwinger}).}
  \label{fig:string-breaking}
\end{figure}

\textit{The Lund string model.}
The dominant hadronization model in \textsc{Pythia} is the Lund string
model~\cite{Andersson:1983ia}. Its physical basis is color confinement: at
large quark-antiquark separations, the chromoelectric field is compressed
into a flux tube approximated as a one-dimensional string with tension
$\kappa\approx 1\;\mathrm{GeV/fm}$, giving a linear potential $V(r)=\kappa r$.
String breaking in the Lund model and in a quantum simulation of a lattice
gauge theory is shown schematically in Fig.~\ref{fig:string-breaking}, together
with the Lund parameters that such simulations are expected to inform.

When the string stretches sufficiently, it breaks via the Schwinger
mechanism~\cite{Schwinger:1951nm}: a quark-antiquark pair tunnels through
the color-electric field with probability
\begin{equation}
  \mathcal{P} \;\propto\;
  \exp\!\left(-\frac{\pi m_{\perp}^{2}}{\kappa}\right)
  =\exp\!\left(-\frac{\pi p_{\perp q}^{2}}{\kappa}\right)
   \exp\!\left(-\frac{\pi m_{q}^{2}}{\kappa}\right),
\label{eq:schwinger}
\end{equation}
where $m_{q}$ is the mass of the produced quark, $p_{\perp q}$ is its
transverse momentum, and $m_{\perp}^{2}=m_{q}^{2}+p_{\perp q}^{2}$. This formula
simultaneously explains the approximately Gaussian transverse momentum
spectrum of produced hadrons and the suppression of heavy flavors:
$u\bar{u}:d\bar{d}:s\bar{s}:c\bar{c}\approx 1:1:0.3:10^{-11}$. The
longitudinal momentum distribution is governed by the Lund symmetric
fragmentation function
\begin{equation}
  f(z)\;\propto\;\frac{(1-z)^{a}}{z}
  \exp\!\left(-\frac{b\,m_{\perp}^{2}}{z}\right),
\label{eq:lund_ff}
\end{equation}
where $z$ is the light-cone momentum fraction carried by the hadron, $a$
and $b$ are free parameters, and the functional form is motivated by the
Wilson area law $e^{-bA}$ for the string worldsheet.

Critically, the Schwinger formula~(\ref{eq:schwinger}) assumes a
spatially and temporally uniform color-electric field. The actual breaking
tube is short, inhomogeneous, and dynamically evolving. Moreover, recent
quantum simulation studies~\cite{Grieninger:2026bdq} demonstrate that the
string state carries nonlocal quantum correlations along its full length
that are absent from this semiclassical picture. Corrections from
non-uniform fields and from the quantum entanglement structure of the
breaking state could modify the pair production rate by order-one factors.

\textit{String-breaking parameters and their uncertainties.}
Several parameters of the \textsc{Pythia} string-breaking tune~\cite{Skands:2014pea} are poorly constrained and correlated with each other
and with the parton shower infrared cutoff. The strangeness suppression
parameter cannot explain the
$\mathrm{K}/\pi$ enhancement observed by ALICE as a function of charged
multiplicity~\cite{ALICE:2016fzo}. The diquark rate leads to a factor of $\sim 5$ underprediction
of the $\Lambda_c/D^0$ ratio at low $p_T$~\cite{ALICE:2020wfu}. The
vector-to-pseudoscalar meson ratio is predicted as $3/4$ from spin
counting but is tuned to $0.33$--$0.50$, a discrepancy with no
theoretical explanation.

\textit{Extensions of the string model.}
Several extensions address multi-string effects in dense hadronic
environments. Rope hadronization~\cite{Bierlich:2014xba} models overlapping
color strings in high-multiplicity events as coherent ropes with enhanced
effective tension, producing enhanced strangeness and collective flow-like
behavior. String shoving~\cite{Bierlich:2016vgw} introduces repulsive
interactions between strings, generating azimuthal anisotropy without
hydrodynamics. The thermal string model adds a Boltzmann factor to the
tunneling probability. All of these extensions introduce additional free
parameters based on semiclassical estimates with no first-principles derivation from
quantum field theory. Quantum simulation of multi-string
systems would contribute genuinely new information that is not accessible otherwise. The discovery that
strings carry nonlocal quantum correlations~\cite{Grieninger:2026bdq} suggests
that these semiclassical models may be missing qualitatively important
physics.

\textit{The cluster model.} \textsc{Herwig} and \textsc{Sherpa} use the cluster model~\cite{Webber:1983if}, in which
color-singlet clusters of adjacent partons form at the shower scale $Q_0$
through preconfinement and decay to hadrons through phase-space-weighted
two-body processes. The preconfinement mechanism has never been tested
non-perturbatively. A quantum simulation of the full
showering-plus-hadronization process could determine whether this
assumption accurately captures QCD confinement dynamics.

\section{Toy models}
\label{sec:toy}

\begin{figure}[!t]
  \centering
  \includegraphics[width=\linewidth]{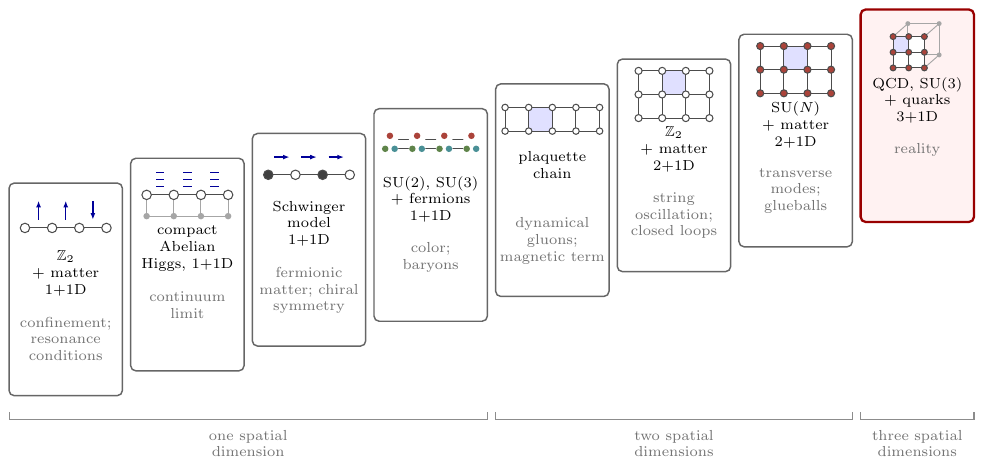}
  \caption{The toy models proposed for studying hadronization physics, ordered by the degrees of freedom
  of QCD each adds. The lattice geometry of each model is drawn above its name.}
  \label{fig:toy-ladder}
\end{figure}

A full quantum simulation of QCD with dynamical quarks in $3+1$ dimensions will be challenging to perform. This goal can be reached by developing techniques in simulations of simplified theories, each of which retains some features of real hadronization physics and discards others~\cite{Banuls:2019bmf,Zohar:2021nyc,Bauer:2022hpo,DiMeglio:2023nsa,Aidelsburger:2021mia,Halimeh:2023lid,Halimeh:2025vvp,Davoudi:2025kxb,Banuls:2018jag,Magnifico:2024eiy}. The value of a given model is set by which assumption of the phenomenological models it can test, which degree of freedom of QCD it adds, and how clearly its artifacts can be identified. This section describes the models in that order, from the theory with the fewest degrees of freedom to the theories closest to QCD, and states for each what physics it provides. Fig.~\ref{fig:toy-ladder} shows the proposed hierarchy of toy models for studying hadronization physics.

\textit{$Z_2$ lattice gauge theories:} A $\mathbb{Z}_2$ gauge theory with matter is the simplest confining lattice theory. The string-breaking dynamics in these theories is set by resonance conditions. This limits the distances at which the string is allowed to break. These conditions can be determined analytically, which enables them to be used as a benchmark for hardware performance~\cite{Alexandrou:2025vaj,Desaules:2024cua}. Simulations on a trapped-ion platform have demonstrated that string breaking begins at the ends of the string rather than through pair creation in its bulk~\cite{De:2024smi}. The Lund model assumes that successive breaks are independent, and that they occur throughout the string with equal probability. The edge effects in this model are the first observed violation of this assumption in a quantum simulation. In one spatial dimension, $\mathbb{Z}_2$ lattice gauge theories can be mapped onto Ising models, which enables their efficient implementation on quantum computers~\cite{Verdel:2019chj,Lerose:2019jrs,Liu:2018fza,Mildenberger:2022jqr,Charles:2023zbl}. $\mathbb{Z}_2$ lattice gauge theories are also amenable to near-term simulation in two spatial dimensions~\cite{Gyawali:2024tpz,Borla:2025gfs}. In multiple spatial dimensions, the gauge fields become dynamical, enabling string oscillations in addition to breaking, both of which have been observed in recent quantum simulations~\cite{Cobos:2025krn,Cochran:2024rwe,Gonzalez-Cuadra:2024xul,DiMarcantonio:2025cmf,Bacciconi:2026qrs}. However, the continuum limit of these models is not a gauge theory. This restricts the connection to hadronization to purely qualitative comparisons of lattice dynamics. Despite this restriction, qualitative features found in $\mathbb{Z}_2$ gauge theories can be used to guide simulations of toy models that share more similarities with QCD. Furthermore, qualitative insights could guide modifications to hadronization models that are then fit to experimental data.

\textit{1+1D U(1) gauge theory with bosonic matter:} The compact Abelian Higgs model is scalar quantum electrodynamics (QED) with compact matter fields. The theory can be truncated down to a single spin-1 degree of freedom per link without breaking the symmetries of the model~\cite{Bazavov:2015kka,Meurice:2020pxc,Meurice:2019ddf,Meurice:2020gcd,Meurice:2021pvj}. The truncated theory can be realized in analog neutral atom simulations, making it amenable to near-term quantum simulation~\cite{Senseman:2022jpu,Zhang:2023agx,Asaduzzaman:2026jyi}. Furthermore, this model is amenable to tensor network simulations~\cite{Senseman:2022jpu,Senseman:2026pif}. The Abelian Higgs model is the simplest confining lattice gauge theory with a continuum limit, which will enable probing qualitative features observed in $\mathbb{Z}_2$ theories in the continuum.

\textit{1+1D U(1) gauge theory with fermionic matter:} The Schwinger model, QED in $1+1$ dimensions, replaces the bosonic matter in the Abelian Higgs model with fermionic matter. It shares several qualitative features with QCD such as confinement, chiral symmetry breaking, and a $\theta$ vacuum~\cite{Zhang:2023hzr,Thompson:2021eze,Grieninger:2026pai}. This theory is the simplest one where the Schwinger-mechanism picture on which the Lund string is built can be tested~\cite{Surace:2019dtp,deJong:2021wsd,Liu:2024lut,Grieninger:2026bdq,Artiaco:2025qqq,Grieninger:2026fcr,Grieninger:2025rdi}. Tensor network simulations with dynamical sources have demonstrated that the hadrons produced by string breaking follow a Boltzmann distribution~\cite{Florio:2024aix,Batini:2024zst,Florio:2025hoc,Janik:2025bbz,Janik:2026foi}. 
Therefore, this model allows a first-principles answer to whether fragmentation is thermal or tunneling-like, at least in one dimension. 
Heavy external charges can be used to probe in-medium energy loss~\cite{Farrell:2024mgu,Lee:2023urk,Barata:2025hgx}. More sophisticated aspects of hadronization can be probed by including multiple flavors. A four-flavor deformed Schwinger model with an $\mathrm{SU}(3)\times\mathrm{U}(1)$ flavor structure can encode light and strange flavors plus an effective spin. Numerical simulations demonstrate that the multi-flavor model generates hyperon spin correlations through multiparton exchange that are absent from the default Lund model~\cite{Barata:2023jgd}. The Schwinger model is the simplest field theory containing gauge fields and fermionic matter~\cite{Martinez:2016yna,Klco:2018kyo,Farrell:2024fit}. This allows qualitative features of hadronization to be probed; however, this model is restricted to only producing mesons.

\textit{1+1D Non-Abelian gauge theory with fermionic matter:}
The next step towards QCD is to simulate theories in $1+1D$ with fermionic matter and non-Abelian gauge groups. $\mathrm{SU}(2)$ with staggered fermions in $1+1$ dimensions is the simplest non-Abelian theory. Preliminary simulations have been performed of this theory and resource estimates have been made for what is necessary to scale up the simulations~\cite{Davoudi:2022xmb,Ciavarella:2024lsp,Gupta:2026tcg,Klco:2019evd,Atas:2021ext,Banuls:2017ena,Li:2025sgo,Das:2025utp,Verde:2026hcg}. An $\mathrm{SU}(3)$ gauge theory with dynamical quarks in $1+1D$ shares the same gauge group as QCD and has been simulated on existing quantum computers~\cite{Atas:2022dqm,Farrell:2022wyt,Ciavarella:2023mfc,Farrell:2022vyh,Than:2024zaj,Chernyshev:2025lil}. This theory contains both baryons and mesons, which will allow probes of the differences between how hadronization produces these types of particles~\cite{Kadam:2022ipf,Rigobello:2023ype,Cogburn:2026aqy}. The simulations performed in these theories can give insight into which qualitative features observed in the simpler models are artifacts of an Abelian gauge group, and which may persist in full QCD~\cite{Popov:2025byi}.

\textit{2+1D gauge theories:}
Theories in one spatial dimension do not contain dynamical gauge fields. A plaquette chain is the simplest lattice geometry in which there are dynamical gluons, and is a good first step towards seeing qualitative changes from a purely $1+1D$ calculation~\cite{Ciavarella:2021nmj,Yao:2023pht,Hayata:2024fnh,Chen:2026tvd,Siew:2025thj,Siew:2026fax}. In a full $2+1D$ simulation, the flux string becomes dynamical, and it is possible for emitted hadrons to carry transverse momentum. Performing simulations in $2+1D$ requires the implementation of a plaquette term, which introduces new computational difficulties~\cite{Grabowska:2022uos}. Despite this, exploratory calculations have been performed for lattice gauge theories in $2+1D$~\cite{Tian:2025mbv,Cao:2026qky,John:2026gut,Ciavarella:2026tau,Paulson:2020zjd,Muller:2023nnk,Turro:2024pxu,Osborne:2024zpx,Crippa:2024hso,Cataldi:2025cyo,Joshi:2025pgv,DePaciani:2025uzj,Rosanowski:2025nck,Miranda-Riaza:2025fus,Majcen:2026sfm,Joshi:2026hfe,Gandon:2026das}. The inclusion of dynamical gauge fields also introduces glueballs into the spectrum of the theory~\cite{Xu:2025abo,Xu:2026ibi,Cao:2026qky,Cataldi:2025cyo,Siew:2025thj,Siew:2026fax}. Glueball production is not present in existing hadronization models, and simulations in $2+1D$ are the simplest place where this assumption can be tested. In addition, the second spatial dimension will allow for the emission of soft transverse radiation, which will allow for probing qualitative features of hadronization that do not exist in a single spatial dimension.

\section{Observables}
\label{sec:observables}

In principle, any observable can be measured in a quantum simulation. In practice, each observable has a cost in state preparation, circuit depth, and shots. This section lists the observables that should be extracted and what extracting them requires.

\textit{Conventional observables.} The quantities that event generators currently fit are the natural first targets. The Lund kernel is fixed by the string tension $\kappa$, the fragmentation function $f(z)$, the tunneling factor, and the breakup proper times, all tuned against $e^+e^-$ event shapes and already in tension with LHC data~\cite{Bierlich:2022pfr,L3:2004cdh,ALICE:2017thy}. Simulations can supply these quantities rather than fit them. The assumptions listed in Section~\ref{sec:intro} can also be tested one at a time. The models contain phenomenological spacetime information, while a Hamiltonian simulation gives the local energy and charge densities along the string, the timescale of hadron formation, and the transition from partonic to hadronic degrees of freedom. This will allow first-principles probing of the assumptions that go into the phenomenological hadronization models currently being used. The form of $f(z)$, the Schwinger-like generation of transverse momentum, the dependence of both on the quark mass, and the assumed independence of successive breaks are each directly testable. Microscopic breaking mechanisms, including pair nucleation inside the flux tube, screening of endpoint charges, and string revivals, can also be directly probed in quantum simulation.

\textit{Dynamical string tension.} Static string tensions can be directly probed with traditional lattice techniques.  The energy of a string grows linearly with its length and then saturates, and piecewise-linear fits give the static tension $\sigma$ and the breaking length $L^{*}$. The dynamical tension, denoted $\kappa$ as in the Lund model, is a different quantity from $\sigma$. It is a fitted parameter in \textsc{Pythia} and differs from the static tension obtained from Euclidean lattice QCD. Rope and close-packing models make the effective tension density-dependent by construction rather than by derivation, and the constant-tension assumption for expanding strings rests on static lattice evidence alone~\cite{Bierlich:2014xba,Fischer:2016zzs}. No agreed definition or extraction protocol currently connects simulated string dynamics to the effective $\kappa$ a generator would consume. Such a definition must be developed before simulated dynamics can be matched to event generators.

\textit{Physics the generators omit.} Hadronization models are based on a few insights into the hadronization mechanism, such as the linearity of the QCD potential. Many other aspects of string breaking are simplifying assumptions. For example, neither string nor cluster models produce glueballs, and in both models, hadron spin enters only through the relative production rates of spin multiplets. Neither default model contains spin dynamics or spin correlations for light hadrons~\cite{Bierlich:2022pfr,Bellm:2025pcw}. String breaking models produce baryons through diquark breaking of the string, popcorn configurations, and string junctions. Direct simulations of string breaking are likely to provide new insights into basic principles in hadronization that can be used to improve the classical modeling. The temperature and chemical potential dependence of hadronization physics will also be accessible from quantum simulation.

\textit{Entanglement and magic.} Entanglement entropy has been identified as a physical diagnostic of string breaking~\cite{Pichler:2015yqa,Alexandrou:2025vaj,Grieninger:2026bdq,Gupta:2026tcg,Cataldi:2025cyo}. However, stabilizer states can be maximally entangled while being classically simulable. Magic provides a measure of how far a state is from a stabilizer state. 
The presence of substantial magic and entanglement in a quantum state is necessary for that state to be hard to simulate classically~\cite{Gottesman:1998hu,Leone:2021rzd}. Magic behaves differently from entanglement in lattice gauge theories where both have been computed~\cite{Falcao:2024msg,Ebner:2025pdm,Froland:2026aff}. In Schwinger-model simulations of static string breaking, all complexity measures peak at the breaking point, with non-local magic located on the string and absent from the screened charges~\cite{Grieninger:2026bdq,Grieninger:2025rdi,Grieninger:2026fcr}. Whether analogous correlations develop in real-time string breaking and survive into the hadronic final state can be answered by simulation. If they produce measurable correlations between hadrons that existing models do not reproduce, the models will need new parameterizations, and the universality of any such parameterization can then be tested across observables.

\textit{Long-distance matrix elements in effective field theories.} Effective field theories (EFTs) separate long- from short-distance physics, and identify long-distance matrix elements as important targets for quantum  simulations. SCET separates hard, collinear, and soft physics~\cite{Bauer:2000ew,Bauer:2000yr,Bauer:2001ct,Bauer:2001yt,Bauer:2002nz,Becher:2014oda}. For an event shape $e$ measured in $e^+e^-$ collisions, the cross section factorizes as $d\sigma/de = [H \otimes J_1 \otimes J_2 \otimes F_e](e)$, where the soft function $F_e$~\cite{Collins:1981uk,Collins:1981ta,Collins:1987pm,Collins:1989gx,Bauer:2008dt,Bauer:2008jx} encodes universal non-perturbative information. It is described by matrix elements of soft Wilson lines $Y_n$ and $Y_{\bar n}$ along the light-like directions of the two jets. Two properties make it a quantum-computable target~\cite{Bauer:2025nzf}. With a lattice regulator, the soft function is a sum of squared amplitudes, so it is positive. It is also normalized, since the vacuum is normalized, the Wilson-line operator is unitary, and the sum runs over a complete set of states. Therefore, it can be sampled by preparing the interacting vacuum, applying the unitary that creates the Wilson lines, and measuring the event shape of the final state as the experiment does. Additionally, it describes only physics below the soft scale $\Lambda_s \sim Q e$, so the lattice spacing is set by $\Lambda_s$ rather than by the collision energy. Defining the soft function fully differentially in the energy flow $\mathcal{E}(\hat n)$~\cite{Sveshnikov:1995vi,Korchemsky:1997sy,Lee:2006nr,Bauer:2008jx} allows a single simulation to be post-processed into the soft functions of all event shapes computed from the measured energy flow, without an assumption of universality~\cite{Bauer:2025nzf}. However, a single simulation fixes the directions and color representations of the Wilson lines. Therefore, soft functions for hadron collisions, multijet final states, and gluon jets each require a separate simulation. The energy flow also does not determine hadron species or baryon number. A three-dimensional soft function of this kind is likely more than a decade away. Additionally, the shower infrared cutoff $Q_0$ must be treated as a factorization scale~\cite{Hoang:2024nqi}. 

\textit{The measurement problem for hadrons.} Extracting information about hadronization requires performing measurements that can identify hadrons in the final state. For scalar field theories, Jordan, Lee, and Preskill proposed performing both state preparation and measurement through adiabatic switching between the free and interacting theories~\cite{Jordan:2012xnu,Jordan:2011ci}. In this construction, the adiabatic switch defines the observables of the simulation, since it maps the interacting final state onto free particles that can be counted and binned in momentum. For confining gauge theories the free limit is the wrong starting point, since in a free theory of strongly interacting particles the degrees of freedom are quarks and gluons while the asymptotic states are hadrons. The strong-coupling limit provides a replacement. In this limit, the vacuum is a product state with no electric flux excited, hadrons are localized excitations, and hadron numbers and positions follow from local measurements. Switching back to strong coupling before readout maps the final state into a basis where these observables are simple to evaluate, but the required circuit depths are large as described in Section~\ref{sec:challenges}. The circuit depth can be reduced by replacing adiabatic switching with a unitary that is variationally optimized to prepare the vacuum and single hadron states~\cite{Ciavarella:2024lsp}. Alternatively, one can variationally construct creation and annihilation operators for hadrons and perform measurements of these operators~\cite{Davoudi:2024wyv,Davoudi:2025rdv}. 

\section{Challenges}
\label{sec:challenges}

\begin{table}[!t]
  \centering
  \footnotesize
  \renewcommand{\arraystretch}{1.15}
  \setlength{\tabcolsep}{4pt}
  \begin{tabular}{l p{3.0cm} l l l l l l}
    \hline\hline
    Theory & Formulation & Lattice & Qubits & T gates & CNOT gates & Algorithm & Ref. \\
    \hline
    \multicolumn{8}{l}{\textit{$1+1$D}} \\
    U(1) with fermions & \raggedright electric, $\eta=4$ & $64$ sites & $316$ & -- & $9{,}450$ & Trotter, second order & \cite{Shaw:2020udc} \\
    U(1) with fermions & \raggedright gauge field integrated out, interaction range $\lambda=1$ & $112$ sites & $112$ & -- & $926$ & Trotter, second order & \cite{Farrell:2024fit} \\
    SU(2) with fermions & \raggedright Schwinger bosons, $\eta=2$ & $L=10$ & $92$ & -- & $26{,}136$ & Trotter, second order & \cite{Davoudi:2022xmb} \\
    SU(2) with fermions & \raggedright LSH, $\eta=2$ & $L=10$ & $40$ & -- & $4{,}176$ & Trotter, second order & \cite{Davoudi:2022xmb} \\
    SU(2) with fermions & \raggedright axial gauge, $N_f=1$ & $L=10$ & $40$ & -- & $3{,}344$ & Trotter, first order & \cite{Farrell:2022wyt} \\
    SU(3) with fermions & \raggedright axial gauge, $N_f=1$ & $L=10$ & $60$ & -- & $8{,}436$ & Trotter, first order & \cite{Farrell:2022wyt} \\
    SU(3) with fermions & \raggedright gauge field integrated out, $N_f=1$ & $20$ sites & $60$ & -- & $36{,}024$ & Trotter, first order & \cite{Atas:2022dqm} \\
    SU(3) with fermions & \raggedright axial gauge, $N_f=2$ & $L=100$ & $1{,}200$ & -- & $3.6\times10^{6}$ & Trotter, first order & \cite{Farrell:2022wyt} \\
    \hline
    \multicolumn{8}{l}{\textit{$2+1$D}} \\
    SU(3) & \raggedright large $N_c$, $(1,1,1)$ & $10\times10$ & $100$ & $6.46\times10^{4}$ & $1{,}600$ & Trotter, first order & \cite{Ciavarella:2025bsg} \\
    SU(3) & \raggedright large $N_c$, $(1,2,1)$ & $10\times10$ & $100$ & $7.27\times10^{4}$ & $2{,}000$ & Trotter, first order & \cite{Ciavarella:2025bsg} \\
    SU(3) & \raggedright large $N_c$, $(2,2,2)$, $\mathcal{O}(1/N_c)$ & $10\times10$ & $500$ & $9.4\times10^{8}$ & $2.5\times10^{7}$ & Trotter, first order & \cite{Ciavarella:2025bsg} \\
    SU(3) & \raggedright irreps, $T_1$ & $2\times2$ & $27$ & $2.5\times10^{5}$ & $2.2\times10^{5}$ & Trotter, first order & \cite{Balaji:2025afl} \\
    U(1) with fermions & \raggedright electric & $100^2$ & $8.1\times10^{4}$ & $4.7\times10^{12}$ & -- & Qubitization, HHKL & \cite{Rhodes:2024zbr} \\
    \hline
    \multicolumn{8}{l}{\textit{$3+1$D pure gauge}} \\
    SU(2) & \raggedright BT, $\hat{H}_I$ & $10^3$ & $15{,}000$ & $2.2\times10^{9}$ & -- & Trotter & \cite{Gustafson:2022xdt,Gustafson:2023kvd} \\
    SU(2) & \raggedright BO, $\hat{H}_{KS}$ & $10^3$ & $18{,}000$ & $4.0\times10^{9}$ & -- & Trotter & \cite{Gustafson:2023kvd} \\
    SU(2) & \raggedright BI, block encoded, $\hat{H}_{KS}$ & $10^3$ & $36{,}000$ & $2.8\times10^{11}$ & -- & Trotter & \cite{Lamm:2024jnl} \\
    SU(3) & \raggedright $\Sigma(36\times3)$, $\hat{H}_{KS}$ & $10^3$ & $24{,}000$ & $7.0\times10^{10}$ & -- & Trotter & \cite{Gustafson:2024kym} \\
    SU(3) & \raggedright $\Sigma(72\times3)$, $\hat{H}_{KS}$ & $10^3$ & $27{,}000$ & $7.0\times10^{10}$ & -- & Trotter & \cite{Perez:2025cxl} \\
    SU(3) & \raggedright large $N_c$, $(1,1,1)$ & $10^3$ & $3{,}000$ & $1.79\times10^{7}$ & $4.0\times10^{5}$ & Trotter, first order & \cite{Ciavarella:2025bsg} \\
    SU(3) & \raggedright large $N_c$, $(1,2,1)$ & $10^3$ & $3{,}000$ & $2.3\times10^{8}$ & $5.2\times10^{6}$ & Trotter, first order & \cite{Ciavarella:2025bsg} \\
    SU(3) & \raggedright large $N_c$, $(2,2,2)$, $\mathcal{O}(1/N_c)$ & $10^3$ & $12{,}000$ & $2.8\times10^{10}$ & $7.5\times10^{8}$ & Trotter, first order & \cite{Ciavarella:2025bsg} \\
    SU(3) & \raggedright irreps, $T_1'$ & $3^3$ & $180$ & $7.2\times10^{6}$ & $6.2\times10^{6}$ & Trotter, first order & \cite{Balaji:2025afl} \\
    SU(3) & \raggedright irreps, $B\le6$ & $10^3$ & $10^{4}+\mathcal{O}(10^3)$ & $1.2\times10^{5}$ & -- & QSVT & \cite{Draper:2026bcj} \\
    \hline
    \multicolumn{8}{l}{\textit{$3+1$D with fermions}} \\
    SU(2) & \raggedright electric & $10^3$ & $5.5\times10^{4}$ & $6.7\times10^{13}$ & -- & Qubitization & \cite{Rhodes:2024zbr} \\
    SU(3) & \raggedright electric & $10^3$ & $1.3\times10^{5}$ & $1.1\times10^{15}$ & -- & Qubitization & \cite{Rhodes:2024zbr} \\
    SU(3) & \raggedright axial gauge, field basis, $n_A=23$, $N_f=2$ & $10^3$ & $3.9\times10^{5}$ & $1.8\times10^{14}$ & $2.8\times10^{13}$ & Trotter & \cite{Yao:2026rya} \\
    \hline\hline
  \end{tabular}
  \caption{Published resource estimates for quantum simulations of lattice gauge theories. Gate counts are for a single time step. The exceptions are the estimates of Ref.~\cite{Rhodes:2024zbr}, which give the total cost of the simulation because qubitization does not use time steps, and the estimate of Ref.~\cite{Draper:2026bcj}, which gives the cost of a single call to the plaquette block encoding. The $3+1$D estimates of Ref.~\cite{Rhodes:2024zbr} are published for lattices of $100^3$ and $1000^3$ sites. The T gate counts given here for $10^3$ sites are obtained by evaluating the cost formulas of Ref.~\cite{Rhodes:2024zbr} with $\epsilon=10^{-3}$, $a=1$, $\Lambda=10$, and $m=g=10$. The counts of Ref.~\cite{Balaji:2025afl} are for the magnetic term only. A dash indicates that the count is not given in the cited reference. Qubit counts are logical qubits for the full lattice. Ancilla qubits are included for Refs.~\cite{Balaji:2025afl,Draper:2026bcj} and for the $2+1$D estimate of Ref.~\cite{Rhodes:2024zbr}, and excluded for the other references. The T gate counts of Refs.~\cite{Ciavarella:2025bsg,Yao:2026rya} assume that each $R_z$ gate is synthesized to an accuracy of $10^{-10}$. The estimate of Ref.~\cite{Yao:2026rya} is evaluated from the bounds given there with energy $E'=10$, accuracy $\epsilon=0.1$, and bare coupling $g=1$ in lattice units. The notation for the truncation parameters follows the cited references.}

  \label{tab:resources}
\end{table}

Several distinct problems separate present demonstrations from input that an event generator could use. They are listed here together with the current state of each.

\textit{Continuum extrapolation.} Every simulation is performed at finite lattice spacing, and a continuum limit must be taken before comparison with QCD. Present real-time quantum simulations mostly sit deep in the strong-coupling regime, far from where lattice dynamics can be matched to physical dynamics~\cite{Cataldi:2025cyo}. Probing the continuum limit is also expensive for the classical methods currently used to validate quantum simulations. As the lattice spacing is reduced, the correlation length in lattice units grows, the entanglement across any cut grows with it, and the bond dimension needed for a fixed accuracy grows at every lattice spacing along the continuum extrapolation~\cite{Banuls:2019bmf,Gupta:2026tcg}. Real-time dynamics can also suffer from lattice artifacts without a static analog. For example, in the Schwinger model, the sub-luminal maximum lattice velocity produces spurious fragmentation of fast hadrons~\cite{Farrell:2024mgu}. The errors due to a finite lattice spacing can be reduced through the use of improved Hamiltonians that systematically improve the scaling of errors with lattice spacing~\cite{Carena:2022kpg,Illa:2025njz,Illa:2025dou}.

\textit{Formulation and truncation dependence.} Many lattice Hamiltonian formulations exist, and each one has its trade-offs. Electric-basis formulations give a mostly local Hamiltonian with discrete basis states that can be mapped onto a quantum computer. The electric basis works best at large values of the bare coupling, but the continuum limit is reached by reducing the bare coupling. The error introduced by an electric basis cutoff falls off as a factorial of the field cutoff, which may enable the use of relatively low truncations in physical simulations~\cite{Ciavarella:2025tdl}. Note that a cutoff that is too small produces real-time dynamics that differ qualitatively from the untruncated theory even where static quantities extrapolate cleanly~\cite{Gupta:2026tcg}. Loop-string-hadron and dressed-site formulations work with gauge-invariant electric basis states, at the cost of more complicated circuits and a larger local dimension~\cite{Raychowdhury:2019iki,Kadam:2025trs,Silvi:2014pta,Cataldi:2023xki}. Mixed or gauge-fixed bases allow one to work at small values of the coupling and with a fully gauge-invariant basis at the cost of locality~\cite{Bauer:2021gek,Kane:2022ejm,DAndrea:2023qnr,Grabowska:2024emw,Burbano:2024uvn,Froland:2025bqf}. Alternatively, one could use discrete subgroups of the gauge group. Quantum simulations with discrete gauge groups can be implemented in a scalable manner, but extrapolation to the continuum requires more care as they do not live in the same universality class as the continuum theory~\cite{Alexandru:2019nsa,Lamm:2019bik,Alam:2021uuq,Ji:2022qvr,Gustafson:2022xdt,Gustafson:2023swx,Gustafson:2023kvd,Carena:2024dzu,Gustafson:2024kym,Assi:2024pdn}. Quantum link models provide yet another option for formulating lattice gauge theory Hamiltonians and approaching the continuum limit~\cite{Chandrasekharan:1996ih,Zohar:2014qma}. Each of these formulations has different systematic errors and computational costs. This is a strength of the program, as the same physics can be simulated in different formulations to determine which features of string breaking are universal and which are formulation artifacts.

\textit{Resource estimation.} Before running a simulation, one should know how large a computer is necessary, and how many gates need to be implemented to perform the simulation. Resource estimates in quantum computing are commonly performed using triangle inequality bounds to rigorously upper bound the error induced in the approximation to the time evolution operator. These estimates have been developed for various compilation strategies for lattice gauge theory dynamics~\cite{Kan:2021xfc,Tong:2021rfv,Davoudi:2022xmb,Kane:2022ejm,Sakamoto:2023cxs,Rhodes:2024zbr,Balaji:2025afl,Davoudi:2026pcu,Shaw:2020udc,Yao:2025uxz,Yao:2026rya}. However, the formalism behind these estimates does not capture the deviation of simulated dynamics from the quantum field theory being simulated. The Trotter error bounds they rely on are obtained from triangle inequalities applied to nested commutators. These bounds are loose, often by orders of magnitude. Furthermore, they bound the deviation from a particular lattice-regularized time evolution rather than the deviation from continuum physics. Alternatively, the finite time step used in quantum simulation can be understood as a temporal lattice spacing. One can then approach the continuum by taking the spatial and temporal lattice spacings to zero simultaneously~\cite {Carena:2021ltu,Carena:2022hpz,Kane:2025ybw}. This will likely require far fewer resources than requiring that the Trotter error vanishes at each fixed lattice spacing. Table~\ref{tab:resources} provides resource estimates from the literature for the toy models that have been proposed in Section~\ref{sec:toy}.

\textit{State preparation.} Extracting continuum physics from lattice field theories requires preparing low-energy states where the description of the lattice physics by an effective continuum theory is valid. In the present quantum simulations of string breaking dynamics, the initial state is a flux tube prepared on top of some approximation of the vacuum state. In the strong coupling regime, the vacuum is close to a product state in the electric basis, and a string can be prepared by exciting the links along a path. As the coupling is reduced toward the continuum, the vacuum becomes an interacting gauge-invariant state with magnetic fluctuations whose correlation length in lattice units grows, and a string placed on the strong-coupling vacuum is no longer a low-energy state of the theory. Two approaches have been developed to reach the vacuum of lattice gauge theories outside of the strong coupling limit. The first is to perform adiabatic switching from a solvable limit of the theory~\cite{Surace:2024bht,Jordan:2012xnu,Jordan:2011ci}. The circuit depth to implement this procedure grows as the minimal gap along the switching path decreases. Alternatively, one can use a variational state preparation procedure~\cite{Paulson:2020zjd,Atas:2021ext,Zhang:2021bjq,Ciavarella:2021lel,Atas:2022dqm,Farrell:2023fgd,Angelides:2023noe,Farrell:2024fit,Davoudi:2024wyv,Guo:2024tnb,Gustafson:2024bww,Crippa:2024hso,Ciavarella:2024lsp,Gaz:2025ptu,Yamanaka:2026opq}. Hamiltonian VQE ans\"atze that alternate electric, magnetic, and matter layers can produce gauge-invariant ground states without penalty terms~\cite{Sala:2018dui,Surace:2023qwo}. Alternative ans\"atze that are chosen to match the symmetries of the vacuum state have also been used. Variational circuits optimized on small systems can be used to inform state preparation on larger systems through the scalable circuit formalism. This approach exploits the fact that, once the lattice exceeds a few correlation lengths, the optimal parameters converge exponentially with system size, so parameters determined on small lattices can be extrapolated to lattices that cannot be simulated classically~\cite{Farrell:2023fgd,Ciavarella:2024lsp}. This procedure has been verified for the vacuum and for single-hadron states, but its depth still grows with the correlation length.

\textit{Extracting observables.} Factorization identifies soft and shape functions as the natural quantum targets, with lattice resolution set by the soft scale rather than the collision energy~\cite{Bauer:2025nzf}. However, no published prescription matches a quantum-simulated string-breaking amplitude to a shower at a defined $Q_0$~\cite{Hoang:2024nqi}, and no procedure maps a lattice calculation onto Lund parameters with controlled errors. A further separation is unsolved. The quarks at the ends of a string move apart with high energy and inject a perturbative collinear short-distance scale. This implies that a naive string simulation mixes hadronization with perturbative collinear physics. The two must be separated, presumably through a factorization theorem using SCET~\cite{Bauer:2000ew,Bauer:2000yr,Bauer:2001ct,Bauer:2001yt}, before a simulation can constrain the models quantitatively. These gaps remain open.

\textit{Device noise.} Quantum computers do not operate perfectly, and each operation on them suffers from some degree of noise or error. This device noise is the primary source of limitations of existing quantum computers. In lattice gauge theory simulations, a fixed physical evolution time requires a number of Trotter steps that grows as the lattice spacing is reduced, and the circuit depth per step grows with the truncation of the gauge group. Therefore, the accumulated error per circuit grows along the continuum extrapolation and along the approach to QCD. In present simulations, the effects of device noise are often addressed with error mitigation techniques such as dynamical decoupling~\cite{Viola:1998jx,Viola:1998gg}, readout mitigation~\cite{Nachman:2019zfw,Berg:2020ibi}, Pauli twirling~\cite{Knill:2004vlm,Wallman:2015uzh}, zero noise extrapolation~\cite{He:2020udd,Pascuzzi:2021mhw}, and operator decoherence renormalization (ODR)~\cite{Urbanek:2021oej,Farrell:2022vyh,Charles:2023zbl,Farrell:2023fgd,Farrell:2024fit,Chai:2025kbi}. These techniques have enabled successful simulation of hadron dynamics in the Schwinger model with $112$ qubits at a CNOT depth of $370$~\cite{Farrell:2024fit}. These mitigation techniques are limited by the circuit depth, as they must amplify a signal that decays exponentially with the depth of the circuit~\cite{Eisert:2020ckk,Gheorghiu:2018eif,Tindall:2023cqi}. As quantum computers improve, the noise can be systematically removed with the use of quantum error correction. Gauge theories offer a unique opportunity as their Hilbert space contains redundancies that can be combined with application-specific error-correcting codes. In the near term, these redundancies can be used for post-selection to remove errors that take the system outside of the physical Hilbert space~\cite{Davoudi:2024wyv,Cobos:2025krn,Ciavarella:2026tau}.

\section{Hardware and program}
\label{sec:program}
Over the past several years, quantum simulations of gauge theories have progressed from few-qubit demonstrations to simulations of confining dynamics on more than $100$ qubits in both one and two spatial dimensions~\cite{Martinez:2016yna,Farrell:2024fit,Ciavarella:2024fzw,Gonzalez-Cuadra:2024xul,Gandon:2026das,Ilcic:2026cac,Davoudi:2025rdv,Ciavarella:2026tau,Xu:2026ibi,Joshi:2026hfe,John:2026gut,Froland:2026aff}. The consistent lesson of these simulations is that state preparation is classically inexpensive while time evolution is not. The computational cost of classical simulation grows rapidly with evolution time~\cite{Farrell:2024fit,Xu:2026ibi}. However, no demonstrated quantum advantage for gauge dynamics exists. Every hardware result to date has been validated against a classical baseline. Going beyond the limits of classical simulation requires advancement in available quantum hardware.

The hardware available over the next several years is set by published development plans. IBM's quantum development plan schedules the \texttt{Kookaburra} module with a quantum low-density parity-check memory for release in 2026, entanglement between modules with \texttt{Cockatoo} in 2027, and the fault-tolerant \texttt{Starling} system in 2029, specified as $200$ logical qubits executing $10^8$ gates~\cite{IBMQuantumHardwareRoadmap,Yoder:2025ooz}. In June 2026, the U.S. Department of Energy announced the Quantum Genesis initiative, whose goal is the deployment of a fault-tolerant quantum computer with logical qubits numbering in the low hundreds by 2028, with high energy physics among the named applications~\cite{DOE:2026genesis}. Note that these plans specify logical qubit counts of order $10^2$ and gate counts of order $10^8$. Against the resource estimates above, such a machine would run the $1+1$-dimensional and small $2+1$-dimensional simulations described in this white paper with error correction, but it would not run a $3+1$-dimensional simulation of QCD with matter, whose costs exceed these numbers by many orders of magnitude~\cite{DiMeglio:2023nsa,Shaw:2020udc}. Therefore, the proposed near-term program is written for mitigated devices of a few hundred physical qubits, and the medium-term program plans for the first error-corrected devices.

\begin{table}
  \centering
  \footnotesize
  \renewcommand{\arraystretch}{1.2}
  \setlength{\tabcolsep}{4pt}
  \begin{tabular}{l p{3.4cm} p{3.6cm} p{9.0cm}}
    \hline\hline
     & Milestone & Models & Completion criterion \\
    \hline
    \multicolumn{4}{l}{\textit{Near term, error-mitigated devices with a few hundred physical qubits}} \\
    1 & \raggedright Benchmark of the four ingredients & \raggedright Schwinger model & \raggedright The vacuum is prepared, a source is inserted, the state is evolved, and hadron multiplicities and positions are measured. The results agree with analytic results~\cite{Iso:1988zi} and tensor network simulations within stated errors. The dependence on the truncation and the lattice spacing is reported. \tabularnewline
    2 & \raggedright Meson and baryon production & \raggedright $1+1$D $\mathrm{SU}(2)$ and $\mathrm{SU}(3)$ with fermions & \raggedright Meson and baryon yields from string breaking are measured. The results agree with tensor network simulations, and convergence in the electric field cutoff is demonstrated. \tabularnewline
    3 & \raggedright Test of the Lund model assumptions & \raggedright $\mathbb{Z}_2$, Schwinger model, $1+1$D non-Abelian theories & \raggedright The independence of successive breaks, the location of breaks along the string, the form of $f(z)$, and their dependence on the quark mass are compared with the Lund model at the same string energy. The comparison is repeated at more than one lattice spacing. \tabularnewline
    4 & \raggedright Qualitative generator modification & \raggedright Features found in more than one of the models above & \raggedright A feature found in the simulations, including a feature that no existing model anticipates, is added to an event generator. The modified generator is fit to data and compared with LHC measurements that were not used in the fit, such as identified-particle spectra and baryon yields. \tabularnewline
    \hline
    \multicolumn{4}{l}{\textit{Medium term, first error-corrected devices}} \\
    5 & \raggedright Dynamical gluons and transverse structure & \raggedright Plaquette chains, $2+1$D $\mathrm{SU}(3)$ & \raggedright Glueball production and the transverse momentum of emitted hadrons are measured. The results agree with tensor network simulations where these are available, and convergence in the electric field cutoff is demonstrated. \tabularnewline
    6 & \raggedright Hadronization in a medium & \raggedright $1+1$D and $2+1$D theories at finite temperature and density & \raggedright The dependence of hadron yields on temperature and chemical potential is measured. The thermal or finite-density initial state is prepared and validated against tensor network simulations on small lattices. \tabularnewline
    7 & \raggedright Interface to event generators & \raggedright No quantum hardware required & \raggedright An operational definition of the dynamical string tension is given. A prescription is published that separates perturbative collinear physics from hadronization and matches a simulated quantity to a shower at a defined $Q_0$. The prescription satisfies the consistency conditions of Ref.~\cite{Hoang:2024nqi}, and the quantity handed to the generator is specified. \tabularnewline
    \hline
    \multicolumn{4}{l}{\textit{Long term, fault-tolerant devices}} \\
    8 & \raggedright Quantitative non-perturbative input & \raggedright $3+1$D QCD with matter & \raggedright A soft function is computed fully differentially in the energy flow~\cite{Bauer:2025nzf} and extrapolated to the continuum. It is used in an event generator through the prescription of milestone 7, with a stated total uncertainty. \tabularnewline
    \hline\hline
  \end{tabular}
  \caption{Proposed milestones for the program described in Section~\ref{sec:program}. Every simulation milestone requires preparation of the vacuum, insertion of a colored source, real-time evolution, and measurement of observables, with controlled truncation errors and continuum extrapolations. Milestone 7 is a theoretical milestone and can be pursued in parallel with the others.}
  \label{tab:milestones}
\end{table}

In the near term, the deliverable is qualitative modification of event generators rather than full event simulation. Since the dynamics of hadronization have never been accessible, there is no reason to expect that the current models contain all of their relevant features. The most valuable results of a simulation may be qualitative features that no model anticipates, and observables and analysis techniques must be developed to find such features rather than only to test the assumptions listed in Section~\ref{sec:intro}. Note that no quantum-simulated quantity has yet been inserted into a production generator, and which object should be handed over, whether a soft function, a fragmentation kernel, or a multiplicity distribution, remains to be settled.

Qualitative features of hadronization can be probed through simulations of the ladder of toy models proposed in Section~\ref{sec:toy}. The Schwinger model comes first, because analytic results allow the systematic errors of a simulation to be validated~\cite{Iso:1988zi}. Non-Abelian theories in $1+1$ dimensions add color, plaquette chains add dynamical gluons and a magnetic term without being two-dimensional, and $2+1$-dimensional $\mathrm{SU}(3)$ adds transverse string dynamics and glueballs~\cite{Ciavarella:2024fzw,Ciavarella:2025bsg}. Each stage of this ladder gets closer to full QCD, and qualitative features present in all of them may be worth incorporating into event generators. Every stage requires the same four ingredients, which are preparation of the vacuum, insertion of a colored source, real-time evolution, and measurement of observables. They must also be performed with controlled truncation error and continuum extrapolations. These simulations should be performed for hadronization both in the vacuum and at finite temperature and density. This is directly relevant to heavy-ion collisions, where the decadal priorities of the hot-QCD community explicitly welcome such work~\cite{Arslandok:2023utm,Farrell:2024mgu,Batini:2024zst}. Table~\ref{tab:milestones} lists proposed milestones for this program and which models need to be simulated to reach the milestone.

In the long term, extension to $3+1$-dimensional QCD with matter depends on fault-tolerant resources~\cite{DiMeglio:2023nsa,Shaw:2020udc}. Realizing any of this at scale will require quantum devices to be integrated as accelerators within existing high-performance computing workflows, together with the classical benchmarking and verification on which every claim will rest~\cite{Bauer:2022hpo,Humble:2022vtm}.

\section{Conclusions}
\label{sec:conclusions}

In this white paper, the case for applying quantum simulation to hadronization has been set out. Hadronization has been identified as a limiting systematic at the LHC. The limitations of classical methods have been stated. Euclidean lattice QCD reaches elastic scattering through finite-volume methods and parton distributions through LaMET, but it does not reach the real-time evolution of a breaking string or the matrix elements of operators with several light-like directions. Tensor-network methods are limited by the growth of entanglement. Quantum computing provides a way around these limitations. Factorization has been used to identify the non-perturbative objects a quantum computer should compute. The toy models used in present simulations have been described together with the physics each provides. The observables to be extracted have been listed. They include the momentum, flavor, and multiplicity quantities that generators currently fit. They also include the entanglement and magic measures that have no counterpart in any generator, but may provide a new viewpoint on the dynamics of hadronization.

The qualitative features of string breaking and hadron production are already accessible on current devices, and qualitative modifications to event generators are a realistic near-term deliverable. Quantitative accuracy will likely require further hardware and algorithmic development. It will also be necessary to develop a prescription that separates perturbative collinear physics from hadronization in a string simulation. The fault-tolerant machines scheduled for 2028 and 2029 are specified at a few hundred logical qubits, which is sufficient for the low-dimensional stages of this program. These low-dimensional simulations will provide useful insights for future simulations of QCD in $3+1$ dimensions.

\begin{acknowledgments}
This white paper is the output of the InQubator for Quantum Simulation (IQuS) workshop \textit{From String Dynamics to Event Generation with Quantum Simulation}, held in April 2026. We thank the workshop participants Ivan Burbano, Roland Farrell, Boyu Gao, Jeremy Hartse, Yutaro Iiyama, Natalie Klco, Marina Krsti\'c Marinkovi\'c, Anne Lashbrook, Sarah Powell, and Prasanth Shyamsundar for discussions that led to the creation of this white paper. The writing of this white paper was performed in part at the Aspen Center for Physics, which is supported by National Science Foundation grant PHY-2210452. We acknowledge the use of Claude Fable 5.1 in generating early drafts of this work and in the formatting of figures in the main text.

\end{acknowledgments}

\appendix

\section*{Appendix: Workshop contributions}

This appendix contains the unaltered original contributions of each author to this whitepaper.

\appendixitem{app:bauer}{Using Quantum Computers for the Long-Distance Physics of Collider Observables}{C.~W. Bauer}
\appendixitem{app:ciavarella}{String Breaking in SU(2) Lattice Gauge Theory: What Can We Measure?}{A.~N. Ciavarella}
\appendixitem{app:cataldi-xu-halimeh}{Real-Time String Breaking in Lattice Gauge Theories: Status and Challenges for Quantum and Tensor-Network Simulation}{G.~Cataldi, K.~Xu, and J.~C. Halimeh}
\appendixitem{app:froland}{Resource Theories in Quantum Simulations of HEP}{H.~Froland}
\appendixitem{app:grabowska}{Comparing Lattice Formulations for Studies of Hadronization}{D.~M. Grabowska}
\appendixitem{app:grieninger}{The Quantum Structure of Confining Strings}{S.~Grieninger}
\appendixitem{app:hite}{Complexity Probes for Choosing Hybrid Simulation Methods}{M.~Hite}
\appendixitem{app:kadam}{Toward First-Principles Quantum Simulation of Hadronization}{S.~V. Kadam}
\appendixitem{app:lamm}{I Can Do It With a Broken String}{H.~Lamm}
\appendixitem{app:li-illa}{Heavy-Hadron Energy Loss and Physics-Meaningful Benchmarks}{Z.~Li and M.~Illa}
\appendixitem{app:pascuzzi}{Resource Estimation, Classical Benchmarking, and HPC Integration}{V.~R. Pascuzzi}
\appendixitem{app:senseman}{Progress towards Hybrid Quantum Hadronization for Event Generators}{B.~Senseman, Z.~Ozzello, K.~Heitritter, Y.~Meurice, and S.~Mrenna}
\appendixitem{app:yao}{String Breaking}{X.~Yao}
\appendixitem{app:zemlevskiy}{Quantum Complexity as Event-Generator Input}{N.~A. Zemlevskiy}
\appendixcontents

\section{Using Quantum Computers for the Long-Distance Physics of Collider Observables}\label{app:bauer}
\contribution
  {Christian W. Bauer}
  {Physics Division, Lawrence Berkeley National Laboratory, Berkeley, CA 94720, USA; Department of Physics, University of California, Berkeley, CA 94720, USA}
  {}
\subsection{Simulating long-distance physics rather than the full theory}

High energy physics is concerned with testing physics at short distances, and therefore at high energies.
However, predictions for any collider observable depend not only on the short-distance physics one is interested in, but also on the physics at all scales down to $\lqcd$ and below.
Due to asymptotic freedom the short-distance physics can be calculated reliably using perturbation theory, while the long-distance physics requires a non-perturbative treatment.
The only known technique to compute non-perturbative quantities in QCD from first principles is Euclidean lattice QCD, and many of the quantities that are required cannot be obtained in this way, since they involve real-time dynamics~\cite{Troyer:2004ge,Alexandru:2016gsd} or matrix elements of operators containing several light-like directions.
In seminal work by Jordan, Lee and Preskill~\cite{Jordan:2012xnu,Jordan:2011ci} it was shown that quantum computers are able to simulate the dynamics of quantum field theories with resources that scale polynomially in the number of lattice sites, such that in principle one can simulate the full scattering process at a collider.
The main problem is that such a simulation requires way too many lattice sites to be feasible on any realistic time scale.
This can be understood from the fact that the lattice spacing has to be inversely proportional to the largest energy scale one aims to capture, while the size of the lattice has to be inversely proportional to the smallest scale, such that the number of lattice sites scales as
\begin{align}
\label{eq:bauer-LatticeSites}
    n_L \sim \left( \frac{E_{\rm max}}{E_{\rm min}} \right)^3
    \,.
\end{align}
Simulating the entirety of a collision therefore requires $n_L \gtrsim 10^{9}$ for LEP and $n_L \gtrsim 10^{15}$ for the LHC, with $\cO(10-100)$ qubits per lattice site~\cite{Bauer:2025nzf}, and it is clear that such complete simulations are out of reach in any foreseeable future.
Note that most of these lattice sites would be spent on physics far above $\lqcd$, which can be calculated perturbatively with sufficient precision.
Given that quantum hardware will likely remain precious for a long time, one should therefore aim to use it only for those effects that are inaccessible to traditional methods.

In Ref.~\cite{Bauer:2025nzf}, building on Ref.~\cite{Bauer:2021gup}, we argued that in the context of collider physics quantum computers are best used to calculate the non-perturbative matrix elements arising in factorization theorems.
Factorization theorems~\cite{Collins:1981uk,Collins:1981ta,Collins:1987pm,Collins:1989gx}, which can be derived systematically~\cite{Bauer:2002nz,Bauer:2008jx,Lee:2006fn,Bauer:2008dt,Jouttenus:2011wh} using soft-collinear effective theory (SCET)~\cite{Bauer:2000ew,Bauer:2000yr,Bauer:2001ct,Bauer:2001yt}, separate the physics at different scales, and for a jet-like observable $e$ one finds schematically
\begin{align}
\label{eq:bauer-Factorization}
    \frac{{\rm d}\sigma}{{\rm d}e} = \big[ H \otimes J_1 \otimes J_2 \otimes F_e \big](e)
    \,,
\end{align}
where the hard function $H$ and the jet functions $J_i$ are perturbatively calculable, while all non-perturbative physics is contained in the so-called soft function
\begin{align}
\label{eq:bauer-SoftFunction}
    F_e(e_s) = \big| \bra{X(e_s)} \, T \big[ Y_n^\dagger Y_{\bar n} \big] \ket{0} \big|^2
    \,.
\end{align}
Here $Y_n$ denotes a soft Wilson line along the light-like direction $n$, and $\ket{X(e_s)}$ denotes a final state with event shape $e_s$.
Traditionally, factorization theorems have proven most useful when the non-perturbative ingredients are universal, in the sense that the same matrix element contributes to more than one observable, such that it can be extracted from one measurement and used to predict another.
However, for most event shapes and essentially all jet substructure observables the soft function depends on the observable itself, and since it contains two light-like Wilson lines it cannot be computed using Euclidean lattice methods.
First-principles predictions for such observables are therefore currently not possible.

An important observation is that the soft function in \cref{eq:bauer-SoftFunction} is the absolute value squared of an amplitude, and is therefore positive definite and normalized to unity.
In other words, it is a probability distribution, and a quantum computer can directly sample from this distribution by preparing the interacting QCD vacuum, applying the unitary Wilson line operator, and measuring the event shape of the resulting final state, in the same way as the observable is measured experimentally.
A second observation is that the soft function only describes physics below the soft scale $\Lambda_s \sim Q \, e$.
This implies that the lattice spacing is set by $\Lambda_s$ rather than the collider energy, such that the number of lattice sites in \cref{eq:bauer-LatticeSites} is no longer parametrically enhanced, and a lattice of order $50^3$ sites should be sufficient.
One can generalize this further by defining a fully differential soft function~\cite{Bauer:2008jx} of the energy flow $\cE(\hat n)$ of the final state~\cite{Lee:2006nr,Sveshnikov:1995vi,Korchemsky:1997sy}.
In a lattice simulation the energy flow is obtained by reading out the energy of the qubits at the boundary of the lattice, integrated over the simulation time, and the soft function for any observable for which a factorization theorem exists can then be obtained by post-processing the measured energy distribution.
Note that this does not require the soft function to be universal.

The general lesson of this discussion is that one should identify the smallest quantum mechanical object that is genuinely inaccessible to classical methods, prepare this object directly rather than producing it in a simulated collision, and measure it in the same way as the experiment does.
Despite these simplifications, the calculation of the soft function of \cref{eq:bauer-SoftFunction} in 3+1 dimensions is likely more than a decade away.
We will now argue that the same logic applies to a long-distance problem that can be addressed on a much shorter time scale, namely the hadronization process.

\subsection{Hadronization as a target for quantum simulation}

The hadronization process, which describes how the quarks and gluons taking part in the short-distance process turn into the observed hadrons, is one of the least understood processes of the Standard Model, and it enters essentially every measurement at a collider.
Since hadronization is an intrinsically real-time process with colored intermediate states, it is inaccessible to Euclidean lattice QCD, and it is currently only described by phenomenological models implemented in exclusive event generators~\cite{Sjostrand:2006za,Bahr:2008pv,Gleisberg:2008ta}.
The two most common models are the Lund string model~\cite{Andersson:1983jt} and the cluster model~\cite{Webber:1983if}.
While the two models differ in their details, both describe at their core how a color string, or a color singlet cluster, produced at the end of the parton shower evolves into a set of hadrons, and both make relatively simple, semi-classical assumptions about this evolution, with several free parameters that are tuned to data.
In the string model, for example, the string has a constant tension $\kappa$, it breaks through the production of $q \bar q$ pairs by a tunneling process with a Schwinger-like probability proportional to $\exp(-\pi m_\perp^2/\kappa)$~\cite{Schwinger:1951nm,Casher:1978wy}, successive breaks are assumed to be causally disconnected such that they can be ordered arbitrarily, the longitudinal momentum of the hadrons is distributed according to a fragmentation function $f(z)$, and the transverse momentum of the hadrons is given by the Gaussian tunneling $p_\perp$, which is locally compensated between neighboring hadrons.
None of these assumptions has been tested against QCD itself.
The only way to test them currently is to compare the resulting hadron distributions to experimental data, and this has several drawbacks.
First, it is not possible to prepare a color string experimentally, such that the hadronization effects can never be separated from the perturbative physics of the short-distance process and the parton shower.
Second, the detectors only observe the final state at essentially infinite time, such that no information about the dynamics of the hadronization process itself is available.
Finally, it is impossible to assign rigorous systematic uncertainties to such models.
The uncertainties are usually estimated by comparing different models, and even with this prescription they dominate the systematic uncertainty of the top quark mass measurement, of the extraction of $\alpha_s$ from event shapes, and of many jet substructure measurements.

Quantum simulation allows one to address this problem in the same way as discussed for the soft function above~\cite{Bauer:2022hpo,Bauer:2021gup}.
Rather than simulating the full collider process, one directly prepares the object the models describe, namely a color string between a heavy quark and antiquark on top of the interacting vacuum, and evolves it in real time using a Hamiltonian formulation of the gauge theory~\cite{Kogut:1974ag,Byrnes:2005qx}.
A key insight that makes this a near-term target is that the main features of the hadronization models rely on dynamics in only two spatial dimensions, namely the directions longitudinal and transverse to the string.
This implies that simulations in 1+1 and 2+1 dimensions with of order 50 lattice sites per dimension already probe the physics the models are built on, and existing resource estimates~\cite{Farrell:2023fgd,Farrell:2024fit,Ciavarella:2025bsg,Ciavarella:2024fzw} indicate that such simulations will be possible on hardware available on a time scale of a few years.
In fact, first simulations of string breaking already exist, in 1+1 dimensions on more than a hundred superconducting qubits~\cite{Farrell:2024fit,Ciavarella:2024lsp}, in 2+1 dimensional $\mathbb{Z}_2$ and U(1) theories on superconducting, Rydberg, and trapped-ion devices~\cite{Cochran:2024rwe,Gonzalez-Cuadra:2024xul,Crippa:2024hso}, on analog simulators~\cite{Banerjee:2012pg,Liu:2024lut}, as well as using tensor network and classical-statistical methods~\cite{Hebenstreit:2013baa,Pichler:2015yqa,Buyens:2016hhu,Kuhn:2015zqa,Spitz:2018eps,Belyansky:2023rgh,Florio:2023dke}.
What is missing so far is a connection between these simulations and the observables predicted by the hadronization models, as well as simulations with quantified systematic uncertainties.
Both are the subject of ongoing work together with M.~Savage.

Note that there is one additional complication that requires further research.
The quarks at the endpoints of the string typically move apart with high energies, which introduces a perturbative collinear scale into the problem.
This implies that a simulation of the string contains both the non-perturbative long-distance physics of hadronization and perturbative collinear physics, and in order to compare with hadronization models, which only describe the former, the two have to be separated.
How to achieve this separation, presumably through a factorization of the kind discussed in the first section, is an open research question, and it has to be answered before the results of quantum simulations can be used to constrain hadronization models quantitatively.

\subsection{Qualitative features of hadronization accessible to quantum simulation}

In this section we discuss which assumptions of the hadronization models can be tested using quantum simulations of gauge theories in one or two spatial dimensions.
Note that the goal is not quantitative agreement between such a theory and QCD.
The hadronization models make a number of structural assumptions, and a quantum simulation can establish whether these assumptions hold or fail, in several cases using hardware that exists today.
We discuss the ones we consider most relevant in turn.

The first feature is the time dependence of the hadronization process.
The hadronization models contain no time evolution, since the hadrons appear fully formed, while a Hamiltonian simulation gives the full time dependence of the local energy and charge densities along the string.
This allows one to study how the transition from partonic to hadronic degrees of freedom proceeds, on what time scale the hadrons form, and how the string breaks into the final hadronic state.
None of this information is accessible experimentally, since detectors only observe the final state.

The second feature is the dependence of the hadron distributions on the longitudinal and transverse momenta.
In the models the longitudinal dependence is parameterized by a fragmentation function, while the transverse momentum is generated through a tunneling process based on the Schwinger mechanism, and both are assumed to arise from independent string breaks.
A quantum simulation allows one to determine the longitudinal and transverse distributions of the produced hadrons directly, as well as their dependence on the mass of the produced quarks, and to test whether the parameterizations used in the models, and the assumed independence of successive breaks, are consistent with the underlying dynamics.

The third feature is the role of dynamical gauge fields and of the non-Abelian nature of the interaction.
Gauge theories in 1+1 dimensions contain no dynamical gluons, while in 2+1 dimensions the string acquires transverse dynamics, and in a non-Abelian theory new hadronic degrees of freedom in the form of glueballs, which are not present in any hadronization model.
Simulations of non-Abelian theories, first in 1+1 dimensions and then on so-called plaquette chains~\cite{Ebner:2024mee} and larger lattices using the large-$N_c$ formulations of SU(3) that make them tractable~\cite{Ciavarella:2025bsg,Ciavarella:2024fzw,Ciavarella:2024cyt}, will establish which features of the hadronic spectra and of the transverse dynamics arise from these effects, and whether they are captured by the models.

Finally, the string is a quantum state, and it was shown recently that quantum correlations such as entanglement and non-stabilizerness (magic) develop in the string and become maximal just before it breaks~\cite{Grieninger:2026bdq}.
Whether such correlations survive into the hadronic final state, in the form of correlations between hadrons that a classical, probabilistic model cannot produce, is a question a quantum simulation can answer, while an experiment cannot.
If they do, the models will require new parameterizations, and a simulation allows one to test the universality of such parameterizations by determining them from one observable and testing them on another.

Note that all of the features discussed so far are motivated by the assumptions of the existing hadronization models, in the sense that one takes an assumption of the model and asks whether it holds in the underlying gauge theory.
It is quite possible, however, that the most interesting results of quantum simulations will be qualitative features of the hadronization process that are not anticipated by any of the current models, and which can therefore not be found by testing their assumptions.
Since the dynamics of string breaking has never been accessible before, there is no reason to expect that the models capture all of its relevant features, and dedicated research is required to develop the observables and the analysis techniques that allow one to discover such features in the results of a simulation, rather than only to confirm or reject known assumptions.
We believe that this might in fact be the most promising route for quantum simulations to have an impact on our understanding of hadronization.

In summary, each of these features is a qualitative question about the structure of the hadronization process rather than a precision measurement, which is why they are well suited for near-term hardware, and the answer to each of them directly translates into a modified or newly justified assumption in the models used by essentially all experiments.
The staged program to answer them, which starts with the Schwinger model where analytical results allow one to validate the systematic uncertainties, continues with non-Abelian theories in 1+1 dimensions, and ends with plaquette chains and 2+1 dimensional SU(3) in the large-$N_c$ expansion, requires exactly the same ingredients as the soft function calculation discussed above, namely the preparation of the vacuum, the insertion of a colored source, the real-time evolution, and the measurement of the local energy flow.
Hadronization is therefore likely the first place where quantum simulation can have a direct impact on the experimental HEP program, while the soft functions of \cref{eq:bauer-SoftFunction} are where it will ultimately allow one to make first-principles predictions for observables for which no such predictions exist today.

\section{String Breaking in SU(2) Lattice Gauge Theory: What Can We Measure?}\label{app:ciavarella}
\contribution
  {Anthony N. Ciavarella\,\orcidlink{0000-0003-3918-4110}}
  {Applied Math and Computational Research Division, Lawrence Berkeley National Laboratory, Berkeley, California 94720, USA}
  {anciavarella@lbl.gov}
High energy physics is directly probed at particle colliders where large numbers of quarks and gluons are produced. As these particles propagate, the confining interactions of quantum chromodynamics (QCD) bind them together into hadrons. This hadronization process is intrinsically real time and non-perturbative. Direct classical simulations of these dynamics suffer from a sign problem and require exponentially scaling resources. Quantum computers offer the potential to directly probe these dynamics, and simulations of string breaking and hadron dynamics in low dimensional field theories have recently been performed on quantum hardware~\cite{De:2024smi,Florio:2024aix,Davoudi:2024wyv,Zemlevskiy:2024vxt,Ciavarella:2024lsp}. However, encoding a theory onto a quantum computer is only the first step in performing a simulation. Physically relevant states such as the vacuum and hadrons need to be prepared, and physically relevant observables such as the number and momenta of produced hadrons need to be measured.

For scalar field theories, Jordan, Lee, and Preskill (JLP) proposed performing both state preparation and measurement through adiabatic switching~\cite{Jordan:2012xnu,Jordan:2011ci}. In this construction, wavepackets of the free theory are prepared and the interaction is turned on adiabatically, mapping them onto wavepackets of the interacting theory. After the real-time evolution of interest, the interaction is adiabatically switched off. The particle content of the free theory is known explicitly, so the occupation number of each free field mode can be measured directly. Therefore, the adiabatic switch does more than prepare states. It defines the observables of the simulation by mapping the interacting final state onto free particles that can be counted and binned in momentum. This construction comes with theoretical guarantees on its accuracy and cost. However, the required circuit depths have mostly restricted its use to theoretical studies.

For confining gauge theories, the free limit is the wrong starting point. At $g=0$, the degrees of freedom are free quarks and gluons, while the asymptotic states of the interacting theory are hadrons. The strong coupling limit provides a natural replacement. In this limit, the vacuum is a product state with no electric flux excited, and hadrons are localized excitations with transparent quantum numbers, e.g. a quark anti-quark pair joined by a short string of electric flux. Observables are correspondingly simple: hadron numbers and positions can be determined from local measurements of the electric field, for example by counting the ends of strings. Adiabatic switching from the strong coupling limit therefore generalizes the JLP construction to gauge theories. Strong coupling hadrons are adiabatically dressed into the hadrons of the interacting theory, and switching back to strong coupling before measurement maps the final state into a basis where hadronic observables are simple to evaluate. However, adiabatic evolution remains expensive. The circuit depths required are large, and the energy gaps protecting the adiabatic evolution can close as system sizes and string lengths grow~\cite{Surace:2024bht}.

This cost can be reduced with a variational procedure~\cite{Ciavarella:2024lsp}. A single unitary $\hat{U}(\vec{\theta})$ is optimized to simultaneously map the strong-coupling vacuum onto the interacting vacuum and strong-coupling single-hadron states of definite momentum onto the corresponding interacting eigenstates. Note that the sum over momentum states in the cost function collapses into a sum over position states, so the optimization can be performed without preparing entangled plane wave states. The resulting circuits can also be made scalable. Provided the lattice is larger than a few correlation lengths, the optimal variational parameters converge exponentially with system size, and parameters determined classically on small lattices can be extrapolated to lattices beyond the reach of brute force classical simulation~\cite{Farrell:2023fgd,Burbano:2025pef}. Measurement is performed by applying $\hat{U}^\dagger(\vec{\theta})$ before reading out the qubits, providing a variational replacement for the adiabatic switch-off. This gives direct access to hadron positions and to the full hadron multiplicity distribution $P(n)$ of the final state. This procedure was demonstrated in a quantum simulation of string breaking in an SU(2) lattice gauge theory performed on \texttt{ibm\_torino}~\cite{Ciavarella:2024lsp}.

These constructions only rely on translation invariance and the presence of a mass gap. It is expected that the same approach can be applied to other gauge groups, higher spatial dimensions, and different matter content. The qualitative features of string breaking and hadron production are already accessible on current devices. Applying this variational procedure to evolved states with a flux string present to perform momentum-resolved measurements would allow analogs of fragmentation functions to be extracted. Reaching quantitative accuracy at scale will likely require improved hardware error rates and error mitigation strategies.

\section{Real-Time String Breaking in Lattice Gauge Theories: Status and Challenges for Quantum and Tensor-Network Simulation}\label{app:cataldi-xu-halimeh}
\contribution
  {Giovanni Cataldi\,\orcidlink{0000-0002-9073-8978}, Kaidi Xu\,\orcidlink{0000-0003-2184-0829}, and Jad C. Halimeh\,\orcidlink{0000-0002-0659-7990}}
  {Max Planck Institute of Quantum Optics, 85748 Garching, Germany; Department of Physics and Arnold Sommerfeld Center for Theoretical Physics (ASC), Ludwig-Maximilians-Universit\"at M\"unchen, 80333 M\"unchen, Germany; Munich Center for Quantum Science and Technology (MCQST), 80799 M\"unchen, Germany}
  {}
\begingroup
\renewcommand{\labelenumi}{(\roman{enumi})}
\subsection{Motivation: from string-breaking to event generation}
In a non-Abelian gauge theory such as quantum chromodynamics (QCD), confinement forces the color-electric field between two static charges to collapse into a flux tube of approximately constant tension, so that the static potential grows linearly with the charge separation~\cite{Philipsen:1998de,Knechtli:2000df,Bali:2005fu,Bulava:2019iut}.
Stretching the tube eventually makes it energetically cheaper to nucleate a dynamical quark--antiquark pair from the vacuum than to keep extending it: the string \emph{breaks}, the potential flattens, and the fragments seed the hadrons observed in a detector.
This real-time conversion of stored field energy into matter is the microscopic face of \emph{hadronization}---the soft, non-perturbative stage that every collider event generator must model phenomenologically, through the Lund string picture of \textsc{Pythia}~\cite{Bierlich:2022pfr,Andersson:1983ia}, or the cluster model of \textsc{Herwig}~\cite{Webber:1983if,Amati:1979fg}.
Both descriptions are tuned rather than derived; a first-principles, real-time account of string fragmentation would benchmark, constrain, and eventually replace their free parameters.
The obstruction is that string breaking is at once far from equilibrium and non-perturbative.
Perturbation theory does not apply, while Euclidean Monte Carlo---the workhorse for static spectra, the linear potential, and lattice-QCD string breaking itself---cannot access real-time evolution because of a sign problem, which also obstructs QCD at finite baryon density~\cite{Kieu:1993gw,Troyer:2004ge,Nagata:2021ugx}\footnote{Two-color theories with an even number of flavors evade the finite-density sign problem~\cite{Hands:1999md,Kogut:2000ek}, but not in real time.}.
The Hamiltonian formulation of lattice gauge theories (LGTs) sidesteps both limitations and has become the natural route, pursued in parallel on classical tensor networks (TNs) and on quantum hardware~\cite{Bauer:2023qgm,Zohar:2021nyc}.
Over roughly the past two years, a wave of $(2{+}1)$D experiments and TN studies has probed string breaking and glueball formation, first in Abelian $\mathbb{Z}_2$ and U(1) theories~\cite{Cochran:2024rwe,Cobos:2025krn,Gonzalez-Cuadra:2024xul,Xu:2026ibi,Joshi:2026hfe,Pichler:2015yqa,Xu:2025abo,Tian:2025mbv} and, most recently, in the non-Abelian SU(2) case, numerically with dynamical matter~\cite{Cataldi:2025cyo} and on hardware in a pure-gauge, truncated setting~\cite{John:2026gut}.
Tables~\ref{tab:halimeh-platforms} and~\ref{tab:halimeh-numerics} summarize this landscape, which sets the backdrop for the challenges discussed below.

\subsection{Model}\label{sec:halimeh-model}
Hamiltonian LGTs discretize space on a lattice $\Lambda$ while keeping time continuous, hosting fermionic matter on the sites and the gauge field on the links~\cite{Wilson:1974sk}.
The matter fields $\hpsi_{\alpha,\vecsite}$ are staggered fermions~\cite{Susskind:1976jm}, which spread the Dirac components over a unit cell and thereby tame---though, for $D>1$, do not fully remove---the fermion doublers at low cost; the color index $\alpha$ runs over the fundamental representation of the gauge group $\gaugegroup$.
Each link carries a parallel transporter $\NApara$ together with the left and right chromo-electric fields that generate the gauge rotations at its two ends.
A local gauge transformation acts jointly on a site and all its attached links, and the physical states are those obeying Gauss's law (the non-Abelian generalization of $\nabla\!\cdot\!\vb{E}=\rho$) up to the static \emph{background charges} we place to pin the two endpoints of a string.
A matter-coupled Yang--Mills theory is then defined by the Kogut--Susskind Hamiltonian~\cite{Kogut:1974ag} ($\hbar\!=\!c\!=\!1$):
\begin{equation}
  \label{eq:halimeh-ham}
  \begin{split}
    \ham = & -w\!\sum_{\vecsite,\latvec}\sum_{\alpha,\beta}\!\big(s_{\vecsite,\latvec}\,\hpsi_{\alpha,\vecsite}^{\dagger}\NApara \hpsi_{\beta,\vecsite+\latvec}+\hc\big) \\
           & + m\!\sum_{\vecsite}\sum_{\alpha}\! s_{\vecsite}\,\hpsi_{\alpha,\vecsite}^{\dagger}\hpsi_{\alpha,\vecsite}                                                  \\
           & + g_E\!\sum_{\vecsite,\latvec}\!\casimir
    \;-\; g_B\!\sum_{\square}\!\mathrm{Tr}(\plaq+\plaq*)\,,
  \end{split}
\end{equation}
where the staggered phases $s_{\vecsite,\latvec},s_{\vecsite}\!\in\!\{\pm1\}$ implement the staggered-fermion construction \cite{Magnifico:2024eiy}.
The four terms are, in order: the matter--gauge \emph{hopping} of amplitude $w$, in which a quark tunnels to a neighbor dressed by the transporter $\NApara$; the staggered \emph{mass} $m$; the \emph{chromo-electric} energy, diagonal in the irreducible-representation (irrep) basis with $\casimir\ket{\spin}=c_{\spin}\ket{\spin}$ the quadratic Casimir ($c_{\spin}=\spin(\spin+1)$ for SU(2), and its Abelian analog for $\mathbb{Z}_n$/U(1)), so a link carrying irrep $\spin$ costs $\sim\! g_E\,c_{\spin}$; and the \emph{chromo-magnetic} energy, a four-link \emph{plaquette} term that exists only for $D>1$ and endows the gauge field with the transverse dynamics separating genuine higher-dimensional physics from an effectively $(1{+}1)$D one.
The couplings scale as $g_E\!\sim\! g^2/a^{D-2}$ and $g_B\!\sim\!1/(g^2 a^{4-D})$ with the bare gauge coupling $g$ and lattice spacing $a$; the majority of the aforementioned numerical and quantum simulations are performed in the strong-coupling regime $g_E,m\gg g_B,w$, where the lowest irreps dominate, and resonances are sharp.

The idea is to quench a far-from-equilibrium electric string, $\ket{\psi(t)}\!=\!e^{-i\ham t}\ket{\psi_0}$, where $\ket{\psi_0}$ is a tube of \emph{active links} (links carrying nontrivial flux, $\spin>0$) joining two static charges, and follow a small set of gauge-invariant probes drawn from the studies in Tables~\ref{tab:halimeh-platforms}--\ref{tab:halimeh-numerics}:
\begin{table*}[ht]
  \footnotesize
  \setlength{\tabcolsep}{1.7pt}%
  \caption{\textbf{Realized (2+1)D string-dynamics quantum simulators.} Two-qubit (two-qudit) entangling depth and gate counts per Trotter step and for the deepest circuit, as reported in each reference. These costs depend on native gate sets and compilation and are not directly comparable across platforms. ``n.r.'': not reported in comparable form.}
  \begin{ruledtabular}
    \begin{tabular}{l l l l l c c c c}
      Platform (ref.)                                                    & Group (matter)                        & Lattice (size)                        & Register                          & Plaquette   & Depth/step    & Gates/step & Steps        & Total                         \\
      \hline
      SC, Google \cite{Cochran:2024rwe}             & $\mathbb{Z}_2$ (dyn.$^{a}$)           & square, $3\!\times\!4$ vertices       & 35 qubits$^{b}$                   & fixed       & 8             & 116        & 9            & $\approx\!1075$               \\
      SC, IBM \cite{Cobos:2025krn}                        & $\mathbb{Z}_2$-Higgs (dyn.)           & heavy-hex, $\le\!7\!\times\!3$        & $\le\!144$ qubits                 & no          & n.r.          & n.r.       & n.r.         & depth $\le\!192^{c}$          \\
      Ion, Quantinuum \cite{Xu:2026ibi}         & $\mathbb{Z}_2$ (dyn.$^{a}$)           & square, $6\!\times\!5$ sites          & 56 qubits$^{d}$                   & tunable     & 6             & 128        & $\le\!8$     & $\le\!1122$                   \\
      Ion, Quantinuum \cite{Joshi:2026hfe}            & U(1) QLM (dyn.)                       & square, $5\!\times\!4$ sites          & 51 qubits                         & tunable     & 28            & n.r.       & n.r.         & $\le\!1540$                   \\
      SC, IBM \cite{Gandon:2026das}                         & U(1) QLM$^{e}$ (static)               & hexafoil, $\le\!114$ plaq.            & $\le\!112$ qubits                 & tunable     & $6{+}8^{f}$   & n.r.       & $\le\!32$    & n.r.                          \\
      Rydberg, QuEra \cite{Gonzalez-Cuadra:2024xul}  & U(1)$^{g}$ (dyn.)                     & kagome, $\le\!5\!\times\!5$ cells     & 31--59 atoms                      & no          & \multicolumn{4}{c}{analog; quench time $\le\!1.6\,\mu$s}                 \\
      Ion qudits \cite{John:2026gut}          & SU(2)$_2^{h}$ (static)                & ladder, 3 plaq.                       & 3 qudits ($d\!\le\!8$)            & yes         & n.r.          & n.r.       & 2            & 67                            \\
    \end{tabular}
  \end{ruledtabular}
  \vspace{2pt}
  {\scriptsize\raggedright
  $^{a}$Ising matter integrated out via Gauss's law (toric-code form); charges remain dynamical.
  $^{b}$17 link + 18 ancilla qubits.
  $^{c}$Total two-qubit layers, not per step.
  $^{d}$49 link qubits + 7 leakage-detection ancillas.
  $^{e}$Pure gauge, dual height variables.
  $^{f}$Entangling layers for the petal and triangle terms, respectively.
  $^{g}$Emergent from the Rydberg blockade.
  $^{h}$$q$-deformed SU(2) at level $k\!=\!2$, pure gauge with static charges.\par}
  \label{tab:halimeh-platforms}
\end{table*}

\begin{table*}[ht]
  \footnotesize
  \caption{\textbf{Selected classical numerical studies of (1+1)D and (2+1)D string dynamics.}}
  \begin{ruledtabular}
    \begin{tabular}{l l l l}
      Study                                                                        & Gauge Group                  & Method      & String-dynamics focus                                \\
      \hline
      \multicolumn{4}{l}{\emph{Abelian,(1+1)D}}                                                                                                                                         \\
      Pichler et al.\ 2016 \cite{Pichler:2015yqa}                    & U(1)$+$matter                 & MPS         & real-time string \& pair production                  \\
      Magnifico et al.\ 2020 \cite{Magnifico:2019kyj}                  & $\mathbb{Z}_n$$+$matter       & t-DMRG      & confinement, real-time dynamics                      \\
      Chanda et al.\ 2020 \cite{Chanda:2019fiu}           & U(1)$+$matter                 & TDVP        & confinement, lack of thermalization (bosonic matter) \\
      Belyansky et al.\ 2024 \cite{Belyansky:2023rgh}         & U(1) Schwinger                & uniform MPS & string in high-energy collisions                     \\
            \hline
      \multicolumn{4}{l}{\emph{Abelian, (2+1)D}}                                                                                                                                        \\
      Di Marcantonio et al.\ 2025 \cite{DiMarcantonio:2025cmf} & $\mathbb{Z}_2$                & MPS         & flux-string roughening \& dynamics                   \\
      Wu et al.\ 2025 \cite{Wu:2025aly}             & Abelian                       & PEPS+VMC    & real-time excitations (method)                       \\
      Xu et al.\ 2025 \cite{Xu:2025abo}                          & $\mathbb{Z}_2$$+$matter       & MPS+TDVP    & string breaking, glueball-like excitations           \\
      Tian et al.\ 2025 \cite{Tian:2025mbv}                           & U(1)/$\mathbb{Z}_2$ $+$matter & MPS+TDVP    & role of plaquette term in string-dynamics            \\
          \hline
      \multicolumn{4}{l}{\emph{Non-Abelian,(1+1)D}}                                                                                                                                     \\
      K\"uhn et al.\ 2015 \cite{Kuhn:2015zqa}                  & SU(2)$+$matter                & MPS         & first non-Abelian string breaking                    \\
      Spitz \& Berges 2019 \cite{Spitz:2018eps}                 & SU(2)$+$matter                & LIH         & pair production \& string breaking                   \\
          \hline
      \multicolumn{4}{l}{\emph{Non-Abelian, (2+1)D}}\\                                 Cataldi et al. 2025 \cite{Cataldi:2025cyo}         & SU(2)$+$matter    &TTN& string-breaking, glueball-like excitations, blockade, tension-reduction \\
    \end{tabular}
  \end{ruledtabular}
  \label{tab:halimeh-numerics}
\end{table*}
\begin{enumerate}
  \item the string-manifold weight $\mathcal{P}(t)\!=\!\sum_k\abs{\braket{\psi_k|\psi(t)}}^2$, summed over all flux configurations $\{\psi_k\}$ of the same length as $\ket{\psi_0}$; its collapse signals string breaking, while the weight $\mathcal{P}_{\gamma\neq\gamma_{\rm i}}$ on configurations \emph{other} than the initial one witnesses genuine higher-dimensional rearrangement;
  \item the electric energy $C=\avg{\casimir}$ per link, averaged over the string region, whose drop tracks the release of confined flux;
  \item the produced-matter content, i.e.\ the meson and (in the non-Abelian case) baryon densities $p_{\rm mes},p_{\rm bar}$;
  \item the bipartite entanglement entropy defined as $S\!=\!-\!\mathrm{Tr}\rho_A\!\ln\rho_A$, a state-complexity and breaking witness that grows continually when the string breaks and saturates otherwise;
  \item closed-loop operators $\hat W_{\square}$ that detect pure-gauge flux loops detached from the string. 
  In the strong-coupling regime, these operators measure glueball-like excitations \cite{Mathieu:2008me}.
\end{enumerate}
A single principle organizes the dynamics: a flux rearrangement becomes \emph{resonant} when the electric energy released by lowering the flux matches the pair-creation cost $2m$.
In SU(2) this promotes the single Abelian condition to a whole family set by the irrep tower (\cref{sec:halimeh-physics}) \cite{Cataldi:2025cyo}.

\subsection{State of the art}
\label{sec:halimeh-soa}

Two trends stand out.
Abelian platforms dominate the hardware column (Table~\ref{tab:halimeh-platforms}): $\mathbb{Z}_2$ and U(1) theories have been run on superconducting, trapped-ion, and Rydberg devices with up to $\mathcal{O}(100)$ qubits, whereas the only non-Abelian realization so far is a trapped-ion qudit simulation of a pure-gauge, $q$-deformed $\mathrm{SU}(2)_2$ truncation on a three-plaquette ladder, in which strings between adjoint static charges break through \emph{gluonic} screening rather than dynamical pair creation~\cite{John:2026gut}.
Meanwhile, numerics (Table~\ref{tab:halimeh-numerics}) have already mapped much of the phenomenology---real-time breaking, roughening, glueball-like loops, and the decisive role of the plaquette term---that hardware is beginning to reach, and have entered the regime hardware has not: the first non-Abelian $(2{+}1)$D real-time simulation of SU(2) Yang--Mills with dynamical matter~\cite{Cataldi:2025cyo}.
The challenges below separate these two columns from a controlled, continuum, non-Abelian description.
\subsection{Challenges}
\label{sec:halimeh-challenges}
\subsubsection{Going beyond (1+1)D}
Genuine higher-dimensional physics requires the magnetic (plaquette) term, which, in the Kogut--Susskind formulation (see \cref{eq:halimeh-ham}), is a four-body operator that is absent in one spatial dimension.
Its presence is a decisive diagnostic: deep in the confined regime and for minimal (Manhattan-distance) strings, real-time breaking \emph{without} a plaquette maps onto an effectively $(1{+}1)$D process independent of the lattice geometry, as shown for both square and honeycomb lattices~\cite{Tian:2025mbv}.
A plaquette generated perturbatively from the hopping arises only at high orders in $w/g_E$ (fourth order on the square lattice, sixth on the honeycomb lattice) and is strongly suppressed in the confined regime, appearing only at very long times~\cite{Tian:2025mbv}.
An explicit magnetic term is therefore required to observe $(2+1)$D dynamics on accessible time scales, and the operational witness is the weight $\mathcal{P}_{\gamma\neq\gamma_{\rm i}}$: for minimal strings, it stays essentially zero without a plaquette and becomes finite once the field propagates transversely.
This bears directly on hardware---runs with an explicit (fixed or tunable) plaquette~\cite{Cochran:2024rwe,Xu:2026ibi,Joshi:2026hfe} access transverse flux spreading and loop formation, whereas plaquette-free realizations can in part be reproduced by a $(1{+}1)$D description.
The step to $(3{+}1)$D compounds the cost through more plaquettes per site, larger local spaces, and---for TNs---an area law that scales with the boundary, and remains essentially unexplored for real-time string breaking.

\subsubsection{Non-Abelian gauge invariance}
Only gauge-invariant states are physical; this imposes local vertex constraints that the simulated state must satisfy exactly~\cite{Zohar:2021nyc}.
One strategy keeps the gauge field explicit and enforces Gauss's law energetically via gauge-protection penalty terms that push gauge-variant leakage to high energy~\cite{Halimeh:2022mct,Halimeh:2021ufh}; this is flexible but incurs coherence and register overhead, and, on hardware, such leakage remains a dominant error channel.
The complementary route works \emph{ab initio} in the gauge-invariant subspace: the dressed-site construction splits each link into rishon modes and fuses matter with the adjacent half-links into gauge singlets through Clebsch--Gordan decomposition, so that Gauss's law holds exactly at every site~\cite{Silvi:2014pta,Cataldi:2023xki}.
What remains is a simpler, Abelian link constraint---the two rishons of each link must carry the same irrep---which is imposed as a two-body symmetry in TN simulations~\cite{Cataldi:2025cyo} but, on hardware, remains a (cheaply checkable) error channel, together with any unused states of a binary encoding of the dressed site.
The distinction matters most in the non-Abelian case, where the SU($N$) constraint is non-commuting and cannot be diagonalized link by link; it is precisely this construction that enabled the $(2{+}1)$D SU(2) simulation while retaining exact gauge invariance, including for the static background charges that define the string endpoints.

\subsubsection{Gauge-field truncation and the continuum limit}
For a continuous group, the link Hilbert space is infinite-dimensional, so beyond one spatial dimension, the gauge field cannot be represented faithfully and must be regularized~\cite{Zohar:2014qma}.
The available recipes---quantum link models~\cite{Chandrasekharan:1996ih}, finite subgroups, gauge-field digitization, fusion-algebra deformations, or capping the electric spectrum at $\spin\!\le\!\jmax$ in the irrep basis~\cite{Klco:2019evd,Cataldi:2023xki}---differ in accuracy, locality, and how they approach the continuum.
The essential trade-off is set by the coupling: the electric (irrep) basis is efficient at strong coupling, where the low-energy strings involve only the lowest irreps, but becomes prohibitive toward the continuum $g\!\to\!0$, where the magnetic term dominates, many irreps are excited, and the minimal cutoff grows as $\sim\!1/g$.
Consequently, a truncation is controlled only within a well-defined regime.
In the strong-coupling window $g_E,m\gg g_B,w$ used by the SU(2) study, a link in irrep $\spin$ is off-resonant by $\sim\! g_E\, c_{\spin}$, so leakage above $\jmax$ is parametrically suppressed---a second-order estimate gives $\mathcal{P}(\spin\!>\!\jmax)\sim(w/g_E)^2\!+\!(g_B/g_E)^2\!\sim\!10^{-3}$ for the couplings employed---and the hardcore-gluon truncation $\jmax\!=\!\jonehalf$ already captures string breaking, with $\jmax\!=\!1$ needed for tension reduction~\cite{Cataldi:2025cyo}.
The open problem is the \emph{continuum-limit location}: current real-time results sit deep in strong coupling, far from where the lattice tension can be matched to a physical one, while pushing $\jmax$ up runs into the explosive growth of the local dimension---$\mathcal{O}(10^4)$--$\mathcal{O}(10^9)$ already for three-dimensional SU(3)~\cite{Magnifico:2024eiy}.
Certifying convergence in $\jmax$ and connecting the strong-coupling dynamics to the weak-coupling continuum where hadronization lives is a central task; a natural first bridge is to compare the string tension and confinement extracted from static (Euclidean) and real-time (Hamiltonian) approaches.

\subsubsection{Initial-state preparation}
In the weak-coupling regime, preparing an appropriate reference state is itself a bottleneck.
At strong coupling, the vacuum is close to a product state in the electric-field basis, from which a string is drawn simply by exciting the links along a path---on the trapped-ion $\mathbb{Z}_2$ experiment, literally a layer of single-qubit gates applied to the string links~\cite{Xu:2026ibi}.
At weak coupling, the physical vacuum is instead an interacting, gauge-invariant many-body state with substantial magnetic fluctuations and long-range correlations; starting a quench from the trivial strong-coupling vacuum would inject spurious energy and obscure the string dynamics.
Adiabatic preparation---slowly ramping the coupling from the strong-coupling limit---is conceptually natural, but its cost is set by the smallest gap along the path and by the need to preserve Gauss's law throughout.
Variational alternatives trade evolution time for a classical optimization that must avoid barren plateaus; gauge-tailored ans\"atze that move only within the gauge-invariant subspace, alternating electric, magnetic, and matter layers, reach gauge-invariant ground states with favorable gradient scaling and without large penalty terms~\cite{Sala:2018dui,Surace:2023qwo}.
Imaginary-time and thermal-sampling schemes extend this to finite temperature and density, while phase-estimation and quantum-signal-processing filters offer near-optimal asymptotic scaling but demand fault-tolerant depths and a starting state with appreciable overlap on the target.
For string breaking specifically, the initial state is not the vacuum but a controlled far-from-equilibrium flux tube on top of it, so a scalable protocol must supply both a faithful vacuum and a clean, gauge-invariant string excitation.

\subsubsection{Fermionic matter and the local-dimension bottleneck}
Discretizing the Dirac field introduces doublers; the staggered prescription used here~\cite{Susskind:1976jm} is standard, but beyond one spatial dimension, the fermionic statistics become the obstacle, since a Jordan--Wigner mapping renders the matter interactions non-local.
Gauge defermionization removes this by absorbing the matter into the same dressed site as the rishons, so that hopping becomes a two-body interaction of bosonic operators and the theory is simulated without fermionic strings, for any U($N$) or SU($2N$) gauge group~\cite{Cataldi:2023xki,Ballarin:2023jpd}.
The price shared by all gauge-invariant encodings is the local Hilbert-space dimension $d$: for the SU(2) dressed site, it is already $30$ at $\jmax\!=\!\jonehalf$ ($42$ at the background-charge sites) and grows to $168$ ($317$) at $\jmax\!=\!1$~\cite{Cataldi:2025cyo}.
This single number drives every cost---TN optimization scales at least quadratically in $d$, while a digital device needs $\lceil\log_2 d\rceil$ qubits (or one qudit) per site---so taming $d$ through optimal local bases and economical encodings is the gateway to larger, near-continuum, non-Abelian simulations~\cite{Magnifico:2024eiy}.

\subsubsection{Observable design, circuits, and validation}
Mapping the dynamics to a digital circuit means Trotterizing $e^{-i\ham t}$ and reading gauge-invariant observables from projective measurements.
Diagonal terms (electric energy, mass) are cheap phase rotations; the hopping is the Clebsch--Gordan two-site entangler; the plaquette is the most expensive term, requiring a cascade of recouplings in the non-Abelian case~\cite{Klco:2019evd,Kavaki:2024ijd}.
Abelian models, such as $\mathbb{Z}_2$ and spin-$\tfrac12$ quantum-link cases, are comparatively easy---one qubit per link, matter often integrated out---and the $\mathbb{Z}_2$ experiment reached a two-qubit depth of $6$ per Trotter step on $56$ qubits with over $1000$ entangling gates~\cite{Xu:2026ibi}; SU(2) is qualitatively harder because of the large dressed-site dimension, the non-diagonal hopping, and the four-link non-Abelian plaquette.
Qudit encodings help---a dressed-site qudit realizes the $(1{+}1)$D SU(2) hopping through native two-qudit gates and cuts entangling-gate counts substantially~\cite{Calajo:2024qrc}, as also shown for a pure-gauge SU(2)$_2$ plaquette~\cite{John:2026gut}---but no hardware realization of $(2{+}1)$D SU(2) with dynamical matter \emph{and} an explicit plaquette has been reported yet, and it is the concrete near-term target, for which hardware-efficient architectures have been proposed~\cite{Zache:2023cfj}.

Two issues then cut across platforms.
Observables must stay informative as the system grows: the number of admissible flux paths proliferates, and the probability of any single configuration becomes exponentially small, so reconstructing individual strings is not scalable.
The workable strategy, shared by the $\mathbb{Z}_2$ experiment and the TN studies, is to group configurations into physically motivated classes---for example by the perturbative order $k$ at which they connect to the initial string---and to sum their probabilities, $\mathcal{P}_k=\sum_{\gamma\in\Omega_k}\mathcal{P}_\gamma$, which retains sensitivity to the mechanism while collecting statistical weight and maps naturally onto the inclusive rates and multiplicities of event-generation language~\cite{Xu:2026ibi}.
Finally, validation: TNs provide the noise-free classical reference for hardware, yet they too run out once entanglement grows---in the trapped-ion $\mathbb{Z}_2$ run only the first two Trotterized steps could be matched by matrix product states before the bond dimension blew up~\cite{Xu:2026ibi}---so error mitigation (Gauss-law and charge-sector post-selection, leakage detection, self-mitigating forward--backward Trotter~\cite{ARahman:2022tkr}, and dynamical decoupling) and TN cross-checks remain complementary and both necessary.

\subsection{Real-time signatures}
\label{sec:halimeh-physics}
In the strong-coupling regime, the dynamics, though far from the continuum, show sharp, interpretable signatures organized by the resonance principle of \cref{sec:halimeh-model}.
We summarize those observed in TN simulations of SU(2)~\cite{Cataldi:2025cyo}, noting their Abelian counterparts~\cite{Xu:2026ibi,Xu:2025abo}:
\begin{enumerate}
  \item \emph{Resonant breaking and its orders.} When the electric energy (proportional to the Casimir) released by removing one active $\spin\!=\!1\!/2$ link, $c_{1\!/2}\,g_E$, matches $2m$---i.e.\ at $m\!=\!c_{1\!/2}\,g_E/2\!=\!\tfrac38g_E$---the string hybridizes with shorter, matter-carrying configurations: $\mathcal{P}$ and $C$ drop, $p_{\rm mes}$ and $p_{\rm bar}$ rise, and $S$ grows continually. The staggered structure makes higher-order (multiple local) breaks resonant at the same point, unlike the $\mathbb{Z}_2$ experiment, where first- and second-order breaking occur at distinct resonances~\cite{Xu:2026ibi}.
  \item \emph{Baryon--antibaryon formation.} SU(2) fusion rules allow spontaneous baryon and antibaryon production at the break, not only mesons---a channel absent in $\mathbb{Z}_2$/U(1).
  \item \emph{Glueball-like loops and self-crossing strings.} 
  Away from resonance, string breaking is strongly suppressed\footnote{At leading order in the strong-coupling expansion. 
  The neglected channels are virtual pair fluctuations, with weight $\sim\!(w/\Delta)^2$ for a detuning $\Delta$ from resonance, and higher-order processes that may become resonant only at much longer times.}.
  Namely, the magnetic term only reshuffles flux among equal-length configurations, generating pure-gauge closed loops---glueball-like excitations in the strong-coupling limit~\cite{Mathieu:2008me}---and self-crossed strings. 
  The non-Abelian case is richer: a four-valent vertex of four $\spin\!=\!\jonehalf$ links supports two linearly \emph{independent} singlet intertwiners, recoupling internally through $\spin\!=\!0$ or $\spin\!=\!1$, which carry the same link-electric energy but couple differently to the magnetic and hopping terms\footnote{The re-coupling label is internal to the vertex and does not correspond to an additional electric link.}---structure with no Abelian analog.
  \item \emph{Baryon blockade.} Because baryon number is a global quantum number that can be fixed exactly (while being out of reach for Monte Carlo in real time), one can add baryons at fixed resonance and watch the break time---the first minimum of $C$---move progressively later, which we attribute to finite density pruning the available final states.
  \item \emph{String-tension reduction.} For higher-energy initial strings, corresponding to higher $\spin$-irrep in the Casimir, a family of representation-changing resonances opens (for SU(2) $2m=g_E(\spin+\tfrac14)$), at which a link lowers its irrep by $\tfrac12$, and the string relaxes to lower tension without fully breaking; it coincides with breaking only at $\spin=\jonehalf$.
\end{enumerate}
Some of these five signatures have no Abelian counterpart and are the concrete non-Abelian targets for the next generation of quantum simulators.

\subsection{Outlook}
\label{sec:halimeh-outlook}
Several directions follow naturally.
\begin{enumerate}
  \item \emph{Toward hadronization.} Promote real-time fragmentation into first-principles input for event generators---fragmentation functions, hadron multiplicities, and the time/length scale of the break, extracted from quench dynamics---to benchmark and constrain the tuned Lund and cluster models~\cite{Bierlich:2022pfr,Webber:1983if}. 
  The grouped, sector-resolved observables above are already generator-facing.
  \item \emph{Close the Abelian$\to$non-Abelian hardware gap.} SU(2) with dynamical matter and an explicit plaquette is the open hardware target~\cite{Cataldi:2025cyo,John:2026gut}; a key deliverable is the minimal set of terms needed for genuine $(2{+}1)$D behavior and a dressed-site qudit circuit that realizes it~\cite{Calajo:2024qrc}.
  \item \emph{Static versus dynamical confinement.} Compare the string tension and confinement obtained from static approaches with those from real-time dynamics, connecting strong-coupling signatures to continuum observables.
  \item \emph{Richer settings.} Concurrent Abelian and non-Abelian gauge fields; strings whose endpoints are pulled apart in real time, injecting energy at a controlled rate so that breakage is triggered on demand; and hadron--hadron scattering, string reconnection, and finite-baryon-density matter, all sign-problem-free in this framework~\cite{Halimeh:2023wrx}.
  \item \emph{Methods co-design.} Hybrid workflows that use TNs for state preparation and validation and hardware for the entanglement-rich real-time stage, with certified gauge truncation, explicit error budgets, and post-selection tailored to the SU($N$) Gauss law.
\end{enumerate}
The through line is that a controlled, non-Abelian, real-time account of string breaking would turn hadronization from a tuned model into a computed process---the concrete bridge from string dynamics to event generation that this workshop set out to build.
\endgroup

\section{Resource Theories in Quantum Simulations of HEP}\label{app:froland}
\contribution
  {Henry Froland\,\orcidlink{0009-0008-4356-0602}}
  {InQubator for Quantum Simulation (IQuS), Department of Physics, University of Washington, Seattle, WA 98195, USA}
  {}

The past several years have seen a convergence between quantum information science and high energy physics (HEP), where an important question has arisen: \textit{where does genuine quantum complexity reside and how does it manifest in the observable physics?} The answer to this question is relevant not only for explaining high-energy phenomenology, but also for designing the algorithms necessary for studying such systems on near-term quantum simulation platforms. Two different frameworks have come to the forefront for assessing quantum complexity, non-stabilizerness and non-gaussianity. The role of non-stabilizerness has been investigated in various contexts like hadronization, neutrino oscillations, and scattering events~\cite{Robin:2026lqp}, whereas it's role in non-abelian gauge theories is underexplored. Non-gaussianity remains an even larger mystery in this regard, despite the fact that most interacting field theories will generate non-gaussian correlations. This motivates the study of both non-stabilizerness and non-gaussianity in other systems relevant for high energy physics.

\emph{Magic} - Nonstabilizerness (colloquially ``magic'') quantifies how far a state lies from the set of stabilizer states, which are efficiently simulable by classical Clifford circuits. An efficiently computable measure is the Stabilizer R\'{e}nyi Entropy (SRE)\cite{Leone:2021rzd}
\begin{equation}
    M_\alpha \;=\; \frac{1}{1-\alpha}\log\!\left(\sum_P \Xi_P^\alpha\right) - n,
    \qquad
    \Xi_P \;=\; \frac{\langle P \rangle^2}{2^n},
\end{equation}
where the sum runs over all $n$-qubit Pauli operators $P$. For pure states, the SRE is a monotone for $\alpha\geq 2$, and for mixed states other monotones such as the Robustness of Magic (ROM) are used. 

\emph{non-gaussianity} - non-gaussianity captures the deviation of a state from the set of Gaussian states, which are classically tractable via Wick's theorem and matchgate circuits. Similarly to the SRE and magic, it may be quantified by a measure known as the Fermionic anti-Flatness (FAF)~\cite{Sierant:2025fax}, defined by 
\begin{equation}
\operatorname{FAF}_{k}(\rho)=n-\frac{1}{2} \operatorname{tr}\left[\left(-\Gamma_{\rho}^{2}\right)^{k}\right]=n-\sum_{j=1}^{n} \nu_{j}^{2 k}
\end{equation}
where $\left(\Gamma_{\rho}\right)_{a b}=-i \frac{1}{2} \operatorname{tr}\left(\rho\left[\gamma_{a}, \gamma_{b}\right]\right)$ for the $2n$ majorana modes $\{\gamma_a\}$. Here the $\{\nu_a\}$ are the $n$ singular values of this matrix.
 
\emph{String Breaking and Non-Stabilizerness} - String breaking, the fragmentation of a color flux tube into hadronic bound states, is among the most physically consequential phenomena in confining gauge theories and underlies hadron production in QCD. Recently, the first systematic characterization of string breaking using quantum complexity measures in the 1+1D Schwinger model (QED$_{1+1}$) was performed using Matrix Product State (MPS) methods~\cite{Grieninger:2026bdq}. A central finding is that entanglement and magic offer genuinely complementary perspectives on string formation and breaking: while entanglement entropy tracks the buildup of long-range vacuum correlations, magic, in particular its \textit{nonlocal} component, rises sharply during the breaking process before relaxing to its vacuum value as the resulting mesons separate. The interpretation of this profile is that the vacuum degrees of freedom rearrange such that there is a departure from representability by stabilizer states which peaks during the string-breaking transition.
 
\emph{Gauge Field Structure} -Parallel work~\cite{Cao:2026qky} examines string breaking in 2+1D quantum link electrodynamics with a three-level representation of the gauge-field, revealing qualitatively richer phenomenology than lower-dimensional analogs. In particular, this gauge-field formulation admits a \textit{two-stage breaking mechanism} not accessible to the two-level truncation: the string passes through an intermediate metastable configuration before complete fragmentation, with distinct glueball-like bound states forming in the real-time dynamics. On the experimental side, Ref.~\cite{Xu:2026ibi} observed glueball excitations and multi-order string breaking in a 2+1D $\mathbb{Z}_2$ lattice gauge theory implemented on the 56-qubit Quantinuum H2 trapped-ion processor via Trotter circuits, providing direct hardware evidence for the richness of higher-dimensional string dynamics.
 
\emph{Summary} - Several threads in the above works, while not yet fully connected in a single treatment, suggest a coherent picture that merits development. First, in Abelian theories like the Schwinger model, the gauge field is largely set by Gauss's law and carries no independent local degrees of freedom. In non-Abelian theories, the gauge sector is dynamical, and plaquette interaction terms $\bigl(\operatorname{Tr}[U_{\mu\nu}] + \mathrm{h.c.}\bigr)$ can generate nonstabilizerness in the gauge field itself. In the context of hadronization, understanding the properties and dynamics of the gauge field through the lens of the non-stabilizerness in the field may reveal structures relevant both for understanding confinement in QCD and quantum simulation. For non-gaussianity, the analysis of non-local magic in Ref.~\cite{Grieninger:2026bdq} offers a particularly sharp case for thinking about the role this resource plays in hadronization. Namely, does the FAF, or some non-local extension, contain a peak at the string breaking transition? Understanding this question would not only better locate string-breaking phenomena within the complexity phase diagram of different resource theories, but may also spur further improvements in classical simulation methods. If certain low-order non-gaussian correlations are found to be dominant during this string breaking transition, then it may be possible to understand this phenomena as a perturbation around a gaussian state~\cite{Froland:2024wxj}. Considering these points, using resource theories like non-stabilzierness and non-gaussianity as probes of nuclear structure in non-abelian gauge fields and strings has the potential to inform future quantum simulations of high energy physics as well as providing new observables for input into event generators for collider experiments.

\section{Comparing Lattice Formulations for Studies of Hadronization}\label{app:grabowska}
\contribution
  {Dorota M. Grabowska}
  {InQubator for Quantum Simulation (IQuS), Department of Physics, University of Washington, Seattle, WA 98195, USA}
  {}
One of the most striking features of Quantum Chromodynamics (QCD) is that the observable states of the theory are different at high and low energies. At high energies, the theory is asymptotically free and the observed states, roughly speaking, are quarks and gluons i.e. the fundamental building blocks of QCD. However, at low-energies, the observable states are instead hadrons, i.e. confined states of quarks and gluons. This transition, from deconfined quarks and gluons to hadronic asymptotic states, plays a central role in the high-energy collisions that occur at particle colliders, including the Large Hadron Collider (LHC) at CERN. Understanding it from first principles requires numerical studies of the dynamics of QCD.  These studies have proven extremely challenging (one might even say impossible) on classical computers. However, quantum computers offer an alternative path for carrying out these studies, one that may be computationally feasible. With the recent rapid advancement in quantum hardware, it is vitally important to map out the reach of quantum simulations for understanding the dynamics of hadronization. 

Studying hadronization, and the related phenomena of string breaking, on quantum devices requires several key steps. One of these is determining an appropriate lattice formulation of the QCD Hamiltonian. Many such formulations exist and each has unique boons and banes. Electric basis formulations, which utilize the irreps of gauge group , result in a (mostly) local Hamiltonian that is easily truncated. However, these formulations tend to suffer from large truncation errors as the bare gauge coupling is reduced. Discrete subgroup formulations, using the irreps of the discrete subgroups of the gauge group be implemented into a quantum circuit in a scaleable manner. However, there exist only a finite number of these subgroups and care must be taken when extrapolating to the continuum due to differing universality classes. The Loop-String-Hadron formulation works with gauge-invariant states, (almost fully) eliminating the possibility of noise-induced Gauss-law violating transitions. However, constructing the circuits for this formulation is complicated. The mixed-basis formulation reduces the Hilbert space to only the total gauge-singlet sector and is resource-efficient at all values of the gauge coupling. However, it results in a highly non-local Hamiltonian that may prove challenging to simulate on quantum devices with limited connectivity. Many other formulations exist, but no one formulation is free of bane. 

In order to best understand the dynamics of hadronization and string-breaking, it is key to carry out an apples-to-apples comparison of the results from quantum simulations of these different formulations. Such comparisons are useful for understanding truncation and finite-volume effects, as well as better controlling the extrapolation to the continuum. The comparisons need be done for simulations of two or three spatial dimensions, in order to quantify the interplay of the electric and magnetic components of the Hamiltonian. Furthermore, even for small-scale simulations, looking at the qualitative behavior of string-breaking in different formulations is key, in order to identify whether an observed property is universal (ie actually true for QCD) or a formulation-dependent artifact.

\section{The Quantum Structure of Confining Strings}\label{app:grieninger}
\contribution
  {Sebastian Grieninger}
  {InQubator for Quantum Simulation (IQuS), Department of Physics, University of Washington, Seattle, WA 98195, USA}
  {}
Phenomenological hadronization assumes independent fragmentation but the
confining string that fragments is a quantum state carrying nonlocal
correlations. As a string in the massive Schwinger model is stretched to its
breaking point, its interior approaches maximal entanglement, its reduced state
becomes thermal, and the string carries nonlocal quantum correlations that
vanish in the outgoing hadrons.

Euclidean lattice methods capture the physics of confinement through the static
potential but cannot resolve the wavefunction of the flux tube, its full
quantum-information content, or its real-time dynamics. Hamiltonian formulations
and tensor-network methods provide direct access to the ground-state
wavefunction and thereby to reduced density matrices, entanglement spectra, and
nonclassical correlations within the flux tube, while also enabling real-time
evolution.

\paragraph{What string breaking looks like from the inside.}
The reduced density matrix of the string subsystem approaches a thermal
form as the string is stretched toward
breaking~\cite{Grieninger:2025rdi}: the overlap with an effective thermal
density matrix peaks sharply near the string-breaking distance, while the
entanglement spectrum undergoes a level crossing: the dominant Schmidt state
transitions from the confining electric string to a screened two-meson
configuration, and the spectral gap narrows as several eigenvalues cluster,
signaling an approach to maximal entanglement~\cite{Kharzeev:2026jkq}. The string fragments just before
this limit is reached. Entanglement-driven
thermalization is also found dynamically in real-time jet
fragmentation~\cite{Florio:2025hoc} where
local observables approach thermal values and the density-matrix overlap with the
thermal state grows.

The nonlocal mutual information and nonlocal Robustness of Magic (RoM) between
disjoint regions of the string~\cite{Grieninger:2026bdq,Robin:2026lqp} directly quantify the
correlations that independent fragmentation discards. These spatially resolved
probes provide a view into the internal structure of the flux tube: the outgoing-meson track vanishes from the nonlocal
component, and longer-range correlations along the string appear predominantly
classical. The nonlocal
nonstabilizerness~$\mathcal{M}_2^{(\mathrm{NL})}$~\cite{Falcao:2024msg} and the
entanglement entropy both peak near the string breaking separation.

In the strong-coupling, large-$N_c$ holographic
dual~\cite{Sonner:2013mba,Jensen:2013ora,Grieninger:2026txt}, the same pair-production mechanism produces the
quark--antiquark pair at maximal entanglement (the Hartle--Hawking
thermofield double), with nonlocal magic intrinsic to the gauge structure~\cite{Grieninger:2026txt}. This is the limit that the string tends to~\cite{Grieninger:2025rdi}.

\paragraph{Observables, benchmarks, and validation targets.}
Quantum simulation of string breaking in the Schwinger model provides concrete
targets and new classes of observables relevant for connecting to hadronization
models:
\begin{itemize}\setlength{\itemsep}{2pt}
\item The nonlocal RoM and mutual information spatially resolve quantum
  correlations along the flux tube~\cite{Grieninger:2026bdq}. These give
  direct insights into whether the independent-fragmentation assumption in the Lund
  string and cluster
  models is justified or how they should be modified~\cite{Andersson:1983ia,Sjostrand:2019zhc}.
\item  Ref.~\cite{Grieninger:2026bdq} provides
  minimal lattice requirements and continuum extrapolations as cross-validation targets that
  quantum-hardware calculations should reproduce.
\item Antiflatness~\cite{Cao:2024nrx}, a lower bound on nonlocal magic,
  confirms the same peak structure and is estimable via randomized measurements
  on quantum hardware as a scalable proxy for nonlocal magic.
\item The local RoM reveals the internal structure of the outgoing
  meson~\cite{Grieninger:2026bdq}: moments of its spatial distribution are
  analogous to charge or axial distributions in hadron structure.
\item Phenomenological models use constant (in time) and homogeneous Schwinger pair production rates. Holographic results~\cite{Grieninger:2023pyb,Grieninger:2023ehb} show 
that the mean number of particles produced depends on the pulse profile of the electric field.  \end{itemize}

\paragraph{Next steps.}
\begin{itemize}\setlength{\itemsep}{2pt}
\item \textbf{Connect static and dynamic pictures.} The complexity
  machinery of Ref.~\cite{Grieninger:2026bdq}, i.e., nonlocal RoM, mutual
  information, and magic, should be adapted to the real-time jet-fragmentation setup
  of Refs.~\cite{Florio:2023dke,Florio:2024aix,Florio:2025hoc,Janik:2025bbz} to directly probe whether outgoing mesons carry nontrivial quantum correlations.
\item \textbf{Non-Abelian theories and higher dimensions.} The central question
  is whether this structure persists in SU(2) and SU(3) with dynamical matter,
  and in $2{+}1$D where transverse fluctuations, a true area law, and glueball
  degrees of freedom appear.
\item \textbf{Event-level observables.} Whether the string's nonlocal quantum
  structure leaves a trace in asymptotic final states is conjectured but not
  established. Identifying such observables,  accessible at the LHC or EIC, is a crucial near-term challenge.
\end{itemize}

\section{Complexity Probes for Choosing Hybrid Simulation Methods}\label{app:hite}
\contribution
  {Michael Hite}
  {Department of Physics, University of Arizona, Tucson, AZ 85721, USA}
  {}

As we scale to larger 1+1D systems and 2+1D systems, efficient use of computational resources is of necessity. Using probes like entanglement and magic (nonstabilizerness), we can get an accurate measure of quantum complexity. Much work has already been done in the context of nuclear and high energy physics \cite{Robin:2026lqp,Hite:2026fsj}, conformal field theories \cite{Cao:2024nrx}, and lattice gauge theories \cite{Jha:2026ror,Froland:2026aff}. This will allow us to better leverage classical computing methods like tensor networks and the stabilizer formalism in parallel with quantum simulations in the NISQ era. For example, for systems in the low magic-high entanglement regime, recent methods like the stabilizer tensor network (STN) formalism \cite{Masot-Llima:2024doz} and STN with magic state injection \cite{Nakhl:2024gfr} could allow for a significant reduction in circuit depth by moving the Clifford portion of a quantum circuit into a classical stabilizer tableau, which scales linearly with system size. Nevertheless, there is still much work to be done, and will rely on the joint efforts of high energy theorists and quantum information science.

\section{Toward First-Principles Quantum Simulation of Hadronization}\label{app:kadam}
\contribution
  {Saurabh V. Kadam\,\orcidlink{0000-0001-9218-1600}}
  {Lawrence Livermore National Laboratory, Livermore, CA 94550, USA}
  {LLNL-MI-2021350}
Formation of color-singlet hadrons from colored quarks and gluons, known as hadronization, is governed by the real time dynamics of quantum chromodynamics (QCD).
Hadronization occurs at the energy scale where the strong coupling becomes large and nonperturbative effects dominate.
Its inherently real time nature makes it inaccessible to Euclidean lattice methods.
Instead, the high energy physics (HEP) community has relied on phenomenological models to relate the partonic scattering amplitudes at high energies to the detection of hadrons in experiments.
Despite their success, these phenomenological models are semiclassical and cannot fully capture quantum mechanical effects, which may become important for precision measurements at future HEP experiments.

Quantum simulation offers a path to computing the underlying physics from first principles.
This can be achieved by mapping lattice gauge theory (LGT) onto a quantum device and performing the time evolution under the LGT Hamiltonian to simulate the quantum dynamics of the confining flux tube between quarks, its breaking via dynamical quark-pair creation, and the subsequent formation of hadrons.
Such first-principles calculations could mitigate the limitations of traditional hadronization models, improving the event generation pipeline used in HEP experiments.
To harness the power of quantum computers to realize these goals, a research workflow must be established to guide the community in the coming years.
This contribution briefly summarizes my perspective on the research directions required to achieve these objectives.

\begin{itemize}
    \item \textbf{Efficient formulations of LGTs:} A non-Abelian LGT can be encoded on qubits in different ways, each offering different advantages and trade-offs for calculating various observables. The field needs to perform a comparative study of how gauge and matter fields are digitized and mapped onto quantum hardware. A primary near-term objective is systematic resource estimation across various non-Abelian LGT formulations to identify the most hardware efficient representations for string dynamics and hadronization.
    
    \item \textbf{Quantum simulation of accessible limits:}
    Before fault-tolerant quantum computers can execute full (3+1)D QCD simulations, the community should target tractable, low-dimensional physical limits that still capture the core features of hadronization. Specifically, this includes simulating the time evolution of flux strings to characterize the string-breaking dynamics, including the evolution of the string tension and breaking timescales in (1+1)D and (2+1)D non-Abelian LGTs, subsequently extending these calculations to include multiple fermion flavors. Crucially, because these scenarios remain within the reach of classical tensor network simulations, they provide an opportunity for validating early quantum hardware while simultaneously driving the advancement of tensor network methodologies.

    \item \textbf{The measurement problem:}
    A near-term objective is the methodological advancement necessary to relate time evolved quantum states on a quantum computer to experimentally meaningful quantities. It is crucial to formulate theoretical frameworks that map quantities naturally accessible on quantum hardware, such as staggered fermion densities, local electric field or energy densities, and the Loschmidt echo, to event level observables including hadron distributions, fragmentation functions, and jet observables.
     
    \item \textbf{Algorithms for fault-tolerant quantum computing:} Beyond the noisy intermediate-scale quantum (NISQ) era, hadronization at physical scales requires fault-tolerant simulation. This involves preparing appropriate string or composite initial states, executing real time evolution across multiple string-breaking timescales, and implementing complex measurement protocols. Realizing this full simulation stack requires the development of resource efficient algorithms accompanied by rigorous cost and sensitivity bounds. Furthermore, the field needs to explore the trade-offs between explicitly enforcing gauge constraints in the simulation versus leveraging the gauge symmetry itself as a resource for error detection and correction.
    
\end{itemize}
Progressing through these objectives will keep near-term hardware and algorithmic development aligned with the long-term phenomenological needs of the HEP community.
Each step is individually well posed, yet together they build toward a first-principles pipeline connecting the real time dynamics of QCD to collider phenomenology, enhancing the predictive power of future HEP experiments.
Beyond improving event generation, these simulations would provide direct insights into the real time mechanism of color confinement.
\subsection*{Acknowledgment}
This work was partly performed under the auspices of the U.S. Department of Energy by the Lawrence Livermore National Laboratory under Contract No. DE-AC52-07NA27344.

\section{I Can Do It With a Broken String}\label{app:lamm}
\contribution
  {Henry Lamm\,\orcidlink{0000-0003-3033-0791}}
  {Fermi National Accelerator Laboratory, Batavia, Illinois 60510, USA}
  {}
\noindent\textit{Connecting quantum simulation of string breaking to event generators is difficult. Tensor networks can already extract the functional forms needed from $1{+}1$D multiflavor $\SU(3)\times U(1)_{\rm em}$ string breaking; an effective theory and, eventually, fault-tolerant hardware could supply the rest. Rigorous matching is the rate-limiting theoretical step, and the baryon-to-meson ratio is the sharp discriminating observable.}
\medskip

\begin{figure}[!t]
  \centering
  \resizebox{8cm}{!}{\begin{tikzpicture}[
  >={Stealth[length=4pt,width=4pt]},
  line cap=round,
  font=\small,
  quark/.style={circle,draw=black!65,line width=0.5pt,minimum size=7mm,
                inner sep=0pt,text=white,font=\footnotesize\bfseries},
  qnew/.style={circle,draw=black!55,dashed,line width=0.5pt,minimum size=6.2mm,
                inner sep=0pt,text=white,font=\scriptsize\bfseries},
  anno/.style={font=\scriptsize,align=center,rounded corners=2pt,
                draw=black!30,fill=black!4,inner sep=2.6pt},
  leader/.style={->,draw=black!45,line width=0.45pt,shorten >=1.5pt,shorten <=1.5pt},
  stepbox/.style={font=\scriptsize,align=center,rounded corners=2.5pt,draw=black!55,
                fill=black!5,inner sep=3pt,minimum height=8mm,text=black!85},
]
\definecolor{rcol}{RGB}{171,67,58}
\definecolor{rbarcol}{RGB}{74,144,150}
\definecolor{tubecol}{RGB}{208,201,158}%
\definecolor{photoncol}{RGB}{150,120,40}

\fill[tubecol,rounded corners=3pt] (1.05,-0.20) rectangle (3.15,0.20);
\draw[decorate,decoration={coil,aspect=0.5,segment length=4.2pt,amplitude=2.4pt},
      draw=black!40,line width=0.5pt] (1.15,0) -- (3.05,0);
\fill[tubecol,rounded corners=3pt] (4.85,-0.20) rectangle (6.95,0.20);
\draw[decorate,decoration={coil,aspect=0.5,segment length=4.2pt,amplitude=2.4pt},
      draw=black!40,line width=0.5pt] (4.95,0) -- (6.85,0);
\draw[black!45,line width=0.5pt,dotted] (4.0,-0.34) -- (4.0,0.34);
\node[font=\scriptsize,text=black!55] at (4.0,0.66) {break};

\node[quark,fill=rcol]    (q)   at (0.7,0) {$q$};
\node[quark,fill=rbarcol] (qb)  at (7.3,0) {$\bar q$};
\node[qnew,fill=rbarcol]  (qbn) at (3.5,0) {$\bar q'$};
\node[qnew,fill=rcol]     (qn)  at (4.5,0) {$q'$};

\node[font=\scriptsize,text=black!55] at (0.7,-0.5) {$r$};
\node[font=\scriptsize,text=black!55] at (7.3,-0.5) {$\bar r$};

\node[font=\scriptsize,text=black!70] at (4.0,-0.5) {flavour $u/d/s$};
\node[font=\scriptsize,text=rbarcol!55!black] at (3.5,0.45) {$-$};
\node[font=\scriptsize,text=rcol]             at (4.5,0.45) {$+$};

\draw[decorate,decoration={snake,amplitude=1.3pt,segment length=4.5pt},
      draw=photoncol,line width=0.6pt] (4.66,0.28) -- (5.4,0.92);
\node[font=\scriptsize,text=photoncol] at (5.55,1.02) {$\gamma$};

\draw[black!50,line width=0.5pt] (0.7,-0.74) -- (0.7,-0.86) -- (3.5,-0.86) -- (3.5,-0.74);
\node[font=\scriptsize,text=black!70] at (2.1,-1.06) {meson};
\draw[black!50,line width=0.5pt] (4.5,-0.74) -- (4.5,-0.86) -- (7.3,-0.86) -- (7.3,-0.74);
\node[font=\scriptsize,text=black!70] at (5.9,-1.06) {meson};

\node[font=\scriptsize\itshape,text=black!70,align=center] at (1.05,-2.05)
      {or diquark /\\ junction};
\coordinate (jc) at (2.35,-2.45);
\node[circle,fill=rcol,minimum size=3.2mm,inner sep=0pt] (b1) at (2.15,-2.20) {};
\node[circle,fill=rcol,minimum size=3.2mm,inner sep=0pt] (b2) at (2.15,-2.70) {};
\node[circle,fill=rcol,minimum size=3.2mm,inner sep=0pt] (b3) at (2.67,-2.45) {};
\draw[black!55,line width=0.5pt] (b1) -- (jc) -- (b2) (b3) -- (jc);
\node[font=\scriptsize,text=black!70] at (3.78,-2.45) {baryon $(qqq)$};
\draw[->,draw=black!45,line width=0.5pt,dashed]
      (3.35,-0.18) .. controls (2.70,-1.10) and (2.30,-1.75) .. (2.25,-2.00);

\node[anno,text width=2.0cm] (cK) at (1.45,1.72)
      {$\kappa(\tau)$\\ real-time\\ string tension};
\draw[leader] (cK.south) to[bend right=8] (1.95,0.22);

\node[anno,text width=2.2cm] (cEM) at (4.2,2.15)
      {EM mass splitting\\ ($U(1)_{\mathrm{em}}$ charges)};
\draw[leader] (cEM.south) -- (4.52,0.62);

\node[anno,text width=1.7cm] (cV) at (6.55,1.72)
      {$V/PS$\\ (meson spin)};
\draw[leader] (cV.south) to[bend left=8] (6.1,0.22);

\node[anno,text width=2.6cm] (cGs) at (4.0,-1.62)
      {$\gamma_s = P(s\bar s)/P(u\bar u)$\\ strangeness (mass)};
\draw[leader] (cGs.north) -- (4.0,-0.66);

\node[anno,text width=3.0cm] (cB) at (1.85,-3.55)
      {baryon/meson ratio,\\ $B\bar B$ correlation};
\draw[leader] (cB.north) -- (2.15,-2.97);

\node[font=\scriptsize,text=black!60] at (5.1,-3.85) {increasing capability $\longrightarrow$};
\node[stepbox,text width=1.5cm] (s1) at (1.1,-4.55) {TN\\ today};
\node[stepbox,text width=1.9cm] (s2) at (4.0,-4.55) {1+1D EFT\\ ($\sim\!10^{2}$ qubits)};
\node[stepbox,text width=1.5cm] (s3) at (6.8,-4.55) {FT\\ 2+1D};
\draw[->,draw=black!70,line width=0.8pt] (s1.east) -- (s2.west);
\draw[->,draw=black!70,line width=0.8pt] (s2.east) -- (s3.west);

\draw[->,draw=rcol,line width=0.8pt,dashed] (0.55,-5.17) -- (7.35,-5.17);
\node[font=\scriptsize\itshape,text=rcol] at (4.0,-5.41) {matching theory --- the rate-limiter};
\end{tikzpicture}}
  \caption{The difficulty ladder: multiflavor $\SU(3)\times U(1)_{\rm em}$ string breaking yields the functional forms event generators need, as simulation capability climbs from tensor networks to fault-tolerant $2{+}1$D.}
  \label{fig:lamm-ladder}
\end{figure}
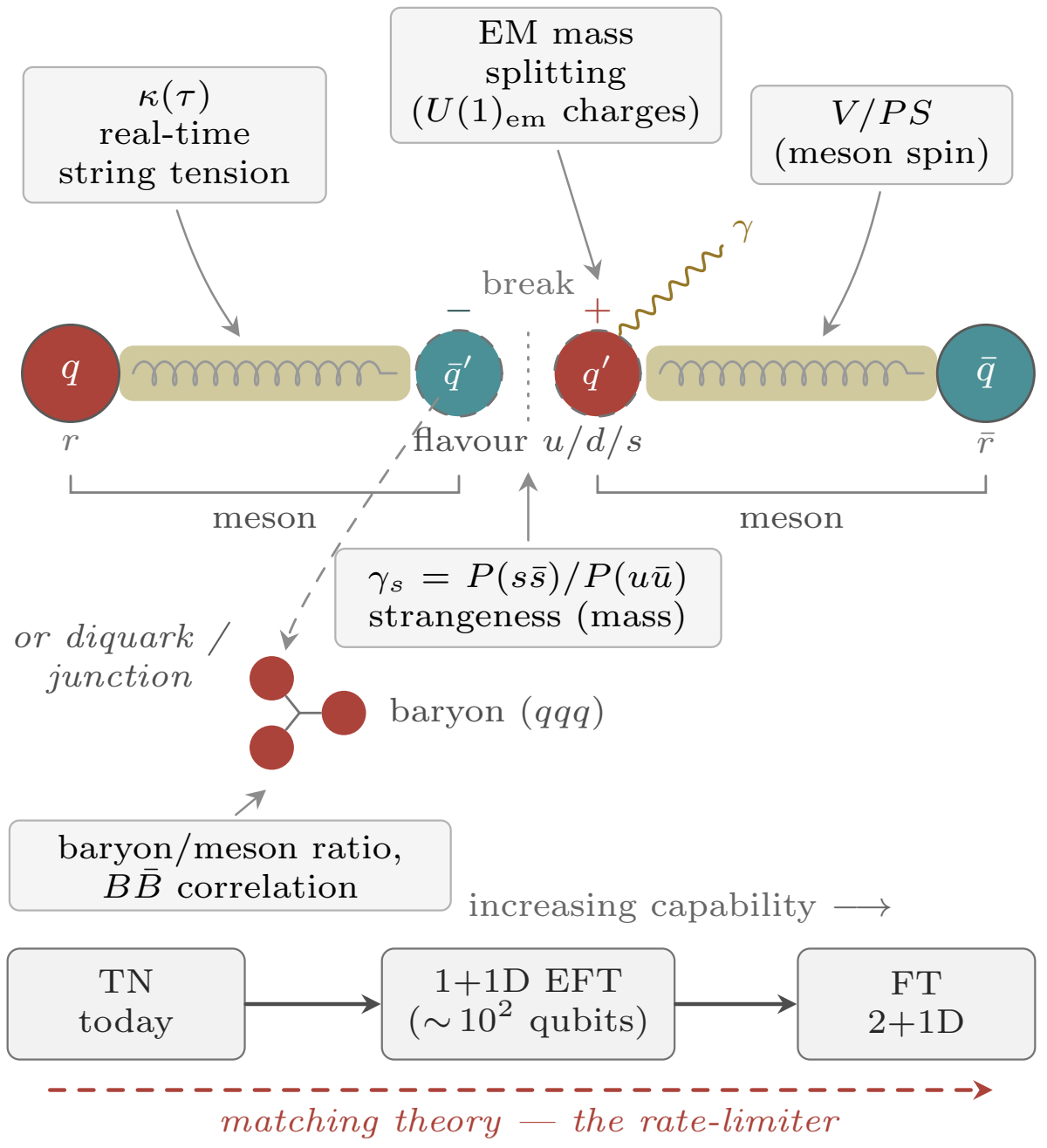

Simulating an end-to-end $pp\!\to\!X$ collision will remain infeasible for the foreseeable future, because the scale hierarchy reaches $\sqrt{s}/\LQCD\!\sim\!10^5$, there is no lattice formulation of the full Standard Model, and electroweak and QED radiation leak down to the hadronization scale. Instead, factorization and soft-collinear effective theory (SCET)~\cite{Becher:2014oda} isolate the few non-perturbative matrix elements that enter event generators as inputs~\cite{Bierlich:2022pfr,Bellm:2025pcw,Sherpa:2024mfk}, which sharply reduces the quantum resources required~\cite{Bauer:2025nzf}. Two constraints apply. First, hadronization is real-time evolution, not a single local matrix element. Second, the $Q_0$ interface between the parton shower and hadronization is not operator-defined~\cite{Hoang:2024nqi}, so any input that fails $Q_0$-invariance is implicitly absorbed into the tune. This is not hypothetical: even the best Lund~\cite{Andersson:1997xwk} and cluster~\cite{Webber:1983if} tunes still miss the identified-particle spectra of some species at the ${\sim}50\%$ level~\cite{Divisova:2025xqp}.

\minisec{Tensor networks today}
The nearest-term calculations are feasible now (\cref{fig:lamm-ladder}), and should be extended to full $\SU(3)$. String breaking in $1{+}1$D multiflavor $\SU(3)\times U(1)_{\rm em}$ is within reach of present-day MPS and DMRG methods for the ground state and first breaks, and this is the baryon-capable case that lies beyond the usual Abelian $U(1)_{\rm strong}\times U(1)_{\rm em}$ proxy, where real-time multiflavor fragmentation has already yielded hyperon spin correlations~\cite{Barata:2023jgd}. What such a calculation delivers is not a set of tuned constants but the functional forms generators need, together with the variables on which those forms depend, read off from how string breaking changes with time, flavor and mass, and charge~\cite{Gonzalez-Cuadra:2024xul,De:2024smi,Li:2025sgo}. The time dependence yields a dynamical string tension $\kappa(\tau)$~\cite{Hunt-Smith:2020lul,Fischer:2016zzs} that the constant-field Schwinger ansatz cannot capture~\cite{Hebenstreit:2010vz}. Sweeping the quark mass continuously tests whether the Schwinger suppression law $\exp[-\pi\Delta m^2/\kappa]$, with $\Delta m^2=m_s^2-m_u^2$ for strangeness and $m_{qq}^2-m_q^2$ for diquarks, is even the correct functional form~\cite{Skands:2014pea}. Varying the charge exposes the electromagnetic mass splitting and the Coulomb bias that current tunes quietly absorb into the strong sector. Only the long-time evolution and the multiplicity tail lie beyond reach~\cite{Batini:2024zst}, and that is where the quantum computer becomes necessary.

\minisec{Effective $1{+}1$D theory}
Transverse physics does not require fault-tolerant $2{+}1$D hardware. An effective $1{+}1$D theory, obtained by integrating out the transverse directions of an anisotropic Kogut--Susskind lattice and matching the result to the SCET collinear hierarchy~\cite{Bauer:2025nzf}, can carry the \emph{scalar} transverse information into roughly $10^2$ qubits. This includes the $3{+}1$D string tension, the transverse-momentum width $\sigma_{p_T}^2=\kappa/\pi$~\cite{Andersson:1997xwk}, and Casimir scaling, all encoded in spatially and temporally dependent couplings---which may be nonlocal, though clustering keeps any such nonlocality short-ranged---fixed by a small Monte Carlo spectral set as outputs rather than tunes. What this theory \emph{cannot} reach is the differential and topological physics, namely the Collins--Soper kernel and multi-string configurations. The same division recurs within the EFT: its static properties can be validated with tensor networks today, and only the long-time soft function requires a quantum computer.

\minisec{Matching is the rate-limiter}
Simulation is necessary but not sufficient, because without a matching procedure even a perfectly computed number cannot be interpreted. Three layers of matching remain open. The first connects the physical hadrons that a simulation produces to a renormalized $\overline{\rm MS}$ object. The second is $Q_0$-invariance of the shower interface~\cite{Hoang:2024nqi}, which any consistent input must respect. The third is the absence of any operator-level definition of the Lund parameters, a gap that is sharpened by the orphan-quark obstruction to the usual fragmentation sum rules~\cite{Collins:2023cuo}. A realistic first milestone is a closed toy calculation in $1{+}1$D, in which the soft function is computed directly, matched to SCET at leading power~\cite{Becher:2014oda}, and checked perturbatively.

\minisec{Baryon vs.\ meson}
Observables should be ranked by discriminating power weighted by tractability, and on that scale baryon-to-meson production comes first. The competing models here disagree about the underlying mechanism rather than about a single number, since the diquark and popcorn pictures, cluster fission, and the $Y$-junction picture make genuinely different predictions, so a first-principles calculation can falsify one of them. In $1{+}1$D $\SU(3)$ with two or more light flavors a baryon is a Hilbert-space label, the color singlet $qqq$ as opposed to $q\bar q$~\cite{Li:2025sgo}, so the baryon-to-meson ratio follows from the wavefunction with no free parameters, exactly as the vector-to-pseudoscalar ratio $V/PS$ does. The even more discriminating $B\bar B$ rapidity correlation, which places the partners adjacently for a diquark but separates them for popcorn, follows in the same parameter-free way. The observable also spans the difficulty hierarchy: $1{+}1$D already distinguishes diquark from sequential production, whereas the junction topology that requires ${\geq}2{+}1$D drives the $\Lambda_c/D^0$ puzzle~\cite{ALICE:2017thy} and the color-reconnection and rope effects~\cite{Altmann:2024odn,Bierlich:2015rha} seen in the data. We should compute what is tractable now and treat the multi-string puzzles as a later target, not a starting point.

\section{Heavy-Hadron Energy Loss and Physics-Meaningful Benchmarks}\label{app:li-illa}
\contribution
  {Zhiyao Li and Marc Illa}
  {InQubator for Quantum Simulation (IQuS), Department of Physics, University of Washington, Seattle, WA 98195, USA (Z.~Li); Physical Sciences Division, Pacific Northwest National Laboratory, Richland, WA 99354, USA (M.~Illa)}
  {}

The non-equilibrium dynamics of hadronization and the energy loss of heavy hadrons are important areas of study in high energy physics for understanding the aftermath of collisions in particle accelerators.
Quantum simulations of these processes have advanced systematically, from the Schwinger model (1+1D QED)~\cite{Farrell:2024mgu} to the non-abelian SU(2) lattice gauge theories~\cite{Li:2025sgo}.

Reference~\cite{Farrell:2024mgu} works in the Schwinger model and studies how a heavy hadron transfers energy when propagating through a vacuum and a dense medium. As expected, having a finite lattice spacing breaks Lorentz symmetry, which artificially causes energy loss and light-hadron production even when a heavy meson is moving through vacuum at a constant velocity, as well as destroy the entanglement structure.
After mitigating lattice artifacts, the authors studied the evolution of a heavy charge moving across a dense medium, and observed in-medium quantum coherence.

In Ref.~\cite{Li:2025sgo}, the authors studied energy loss and hadronization on a one-dimensional SU(2) lattice. Insights from traditional lattice QCD are heavily used to enable efficient quantum simulations. Hadronic interpolating operators are introduced to address the color entanglement between the heavy and light degrees of freedom. Domain Decomposition is used in the Adapt-VQE state preparation of the heterogeneous initial state with a heavy quark present. The simulations, performed classically and on IBM's quantum computer using 18 qubits, captured the energy loss of heavy quarks moving through a light-quark field and the evolution of local charge densities and multi-partite entanglement.

To transform these foundational steps into robust predictive frameworks, the field should comprehensively study these systems through systematic scans over the parameter space of each theory. It is important to discern physical phenomena from simulation-induced lattice artifacts and make reliable extrapolations to the continuum. 

Furthermore, the community should establish benchmarking problems of physical interest to evaluate various lattice gauge theory formulations, rather than problems specifically designed to maximize qubit counts or Trotter steps on quantum hardware without capturing meaningful physics. For example, energy loss and hadronization could be simulated in other formulations with the same parameter set and compared against the above references, both numerically and resource-wise. 

Moving towards higher spatial dimensions introduces the critical challenge of managing dynamical gauge fields. A progression analogous to that of the 1D quantum simulations can be followed, first starting with Abelian theories to establish how to map and evolve gauge degrees of freedom in compact QED, before constructing non-Abelian theories on digital quantum hardware.

\section{Resource Estimation, Classical Benchmarking, and HPC Integration}\label{app:pascuzzi}
\contribution
  {Vincent R. Pascuzzi}
  {IBM Quantum, IBM T. J. Watson Research Center, Yorktown Heights, NY 10598, USA}
  {}

\subsection{Quantum resource estimation requirements and challenges}

For lattice gauge theory and real-time string dynamics, the relevant resources are not simply the number of qubits, but the complete chain connecting a physics target to an executable calculation: lattice volume and spacing, truncation, encoding, state preparation, Hamiltonian simulation algorithm, circuit depth, observable-estimation cost, and the techniques for error mitigation and correction. This is particularly important for real-time dynamics, string evolution and breaking, and particle production, which are often identified as promising targets for quantum simulation~\citep{Bauer:2022hpo,DiMeglio:2023nsa}. A useful resource estimate must therefore be physics-facing: it should specify the observable, the target uncertainty, the regime of couplings and volumes, and the classical benchmark against which success or failure will be assessed.

The most useful near-term strategy is to organize estimates around a hierarchy of models that interpolate between hardware feasibility and QCD relevance. In (1+1)D, the Schwinger model provides a controlled way to probe confinement, pair production, string breaking, and real-time dynamics; explicit NISQ and fault-tolerant resource estimates for this model were given by~\citet{Shaw:2020udc}. $Z_2$ gauge theory with matter and Abelian Higgs-type models add useful structure because they allow one to test gauge constraints, dynamical matter, and string-breaking observables with compact Hilbert spaces; broader lattice gauge theory simulation approaches are reviewed in~\citet{Banuls:2019bmf}. Moving to (2+1)D introduces plaquette dynamics, transverse string motion, multi-string interactions, and glueball-like excitations, making compact QED, $Z_2$-Higgs, and related quantum link models valuable stepping stones~\citep{Zohar:2021nyc}. For non-Abelian theories, the resource problem becomes more crucial because Gauss-law constraints, local gauge-field truncations, and color structure must be handled without destroying the physics. Loop-string-hadron formulations for SU(2)~\citep{Raychowdhury:2019iki} and related SU(3) constructions~\citep{Ciavarella:2024fzw} illustrate how encoding choices can strongly affect both resource requirements and physical interpretability.

A major challenge is that each layer of approximation contributes to the final physics result. A small gauge field cutoff may reduce qubit count but distort the continuum limit; a shallow Trotter circuit may fit current devices but introduce algorithmic errors; an aggressive error-mitigation strategy may reduce bias for one observable while failing for another. The original Hamiltonian formulation of lattice gauge theory~\citep{Kogut:1974ag} provides the conceptual foundation, but practical quantum implementations require finite-dimensional truncations and hardware-aware mappings. The field therefore needs resource estimates that report not only asymptotic scaling but also the “physics cost” of truncation, digitization, measurement, and noise. The evolution of current superconducting devices toward larger pre-fault-tolerant processors around the 2028 timeframe, using IBM’s public roadmap~\citep{IBMQuantumHardwareRoadmap} as one concrete planning reference, should be used as a guide for what levels of depth, connectivity, and error suppression may become accessible. The strategy should be to use current devices to validate encodings, observables, and error budgets, and update resource estimates as hardware and algorithms improve. Building and maintaining a portfolio of experiments across the model hierarchy will complement analytical estimates with empirical findings.

\subsection{Classical benchmarking and quantum experiments}

Classical benchmarking is the standard candle for quantum simulation. Exact diagonalization remains indispensable for small systems, tensor networks provide powerful baselines for (1+1)D and moderately entangled dynamics, and Euclidean lattice gauge theory remains the standard for many static and equilibrium quantities. Tensor network methods have already produced important results for lattice gauge theories~\citep{Banuls:2019rao}, but their cost can grow rapidly with real-time entanglement and spatial dimension. Time-evolution methods for matrix-product states~\citep{Paeckel:2019yjf}, are therefore both a benchmark and a diagnostic: when they fail, one learns something about the onset of classical difficulty. Recent analyses of large superconducting-qubit experiments also show that classical baselines can improve quickly; the quantum utility experiment of \citet{Kim:2023bwr}, motivated subsequent tensor network studies such as~\citet{Tindall:2023cqi}, underscoring that quantum results must be benchmarked against the best classical methods available at the time of publication, not against static estimates or gross assumptions.

The hardest validation problem arises when a quantum calculation enters a regime with no efficient classical analog. In this regime, one cannot rely on full state-vector comparison or exact reproduction of the measurement distribution -- verification must be layered. At small volumes, one should compare directly to exact diagonalization. At intermediate volumes, one should compare to tensor networks, Pauli propagation, semiclassical limits, or controlled truncations, etc. where possible. At all volumes, one should test internal consistency, e.g., Gauss-law preservation, charge conservation, unitarity, convergence with Trotter step and cutoff, and agreement among independent observable families. For general issues in quantum certification and benchmarking, see~\citet{Eisert:2020ckk} and on the difficulty of verifying computations that are themselves classically hard,~\citet{Gheorghiu:2018eif}. For HEP and NP simulations, observables such as entanglement growth, operator spreading, and quantum magic can serve a dual role: they are physics-motivated diagnostics and indicators of when tensor network or stabilizer-based methods may cease to be efficient. The absence of a complete classical analog should therefore raise the standard for validation.

Existing quantum experiments provide useful templates, but they should be interpreted as methodological milestones rather than final physics applications; e.g., the trapped-ion Schwinger model experiment of~\citet{Martinez:2016yna} demonstrated real-time lattice gauge theory dynamics in a controlled few-qubit setting, and the self-verifying variational quantum simulation of lattice gauge theory by~\citet{Kokail:2018eiw} emphasized the importance of verification built into the experimental design. On superconducting devices, Schwinger model and $Z_2$ gauge theory studies~\citep{Pomarico:2023png,Charles:2023zbl} have explored real-time dynamics, noise sensitivity, and mitigation strategies. Larger recent demonstrations~\citep{Farrell:2024fit,Cobos:2025krn,Ilcic:2026cac} show how the field is beginning to connect experiments with gauge-theory observables at scale. These experiments add value as they test encodings, hardware mappings, mitigation, measurement protocols, and classical validation methods that will be needed as devices approach the pre-fault-tolerant regime.

\subsection{Integration of quantum into classical software and HPC}

Quantum computing for HEP and NP should be developed as an accelerator embedded in the classical scientific-computing ecosystem, not as a replacement for it. Event generation, lattice field theory, perturbative calculations, Monte Carlo sampling, detector simulation, and uncertainty quantification are already mature in experimental workflows. Quantum processors will be useful only if they can enter those workflows as specialized resources for subproblems involving coherent real-time evolution, highly entangled amplitudes, or sign-problem-limited dynamics. The Snowmass white paper from~\citet{Humble:2022vtm} identifies this ``systems and software'' layer as a central requirement for making quantum computing useful to HEP rather than merely demonstrative, while ~\citet{DiMeglio:2023nsa} similarly emphasizes benchmark applications, resource estimates, and integration with existing HEP computational practice.

The software requirements are substantial; HEP and NP needs common representations for Hamiltonians, gauge constraints, truncation schemes, observables, and error budgets. Circuit generation tools must interoperate with exact diagonalization, tensor network, Pauli-propagation, noisy-simulation, and hardware-execution backends. Experimental data products must preserve enough provenance to be reproducible: calibration data, transpilation choices, hardware topology, mitigation procedures, shot counts, uncertainty estimates, and validation tests. These requirements mirror broader HEP software challenges, where reproducibility, workflow management, and long-term maintainability are already central concerns~\citep{HEPSoftwareFoundation:2017ggl}. For quantum simulation, the additional challenge is that the “accelerator” is probabilistic, noisy, and rapidly evolving, so the software stack must record not only the requested physics calculation but also the device-dependent path by which the result was obtained.

The period leading to pre-fault tolerance should be used to build these workflows before large logical resources are available. A useful end-to-end pipeline would begin with a physically-motivated Hamiltonian, generate a resource estimate, compile circuits for a target architecture, run classical baselines, execute on quantum hardware, apply verification and mitigation procedures, and publish the resulting observables with uncertainties. As hardware improves, the same pipeline can be used to increase volume, dimension, gauge-group complexity, and real-time depth. IBM’s roadmap~\citep{IBMQuantumHardwareRoadmap} provides one concrete planning horizon for this evolution toward more capable noisy and pre-fault-tolerant processors around 2028, but the infrastructure should be portable across hardware platforms. The long-term impact of near-term experiments will therefore come not only from individual results, but from establishing reusable quantum-HPC workflows that allow HEP and NP applications to move seamlessly from today’s devices to early fault-tolerant systems.

\section{Progress towards Hybrid Quantum Hadronization for Event Generators}\label{app:senseman}
\contribution
  {Blake Senseman, Zane Ozzello, Kenneth Heitritter, Yannick Meurice, and Stephen Mrenna}
  {The University of Iowa, Iowa City, IA 52242, USA; Fermi National Accelerator Laboratory, Batavia, IL 60510, USA; qBraid Co.}
  {FERMILAB-PUB-22-912-SCD}
\subsection{Hadronization Models in Event Generation}

The power of the standard model of particle physics is that it provides a systematic mathematical description of physics at short distances, 
where the relevant degrees of freedom are partons: quarks, gluons, leptons, and photons.    
However, experiments observe strongly-interacting particles: mesons, baryons, exotic hadrons, and nuclei.
To connect to our experimental world, the standard model calculations are merged with phenomenological models that predict hadronic properties from partonic ones (hadronization).
While it is believed that an ab initio (numerical)
QCD calculations at long distance scales will eventually provide a theory-based, rather than a phenomenological, model of hadronization, quantitative predictions are still not available.
To date, the phenomenological models are of two types: one \cite{Bellm:2015jjp,Divisova:2025xqp} that collapses local, color singlet combinations of partons into clusters that undergo simple decays into hadrons, and another \cite{Andersson:1983ia,Sjostrand:2006za,Bierlich:2022pfr} that connects runs of partons into strings that have complex fragmentation properties.   Neither approach is entirely satisfactory:   cluster models over time have acquired string-like properties, while string models have acquired cluster-like properties.

By construction, the cluster model is simple.   On the other hand, we hope to gain some insight into the complexity of long-range objects like the string by studying confining field theories.
The Lund string is a classical, massive, relativistic string that is allowed to decay.   Quantum mechanics motivates this
decay process through 
the tunneling of quark-antiquark pairs from the vacuum in the presence of a strong field (the Schwinger mechanism).   In the model, the decay is stochastic, driven by a fragmentation function, whose form is fixed by demanding that the fragmentation process is symmetric with respect to the quark or antiquark end of the string.  In its simplest form, the fragmentation function has two parameters, one that exponentially suppresses string decays proportional to the space-time area of the string and another that penalizes large steps in the decay.   
Quantum mechanics is used to motivate two other features:  first,
only light quarks $(u,d,s)$ -- and diquarks for baryon production -- are allowed to appear as end points in the string decay, with a tunable suppression of $s$ quarks and diquarks; and, second, momentum transverse to the longitudinal field is introduced to drive the tunneling process.

In practice, the Lund string model is embedded in a Monte Carlo
simulation of particle collisions (or decays) at high momentum transfer.   The full simulation includes fixed-order perturbative calculations, resummation of soft and collinear gluons, and models of various long-range phenomenon.   
The perturbative elements (including resummation) are under
quantitative control and provide the standard model predictions that are compared to data.   However, before those comparisons
can be made, a model is needed to translate the partons to
hadrons.  At this stage of the simulation,
string decay is described by a probability model
that parametrizes our ignorance about the underlying dynamics.
While motivated by quantum mechanics, the description is
classical -- each splitting is assigned a conserved energy momentum and production vertex.   The hadronization models have evolved over the past forty years, but they still lack a 
grounding in quantum calculations.

\subsection{What do we need from quantum simulation?}
Building an ab-initio quantum simulator dealing with the entire time evolution of partons is an overambitious program with currently available resources. In a more feasible near-term approach, one should work separately on the the different modules of the event generators which appear chronologically as: 1) short-distance cross sections, 2) parton showers, 3) multiple interactions, 4) hadronization/string breaking and 5) particle decays. Here we focus on part 4). Quantum aspects of 2) and 3) are discussed in \cite{Bauer:2019qxa} and SCET contributions to this white paper.

The overarching goal is to be able to fully capture the effects of string breaking with quantum device implementation.  Furthermore, this would ideally be able to capture reality better than current methods of hadronization in event generators.  As of now, there is promising work which is showing the ability to capture string breaking on quantum device implementation; however, it is at the small scale.  For full integration into event generators, there will need to be larger system sizes for better capturing physical phenomena.  With current devices, quantum hardware imposes limitations on accuracy and system size, which makes the reliability of practical integration with an event generator infeasible at this time.  
So, something different is needed in the intermediate time to still push progress in quantum simulation\cite{Cochran:2024rwe,Gonzalez-Cuadra:2024xul,De:2024smi,Ciavarella:2024lsp,Gupta:2026tcg,Mildenberger:2022jqr,Alexandrou:2025vaj,Cobos:2025krn,Xu:2025abo,Borla:2025gfs,Cataldi:2025cyo,Liu:2024lut}, and provide an accessible play ground to ready researchers for the integration.  

Arguably the best approach at current times for models composed of chains and rectangular arrays are tensor methods for quantum simulation. 
This is an incredibly diverse class of computational methods for physics that can be tailored and chosen based on the needs of simulation goals.  This actually has twofold ramifications.  One, are the aforementioned computational algorithm considerations.  Second, this plays in directly with the need to mindfully select the model based on the sort of phenomena or limitations that are involved.  Naturally, choosing different models will yield different physical results with different manifestations of their phenomena.  Beyond model selection, there are also necessary decisions for overall system size, representations of a model, the spatial dimensionality of a model, and the parameter range of the model to be simulated under.  All of these can impact the selection of simulation methods.  Minimal system size can motivate a selection of an algorithm that maximizes speed with accuracy based on system size, i.e. perhaps it is easier to just stick with exact diagonalization over tensor methods if the system size is small enough.  Further more, such methods provide excellent means of validating approximate methods.  The jump from one to two spatial dimensions has a huge impact on tensor method selection.  In 1D it is easy enough to stick with DMRG methods\cite{White:1992zz,Schollwock:2005zz}, but these methods are strained in 2D, so then something more friendly for 2D PEPS\cite{Verstraete:2006mgt} may be a more fruitful avenue.  Different parameter ranges cause different entanglement spectra for models, and entanglement is often a limiting factor in tensor method simulation, so algorithmic choice can often be chosen to better suit the entanglement situation of the model.

\subsection{Background to Current Progress}
The physics input necessary to design and understand event generator codes could a discouraging factor to introduce quantum aspects in a modular way. However, it should be appreciated that the Lund model is relatively simple.   Trying to replace it by an ab-initio lattice model in 1+1 or 2+1 dimensions should be achievable in the near term. In the following we consider a very simple example of model involving gauge and bosonic matter.

The lattice compact Abelian Higgs model is a non-perturbative regularized formulation of scalar quantum electrodynamics 
(scalar electrons-positrons  and photons, all with compact fields). It is very simple model with gauge and bosonic  matter where a linear potential develops between a pair of particle-antiparticle, and where Gauss's law can be implemented in a trivial way, namely by eliminating the matter degrees of freedom in terms of the gauge degrees of freedom. In the continuum, this corresponds to the unitary gauge. 

In the rest of this section, we define the lattice model, discuss its exact tensor reformulation in terms of plaquette quantum numbers and introduce the Hamiltonian with the transfer matrix. 
\subsubsection{The compact Abelian Higgs Model}
Using an Euclidean space-time lattice, the partition function reads
\beq
Z_{CAHM}=\prod_x\int_{-\pi}^{\pi}\frac{d\varphi _x}{2\pi}\prod_{x,\mu}\int_{-\pi}^{\pi}\frac{dA_{x,\mu}}{2\pi}
e^{-S_{gauge}-S_{matter}}, 
\label{eq:senseman-gaugemeasure}
\enq
with
\beq
S_{gauge}=\bpl\sum_{x,\mu<\nu} (1-\cos(A_{x,\mu}+A_{x+\hat{\mu},\nu}-A_{x+\hat{\nu}, \mu}-A_{x,\nu})),
\label{eq:senseman-gauge}
\enq
and 
\beq
\label{eq:senseman-smatter}
S_{matter}=\beta_{link} \sum\limits_{x,\mu} (1-\cos(\varphi_{x+\hat{\mu}}-\varphi_x+A_{x,\mu})).\enq
The second term is a gauged version of the O(2) model. We have  the local invariance: 
$\varphi_x'=\varphi_x+\alpha_x$ and 
$A_{x,\mu}'=A_{x,\mu}-(\alpha_{x+\hat{\mu}}-\alpha_x)$.
It is clear that we can fix the gauge in such a way that $\varphi$ is eliminated (this is the unitary gauge), however, we will not need to fix the gauge when we perform a full integration of the compact variables. 
Note that 
$\varphi$ is the Nambu-Goldstone mode of the original model where a complex scalar field is writen as $\Phi=|\Phi|{\rm e}^{i\varphi}$ with 
Brout-Englert-Higgs (BEH) mode is $|\Phi|$ is frozen to its expectation value because we assume that it has a very large mass and decouples from the low energy effective theory. The limit where the BEH mode decouples is discussed in \cite{Bazavov:2015kka}.

\subsubsection{Tensor reformulation}
Following the Tensor Lattice Field Theory strategy \cite{Meurice:2020pxc} we will replace the continuous integration by character expansions (here Fourier series): partition function and averages become {\it discrete} sums of contracted tensors. The ``hard" integrals are done exactly and field integrations provide Kronecker deltas that encode the symmetries.
For continuous field variables, the sums are infinite, but truncations to finite sums do not break the symmetries \cite{Meurice:2019ddf,Meurice:2020gcd}.

We now use Fourier expansions and perform the field integrations explicitly. 
On the links, we use \begin{eqnarray}{\rm e}^{\bl  \cos(\varphi_{x+\hat{\mu}}-\varphi_x+A_{x,\mu})}& = \\
           \sum\limits_{n_{x,\mu}=-\infty}^{+\infty} {\rm e}^{i n_{x,\mu}(\varphi_{x+\hat{\mu}}-\varphi_x+A_{x,\mu})} I_{n_{x,\mu}}(\bl)\  , \end{eqnarray}
The $\varphi$ integration provides the O(2) selection rule: \beq
\sum_\mu[-n_{x,\mu}+n_{x-\hat{\mu},\mu}]=0.\   \  
\label{eq:senseman-noether}
\enq
Plaquettes: ${\rm e}^{\bpl  \cos(A_{x,\mu}+A_{x+\hat{\mu},\nu}-A_{x+\hat{\nu}, \mu}-A_{x,\nu})} = 
          \sum\limits_{m_{x,\mu,\nu}=-\infty}^{+\infty} {\rm e}^{i m_{x,\mu,\nu}(A_{x,\mu}+A_{x+\hat{\mu},\nu}-A_{x+\hat{\nu}, \mu}-A_{x,\nu})} I_{m_{x,\mu,\nu}}(\bpl),\  \nonumber     $
\vskip3pt

$A_{x,\mu}$ integration provides the selection rule:

 \beq \label{eq:senseman-maxwell} \sum_{\nu>\mu}[m_{x,\mu,\nu}-m_{x-\hat{\nu},\mu,\nu}]+\sum_{\nu<\mu}[ -m_{x,\nu,\mu}+m_{x-\hat{\nu},\nu,\mu}]+n_{x,\mu}=0.\  \  \enq
 Note that (\ref{eq:senseman-maxwell}) implies (\ref{eq:senseman-noether}), this is a discrete version of $\dmu \partial_\nu F^{\mu \nu}=\dmu J^\mu= 0$.). This means that the gauge quant. numbers $m$ determine matter $n_{x,\mu}(\{m\})$. 
 This is consistent with the idea that in the unitary gauge, $\varphi$ disappears.

We have now electric quantum numbers attached to the space-time plaquettes and magnetic quantum numbers associated with the space-space plaquettes (for dimensions 2+1  and higher). In addition we have tensors attached to links which connect these quantum numbers according to Eq. (\ref{eq:senseman-maxwell}). The partition function can be written  as the sum over the gauge indices with weights that are products of Bessel functions. 

\subsubsection{Transfer matrix and Hamiltonian}
We can organize the trace (sum over all discrete indices) in the expression of $Z$ by assembling ``time layers"  corresponding to ``magnetic" time slices and  ``electric" slices half-way between the magnetic time slices. 
For a graphical illustration of this ``lasagna" stucture in 2+1 dimensions, see Ref. \cite{Meurice:2020gcd}. 
This construction singles out a time direction as for the Hamiltonian treatment. The partition function reads
\beq    
   Z=   Tr { \mathbb T}^{N_\tau},
      \enq
with $N_\tau$ the number of sites in the temporal direction. 
For numerical study of deconfinement from thermal tensor networks using this construction see Ref. \cite{Naravane:2026zjk}. By taking the time continuum limit of the transfer matrix, one can derive the Hamiltonian \cite{Bazavov:2015kka,Meurice:2020pxc,Meurice:2020gcd}.

\subsection{Current progress}

\subsubsection{Static string phenomenology in the Compact Abelian Higgs Model \cite{Senseman:2026pif}}

\textbf{Model.} On a 1D chain of spin-1 (qutrit) links with truncated electric-field quantum number $m\in\{-1,0,1\}$, the effective Hamiltonian \cite{Meurice:2021pvj,Zhang:2018ufj,Asaduzzaman:2026jyi} is
\[
H_{CAHM}=\frac{U}{2}\sum_i (L_i^z)^2+\frac{Y}{2}\!\!\sum_{\langle ij\rangle}\!(L_i^z-L_j^z)^2-\frac{X}{\sqrt2}\sum_i L_i^x ,
\]
with $X\!=\!1$ fixing the energy scale, $U$ the per-site field (electric) energy, and $Y$ a mass-coupling that penalizes domain walls (the implicit charged matter); $L^x$ alone drives string motion and breaking. Restricting to the two-domain-wall sector and appropriate boundary conditions isolates a 1D lattice-string Hilbert space, block-diagonalized exactly in a string-meson momentum basis $(k,\ell)$ via Bessel-function eigenstates in the confining ($Y\!\to\!\infty$) limit \cite{Surace:2020ycc}.

\textbf{Method — chemical-potential state preparation.} Rather than inserting explicit external charges \cite{Liu:2024lut,Gonzalez-Cuadra:2024xul,Gupta:2026tcg,Surace:2024bht}, we coax the field into a string configuration with a local perturbation, $H'=H_{CAHM}-\mu L^z_{\{s\}}$, applied at two sites separated by a distance $L$, and minimize with DMRG ($\sim\!10^2$ sites). This is a generic, Gauss-law-compatible protocol for any truncated gauge-link model: it isolates a controllable order parameter (here $\langle L^z\rangle$) without altering the Hilbert space. Critically, $\mu$ can be adiabatically ramped down through successive DMRG rounds to relax the hard-wall string into a physically dressed, minimum-energy state (edge peak suppressed, virtual domain-wall pairs restored) suitable as an initial state for future real-time string-breaking simulations. This minimizes spectral distortion from the perturbation and generalizes directly to other 1+1D (and prospectively non-Abelian) truncations targeted elsewhere at this workshop.

\begin{figure}[t!]
    \centering
    \includegraphics[width=0.47\linewidth]{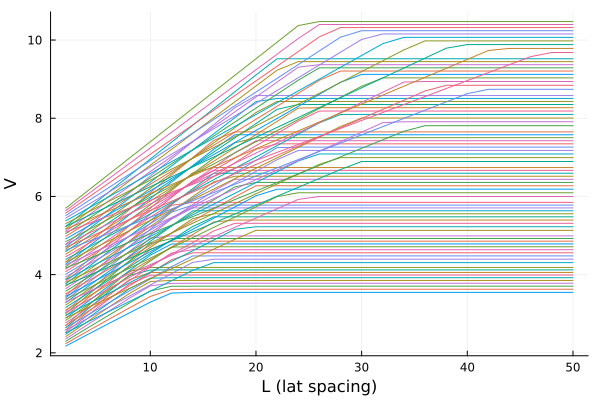}
    \hfill
    \includegraphics[width=0.47\linewidth]{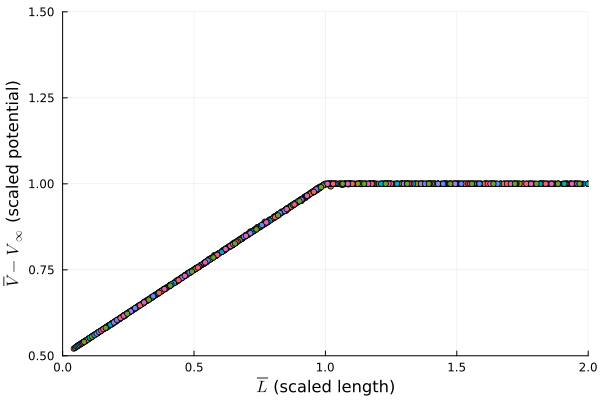}
    \caption{\footnotesize (a) Measured string potentials $V(L; U,Y)$ (left) for $Y \in (2.0, 2.5, ..., 6.0), U \in (0.1, 0.125, ..., 0.35)$ obtained with $\mu=100$. The ground state energy $E_0(U,Y)$, calculated with $\mu=0$, has been subtracted from each sample. (b) Rescaling potentials (right) demonstrates the universal nature of the string behavior once $\sigma$ and $L^*$ are observed. A suitable vertical shift has also been performed to account for the discrepancy between $M_{eff}$ and $\sigma L^*$. From \cite{Senseman:2026pif}.}
    \label{fig:senseman-potentials}
\end{figure}

\textbf{Results.} Scanning $Y\!\in\![2,6]$, $U\!\in\![0.1,0.35]$, the excess energy $V(L;U,Y)$ above the unperturbed ground state grows linearly with distance before saturating, as in \cite{Bergner2026AdjointQCD,Crippa:2024hso,Buyens:2015tea,Surace:2024bht}. Piecewise-linear fits yield the string tension $\sigma$ (slope) and breaking length $L^*$ (knee) [Fig. \ref{fig:senseman-potentials}(a)]. Rescaling $\bar V\equiv V/(2\sigma L^*)$, $\bar L\equiv L/L^*$ collapses \emph{all} sampled $(U,Y)$ onto a single universal curve [Fig. \ref{fig:senseman-potentials}(b)], indicating the linear-potential/string-breaking transition is governed by a single emergent scale over the full parameter range studied. Three independent mass measures — the finite-volume gap $\Delta_m$, the localized (chemical-potential) excitation energy $M_{eff}$, and $M_\sigma\equiv\sigma L^*$ — are each linear in $Y+U/2$ and converge onto one another as $Y$ grows, with clean power-law residuals, $M_{eff}/\Delta_m\approx 1+1.22\,Y^{-1.5}$ and $M_\sigma/M_{eff}\approx 1-1.41\,Y^{-2.5}$. Equivalently, the semiclassical string-breaking criterion $M_{eff}+\sigma L^*=2M_{eff}$ is satisfied to $<\!1\%$ by $Y=6$ (from 12\% at $Y=2$), \emph{independent of $U$}. This demonstrates a quantitative, microscopically-derived bridge between the lattice Hamiltonian and Lund-type fragmentation phenomenology.

\subsubsection{Real-Time Evolution of Multiplicity and Entanglement in a Rydberg Simulator \cite{Senseman:2022jpu}}
 
\textbf{Model.} Neutral atoms with ground/Rydberg states $|g\rangle,|r\rangle$ evolve under 
\begin{equation*}
    H=\tfrac{\Omega}{2}\sum_j(|g_j\rangle\!\langle r_j|+\text{h.c.})-\Delta\sum_j n_j+\sum_{j<k}V_{jk}n_jn_k
\end{equation*}
with Rabi frequency $\Omega$, detuning $\Delta$, and van der Waals repulsion $V_{jk}=C_6/|r_j-r_k|^6$ (Rb-87, $|r\rangle=|70S_{1/2}\rangle$). On a $2\times N_s$ ladder (inverse aspect ratio $\rho=2$), a staggered electric field $E_j\equiv(-1)^{j+1}(n_{j,1}-n_{j,0})\equiv \hat L_j^z$ maps the Rydberg blockade dynamics onto a spin-1 chain effective Hamiltonian structurally close to the truncated CAHM above \cite{Meurice:2021pvj,Zhang:2023agx}: a Rabi/hopping term, a $(\hat L_i^z)^2$-type detuning term, and long-range $\hat L_i^z\hat L_{i+k}^z$ and $(\hat L_i^z)^2(\hat L_{i+k}^z)^2$ couplings set by the on-leg and cross-leg Rydberg interactions. A residual cross-Kerr projector term \cite{Malekakhlagh:2020dgw,Blok:2020may} is the sole obstruction to an exact CAHM correspondence, inducing mild string-potential nonlinearity and weak inter-string repulsion, which are not \textit{prima facie} obstacles to our event generation goals. Gauss's law follows from the blockade itself: $Q_{i,i+1}=E_{i+1}-E_i=\pm1$ defines charge/anticharge pairs, automatically separated by constant-field string regions, since blockade forbids the $Q=\pm2$ configurations.

 \begin{figure}[h!]
    \centering
    \vskip3pt
    \includegraphics[width=0.6\linewidth]{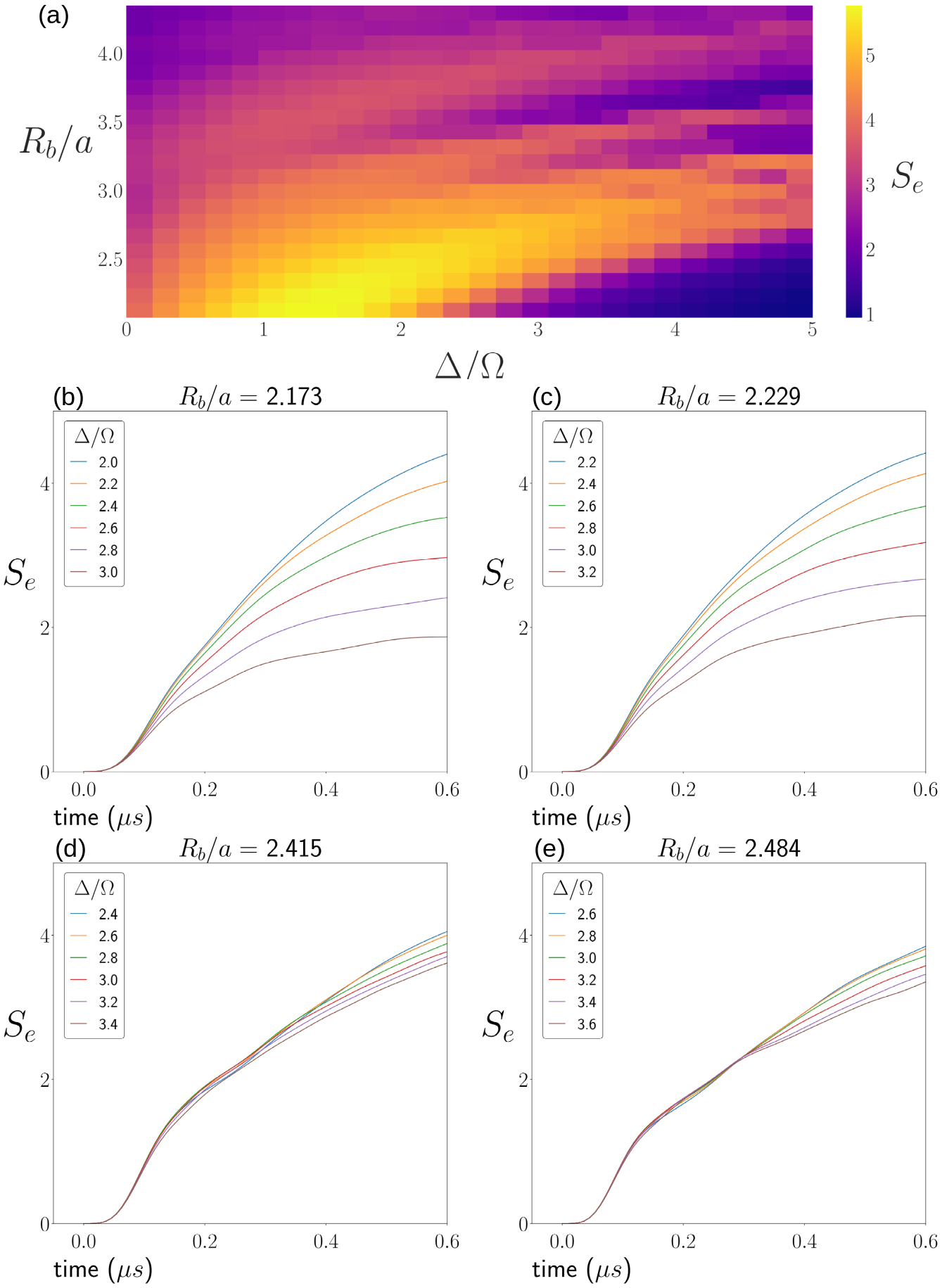}
    \caption{\footnotesize Maximum half-chain entanglement entropy $S_e$ over a $1.8~\mu$s evolution of a 13-rung ladder, swept over blockade ratio $R_b/a$ and detuning $\Delta/\Omega$; suppression toward larger $\Delta/\Omega$ signals confinement. From \cite{Senseman:2022jpu}.}
    \label{fig:senseman-entropy}
\end{figure}

\textbf{Method - Exact-blockade Rydberg simulation.} Initial states (single central $E=-1$ site, a minimum-size hadron with large energy variance) are prepared by adiabatic local detuning of one atom, then quenched to global dynamical parameters $(\Omega,\Delta)$ and evolved via exact real-time simulation enforcing the blockade as a hard constraint, with $\Omega=4\pi$~MHz, $R_b/a=2.173$ chosen to suppress doubly-occupied rungs while remaining Aquila-compatible ($\Omega\le5\pi$~MHz, spacing $\ge4~\mu$m). Three diagnostics, each previously correlated with confinement in spin-chain/LGT studies, are tracked post-quench: (i) local field $\langle E_j\rangle$ spreading, (ii) half-chain von Neumann entanglement entropy $S_e$, and (iii) charge-pair multiplicity, i.e.\ the number of constant-$E$ string domains — the model's direct analog of hadron count.
 
\begin{figure}[h!]
    \centering

    \includegraphics[width=0.6\linewidth]{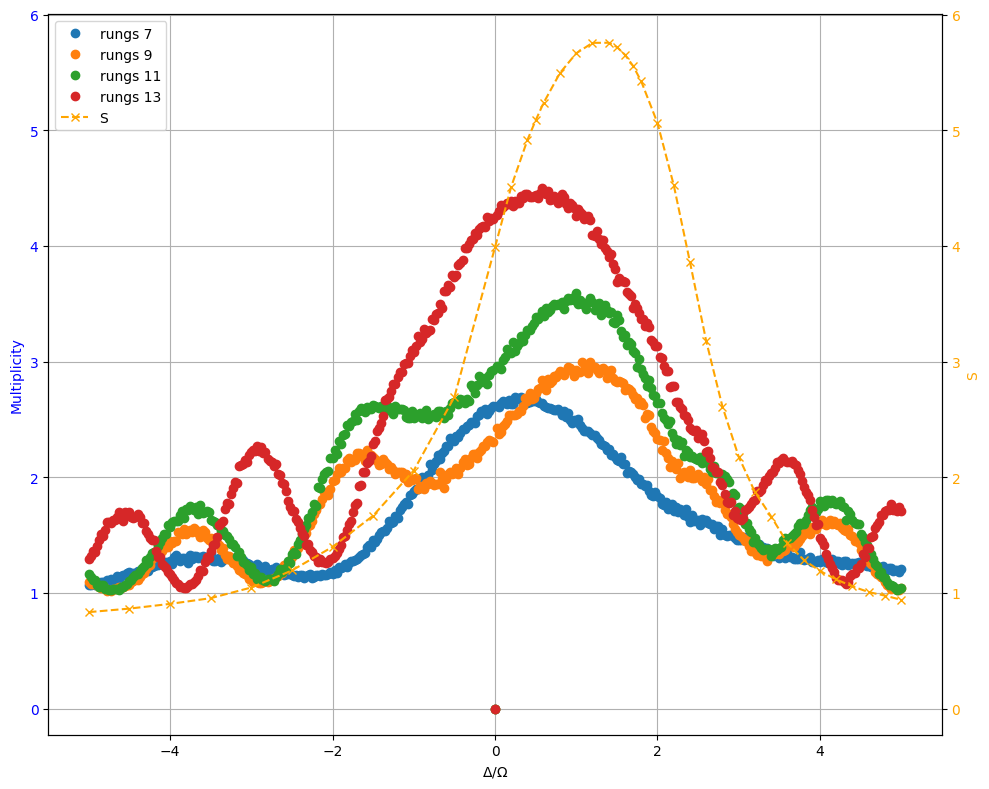}
    \caption{\footnotesize Average hadron multiplicity vs.\ $\Delta/\Omega$ for $N_s=7$--13 rungs at $R_b/a=2.173$, $\Delta t=0.35~\mu$s (dashed: 13-rung entropy, right axis), showing multiplicity growth tracking entanglement proliferation. From \cite{Senseman:2022jpu}.}
    \label{fig:senseman-multiplicity}
\end{figure}

\textbf{Results.} Field spreading is ballistic (constant-velocity, $\Delta/\Omega$-independent) at low detuning and increasingly localized as $\Delta/\Omega$ grows. A broad sweep over $(R_b/a,\Delta/\Omega)$ reproduces phase structure resembling known Rydberg-ladder phase diagrams [Fig. \ref{fig:senseman-entropy}], with entropy-growth suppression at fixed $R_b/a$ as $\Delta/\Omega$ increases — the established entanglement signature of confinement (Calabrese--Cardy-type behavior) \cite{Calabrese:2005in,Fagotti:2008mlc,Kormos:2016osj,Vovrosh:2020nwb}. Multiplicity is non-decreasing from its minimal initial value of 1, saturating well below the ladder's $(N_s+1)/2$ capacity, and depends non-monotonically on $\Delta/\Omega$ with a broad central maximum that co-localizes with the entropy-growth peak across $N_s\in\{7,9,11,13\}$ [Fig. \ref{fig:senseman-multiplicity}] — direct evidence linking field/charge mobility to particle proliferation, and a tunable knob (via $\Delta/\Omega$) for controlling simulated multiplicity, which is one key phenomenological target of Lund-model tuning in \textsc{Pythia}.

\section{String Breaking}\label{app:yao}
\contribution
  {Xiaojun Yao}
  {InQubator for Quantum Simulation (IQuS), Department of Physics, University of Washington, Seattle, WA 98195, USA}
  {}

It will be interesting to compute Lund model parameters that have been used in Pythia via quantum simulation. For example, one parameter mentioned during a workshop discussion is the dynamical string tension, which is a fitted parameter in Pythia and differs from the static string tension calculated via the lattice QCD method. It will also be interesting to study string-breaking dynamics in a nuclear environment. This can help us understand jet and hadron productions in heavy-ion and electron-ion collisions.

\section{Quantum Complexity as Event-Generator Input}\label{app:zemlevskiy}
\contribution
  {Nikita A. Zemlevskiy\,\orcidlink{0000-0002-0794-2389}}
  {InQubator for Quantum Simulation (IQuS), Department of Physics, University of Washington, Seattle, WA 98195, USA; Institute for Quantum Information and Matter, California Institute of Technology; Department of Physics, California Institute of Technology}
  {}

Conventional event generators encode fragmentation through phenomenological functions calibrated to data. 
While these models have proven to be successful across a range of observables, the incorporation of quantum information inputs may increase their predictive power.
Recent work demonstrates that measures of quantum complexity, such as entanglement and non-stabilizerness, offer a complementary, model-independent angle for studying this content~\cite{Robin:2026lqp,Grieninger:2026bdq}. 
Quantum complexity has been shown to act as a diagnostic of truly quantum effects absent from traditional models of hadronization and string breaking.
It has been suggested to be an organizing principle for the structure and dynamics of hadrons and matter. 
A natural extension of these works is to study similar quantities in other processes and models relevant to the physics of event generation.
If classical simulations of tractable models (such as Matrix Product State simulations) show similar trends revealed by quantum complexity, this would strongly suggest that these quantities offer complementary information not captured in current event generation algorithms.
For instance, if quantum complexity measures turn out to be sensitive to string tension or fermion mass, or sensitive to dynamics, they can be used as tunable inputs to event generators. 
In addition, certain quantum complexity measures can be extracted directly from experimental data as already demonstrated for collider final states (see e.g., Ref.~\cite{White:2024nuc}, and the references within Ref.~\cite{Grieninger:2026bdq}). 
Generalizing this kind of analysis from simple spin degrees of freedom to high-multiplicity hadronic final states would, in principle, let measured complexity observables inform novel inputs to event generation directly. 
While precise quantitative inputs currently remain out of reach, the qualitative behavior of quantum complexity could motivate structural choices that may improve predictive power against future data.

A specific instantiation of this program is to use entanglement structure not merely as a global diagnostic but as a tool for dissecting individual fragmentation channels.
A recent real-time approach isolates exclusive scattering channels in Matrix Product State simulations directly from the entanglement structure of the late-time wavefunction~\cite{Zemlevskiy:2026kpc}.
Schmidt decompositions at spatial bipartitions of the post-collision state separate elastic from inelastic contributions and enable deterministic detection of outgoing particles of specific species, without relying on prior knowledge of the asymptotic particle wavefunctions. 
Applied to hadronization, this offers a structured way to decompose the coherent superposition produced by a breaking string into its constituent channels and to read off, channel by channel, the relative weights and kinematics that fragmentation models otherwise treat through inclusive parametrization.
Because the technique reconstructs branching information from the simulated quantum state itself rather than from assumed final states, it could supply qualitative guidance on channel composition and correlations produced final states. 
This information could lead to event generation inputs that conventional methods neither expose nor constrain.

Direct quantum simulations of real-time dynamics are becoming practical for models relevant to hadronization and string breaking. 
Substantial first steps toward producing actionable inputs to event generation already exist.
Demonstrations of state preparation methods~\cite{Farrell:2024fit,Davoudi:2024wyv} and  simulations of elastic collisions~\cite{Zemlevskiy:2024vxt,Hite:2025pvb,Chai:2023qpq,Chai:2025kbi} have been followed by the first quantum simulations of particle production~\cite{Schuhmacher:2025ehh,Farrell:2025nkx}.
Alongside these, early demonstrations of string breaking on quantum hardware and simulators establish that the qualitative dynamics of confinement and pair creation are now accessible~\cite{Crippa:2024hso,Liu:2024lut,De:2024smi,Surace:2024bht,Ciavarella:2024lsp,Alexandrou:2025vaj,Luo:2025qlg,Cataldi:2025cyo,Xu:2025abo,Xu:2026ibi,Joshi:2026hfe,Cochran:2024rwe,Gonzalez-Cuadra:2024xul,Borla:2025gfs,DiMarcantonio:2025cmf}.
Expanding beyond demonstrations to systems that share properties with continuum physics requires system sizes appreciably larger than those reachable today~\cite{Grieninger:2026bdq}.
The near-term value of these simulations is therefore qualitative rather than quantitative.
They would establish which dynamical features (e.g., the timing and locality of string breaking, the buildup of quantum correlations, the multiplicity and species of produced particles) are robust and physically meaningful. 
As hardware scales and algorithms mature, the same pipeline that today yields qualitative inputs to event generation models is positioned to deliver quantitative fragmentation inputs derived from first-principles dynamics rather than phenomenological tuning.

\bibliographystyle{apsrev4-2}
\bibliography{refs}

\end{document}